%% file: main.tex
\documentclass[10pt,sigconf,letterpaper,nonacm]{acmart}
\usepackage{booktabs}
\usepackage{multirow} 
\usepackage{amsmath}  
\usepackage{tabularx} 
\usepackage{adjustbox} 
\usepackage{caption}  
\usepackage{xcolor}
\usepackage{subcaption}
\usepackage{algorithm}
\usepackage{algorithmic}
\usepackage{xspace}

\usepackage[inline]{enumitem}
\setlist[itemize]{noitemsep, topsep=5pt}

\begin{document}

\title{A Truckload of Satoshis: Detecting and Measuring One-Way Arbitrage in the Wild}

\author{Eugenio Nerio Nemmi}
\email{nemmi@di.uniroma1.it}
\affiliation{%
  \institution{Sapienza University of Rome}
  \city{Rome}
  \country{Italy}
}

\author{Tobias Lauinger}
\affiliation{%
  \institution{New York University}
  \city{New York}
  \country{USA}
}

\author{Paz Grimberg}
\affiliation{%
  \institution{New York University}
  \city{New York}
  \country{USA}
}

\author{Massimo La Morgia}
\email{lamorgia@di.uniroma1.it}
\affiliation{%
  \institution{Sapienza University of Rome}
  \city{Rome}
  \country{Italy}
}

\author{Damon McCoy}
\affiliation{%
  \institution{New York University}
  \city{New York}
  \country{USA}
}

\author{Alessandro Mei}
\email{mei@di.uniroma1.it}
\affiliation{%
  \institution{Sapienza University of Rome}
  \city{Rome}
  \country{Italy}
}

\providecommand{\ie}{\emph{i.e.,} }
\providecommand{\eg}{\emph{e.g.,} }
\providecommand{\etc}{\emph{etc.}} 

\newcommand\mypara[1]{\noindent \textbf{#1}}

\providecommand{\MT}{maker/taker\xspace}
\providecommand{\TT}{taker/taker\xspace}

\providecommand{\ione}{\emph{(i)} }
\providecommand{\itwo}{\emph{(ii)} }
\providecommand{\ithree}{\emph{(iii)} }
\providecommand{\ifour}{\emph{(iv)} }
\providecommand{\ifive}{\emph{(v)} }
\providecommand{\isix}{\emph{(vi)} }

\input{variables}
\input{sections/00-abstract}

\maketitle

\input{sections/01-introduction}
\input{sections/02-background}
\input{sections/03-data-collection}
\input{sections/04-owa-detection}
\input{sections/05-analysis}
\input{sections/06-discussion}
\input{sections/07-related-work}
\input{sections/08-conclusion}

\begin{acks}
Funding for this work was provided in part by MUR National Recovery and Resilience Plan, SERICS (PE00000014); the ``Young Researchers 2024-SoE'' Program funded by the Italian Ministry of University and Research (MUR) ST3P under Grant B83C24003210001; TRACE: Taint-based Risk Analysis for Crypto Ecosystems, Sapienza University of Rome 2025 (Grant Number: RG125199C3B7A6DA); and National Science Foundation grant 1844753.
\end{acks}

\bibliographystyle{ACM-Reference-Format}
\bibliography{biblio}

\appendix
\input{sections/AA-appendix}

\end{document}

%% file: variables.tex
\newcommand{\binancetrades}{36.5B}
\newcommand{\binancepair}{2236}
\newcommand{\binancefee}{0.1}
\newcommand{\binancebps}{10}

\newcommand{\krakentrades}{881M}
\newcommand{\krakenpair}{611}
\newcommand{\krakenfee}{0.40}
\newcommand{\krakenbps}{40}

%% file: sections/00-abstract.tex
\begin{abstract}
Centralized cryptocurrency exchanges (CEXes) enable fast off-chain conversions between hundreds of coins.
It is an open question which algorithmic trading patterns occur on these platforms.
A major challenge to measuring CEXes is that their public trade data does not contain addresses or trader identifiers allowing linkage.
We propose a novel methodology to infer one-way arbitrage (OWA) trading in anonymized spot trade data from CEXes.
We identify 402\,M likely OWA sequences in 5~years of trading on Binance (and almost 2\,M during 9~years on Kraken), accounting for 0.94\,\% and 0.13\,\% of the total traded volume, respectively.
While we estimate total profits of \$31.2\,M on Binance and \$975\,k on Kraken, profits from individual OWA sequences are less than \$1 on average after accounting for trading fees.
We also observe that OWA has become faster over time,
while the profitability of individual sequences has decreased.
Our findings highlight that pricing discrepancies regularly occur in CEXes, and raise questions for future work to identify the precise circumstances that enable profitable OWA.
\end{abstract}

%% file: sections/01-introduction.tex
\section{Introduction}

As cryptocurrencies have proliferated, traders have sought low-friction, high-speed ways to exchange between them.
On-chain or cross-chain transactions can be relatively slow and require technical sophistication.
In contrast, centralized cryptocurrency exchanges (CEXes) such as Binance and Kraken offer easy-to-use apps, web interfaces and APIs that enable low-latency, high-volume online trading off the chain.
Many CEXes have long surpassed decentralized exchanges (DEXes) or on-chain transactions in terms of volume~\cite{cexes_vs_dexes}.
While there has been extensive work measuring blockchains and DEXes (\eg to trace payment flows~\cite{meiklejohn2013fistful,moser2017empirical,kumar2017traceability,quesnelle2017linkability,kappos2018empirical} or detect arbitrage~\cite{daian2020flash,qin2022quantifying,wang2022cyclic,heimbach2024non,mclaughlin2023large}), we are not aware of prior work that has measured which algorithmic trading patterns have emerged on CEXes.

\begin{figure}[t]
  \centering
  \includegraphics[width=.3\textwidth]{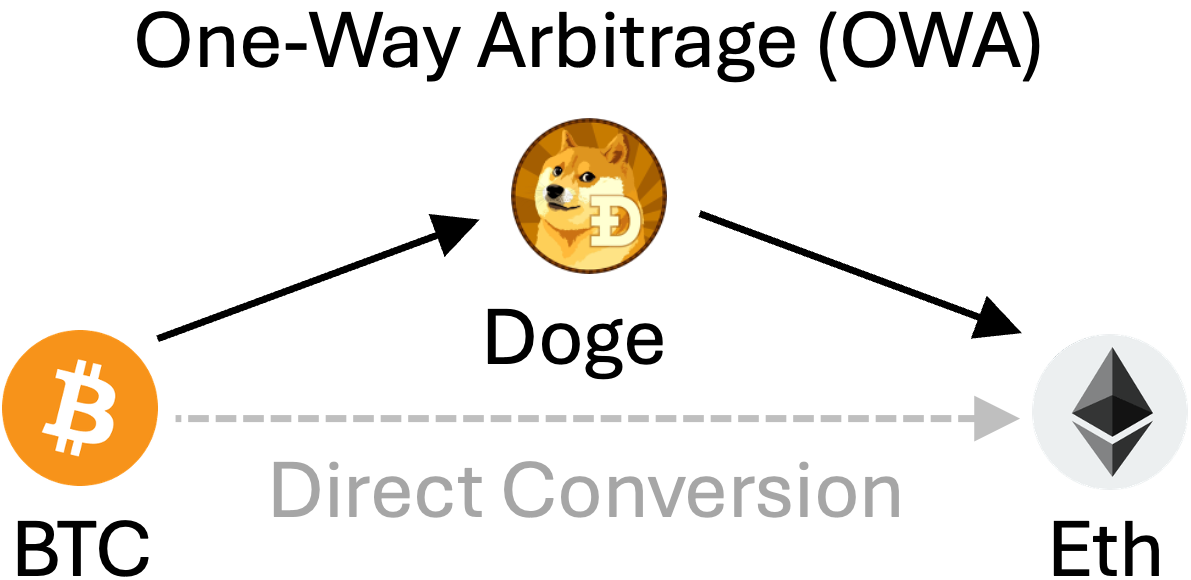}
  \caption{One-way arbitrage (OWA) vs.\ direct conversion. When prices are misaligned, the indirect conversion can yield a larger output than a direct conversion.}
  \label{fig:owa_schema}
\end{figure}

In this paper, we propose a methodology to measure arbitrage in historical spot trade data from CEXes (\ie trades with immediate settlement).
We focus on one-way arbitrage (OWA) because in our preliminary data exploration, we found it much more prevalent than the better known triangular arbitrage.
OWA is a trading strategy that aims to obtain a better price for a conversion $A \rightarrow B$ by executing instead an indirect conversion $A \rightarrow I \rightarrow B$ when there are advantageous price discrepancies.
As an example, consider Figure~\ref{fig:owa_schema}.
When 1~BTC can buy 30~ETH or 582012~DOGE, then a DOGE-to-ETH price of 0.00005155 or above would make it cheaper to convert BTC $\rightarrow$ DOGE $\rightarrow$ ETH instead of BTC $\rightarrow$ ETH directly (excluding transaction fees).
In efficient markets, prices are normally aligned so that such OWA opportunities do not arise (law of one price).

However, no market is perfect, and since Deardorff defined the term as a theoretical concept in 1979~\cite{deardorff1979one}, it has been an open question whether and how arbitrageurs actually carry out OWA in practice.
Prior work has measured the prevalence of other forms of arbitrage.
For example, a line of work has detected arbitrage executed on DEXes (\ie blockchains)~\cite{daian2020flash,qin2022quantifying,wang2022cyclic}.
In the context of CEXes, another line of work has studied the presence of opportunities for triangular arbitrage based on current prices, without measuring however whether any traders took advantage of these opportunities (\eg \cite{muck2025wish}).
The same limitation applies to prior studies of OWA opportunities in foreign exchange markets~\cite{chang2002one}.
Thus, it is still a largely open question how to detect and measure algorithmic trading patterns such as OWA at scale.

This might be due to measurement challenges: the trade data made available by CEXes is anonymized, \ie there are no addresses (or trader identifiers) for linkage. Matching trades by other metadata is fundamentally an $O(n^2)$ problem that can become intractable at the scale of the billions of trades found on large CEXes.

We propose a methodology to detect OWA in public, anony\-mized spot trade data of CEXes.
For practical considerations, we instantiate our methodology in the context of cryptocurrency CEXes, where historical spot trade data is available at no cost.
At the core, however, our algorithm is based on quantities and timing rather than what is being traded, and should therefore be adaptable to other contexts.
We alleviate the computational issues by introducing a set of constraints based on the trading pattern (\ie pairs of trades with closely matching timing and exchanged quantities) and the assumption of rational, profit-driven traders.
In the absence of ground truth, we prioritize precision over recall and fine tune our matching parameters based on the profitability of the resulting OWA sequences.
Our data exhibits distinct patterns, with some parameters resulting in profitable, likely true positive OWA, while others are unprofitable and more akin to background noise.

Using our methodology, we produce two data sets: 580\,M OWA sequences with a total volume of nearly \$190\,B on Binance, and 2.7\,M OWA sequences corresponding to \$1.7\,B of volume on Kraken.
They are based on historical spot trade data encompassing all trades in the international spot exchange of Binance (August 2017 to May 2023) and Kraken (September 2013 to May 2023).
These data sets are a rich source that allows us to measure the prevalence, profitability, and longitudinal evolution of OWA in the wild.

We find that OWA as a trading strategy is carried out at scale.
It amounts to 0.94\,\% of the total volume traded on Binance.
The volume of individual OWA instances is relatively low, in the order of hundreds of U.S.\ dollars each, and so are profits: When taking into account exchange fees, net profits are merely cents.
Due to the scale, however, they accumulate to \$31.2\,M on Binance, and \$975\,k on Kraken.

Overall, OWA activity appears to be more pronounced and more mature on Binance, the larger exchange of the two.
Our results also show that OWA has become faster over time (10\,ms or less between the two trades), likely the result of increased competition among arbitrageurs.
Furthermore, the price discrepancies exploited by OWA have decreased, which may indicate that the cryptocurrency markets are becoming more mature (\ie more balanced).

This paper makes the following contributions:
\begin{itemize}
\item We propose a novel methodology to detect OWA in anonymized spot trade data of large CEXes.

\item We produce (and will publish) two longitudinal data sets of OWA activity on Binance and Kraken.

\item We confirm that OWA is an actively used trading strategy, and show how it has evolved over time to become faster, while price discrepancies (and, consequently, profitability) have declined.
\end{itemize}

%% file: sections/02-background.tex
\section{Background}

Our methodology for detecting and analyzing one-way arbitrage (OWA) relies on spot trade data and is generally applicable to spot trading exchanges.
Because of data availability, we conduct our analysis in the context of centralized cryptocurrency exchanges, and explain the necessary background in the following paragraphs.

\subsection{Exchanges}
\label{subsec:exchanges}
Cryptocurrency exchanges are marketplaces where investors can trade cryptocurrencies. They can be broadly categorized into \emph{centralized exchanges (CEXes)}, where a central entity holds the deposits and manages order execution, and decentralized exchanges (DEXes), where the service runs in the blockchain and is powered by smart contracts.
In this work, we use data from CEXes because of the significantly higher trade volume~\cite{cexes_vs_dexes}.
We focus on \emph{spot trading}, where assets are exchanged with immediate settlement at the current market price through pair quotation.
In the pair $X/Y$, $X$ is the \emph{base} asset, the object of the trade that can either be bought or sold, and $Y$ is the \emph{quote} asset, which is used to measure the price (value) of the base asset.
In our context, the assets are (crypto)currencies, also called \emph{coins} or \emph{tokens}.
For example, if one trader places a ``sell'' order in BTC/USD for 2~BTC at a price of 126,000~USD per BTC, they are looking for a buyer who will give them 252,000~USD in exchange for 2~BTC.

In a CEX, offers to buy or sell are recorded in the \emph{order book}.
The order book is a real-time list that contains pending buy and sell orders (bids and asks), ranked by price level.
The orders specify the highest (or lowest) price a buyer (or seller) is willing to pay (or accept) for the base asset.
The price difference between the lowest ask and the highest bid is called the \emph{spread}.
At a high level, CEXes support two types of order.
\emph{Market orders} buy or sell immediately at the best available price.
They guarantee instant execution while not controlling the price.
In contrast, \emph{limit orders} specify a maximum (for buys) or minimum price (for sells) at which the trader is willing to transact.
An incoming limit order can be executed instantly when its specified price ``crosses the spread'' and matches an existing limit order from the order book.
Traders who cross the spread with such limit orders (or place market orders) are called market \emph{takers}, as they remove liquidity from the order book.
Limit orders not reaching the current price level are added to the order book and will be executed in the future when the price changes (unless the order is canceled or expires).
Traders who place such limit orders contribute to market depth by adding liquidity to the order book, and are called market \emph{makers}.

Cryptocurrencies can be categorized based on their characteristics and common roles on exchanges.
We define \emph{anchor coins} as the union of quote coins and stable coins.
In practice, these are the most actively traded and high-volume coins (BTC, ETH, \etc).
These coins often serve as the quote currency in trading pairs due to their liquidity and relatively stable market value.
Our definition of anchor coins also encompasses so-called \emph{stable coins}, which are designed to maintain a stable value by being pegged to external assets, such as fiat money or commodities like gold (PAX GOLD).
Prominent examples of stable coins pegged to the U.S.\ dollar (USD) include USDC, BUSD, and USDT.
In contrast, \emph{non-anchor coins} are coins only used as base currencies, meaning that on CEXs, they cannot be exchanged directly for other non-anchor coins.
They are typically less liquid.
Finally, some CEXes issue their own \emph{utility tokens} to be used throughout the exchange for access to specific services or functionalities (\eg fee payment).

\mypara{Fees.} CEXes apply fees to each trade, typically as a fraction of the exchanged amount, which is denominated in the base coin.
Because these fractions tend to be small, they are customarily measured in \emph{basis points (bps)}, equivalent to 0.01\,\%.
While different CEXes set varying fee levels, high-volume traders generally benefit from significantly lower fees compared to low-volume traders.
Additionally, takers often incur higher fees, while makers may receive rebates or even negative fees as an incentive to support market depth.
If fees are directly deducted from the trade, they are referred to as \emph{in-band} fees, whereas fees paid separately (using USD or an exchange-issued utility coin) are called \emph{out-of-band} fees.

\subsection{One-Way Arbitrage (OWA)}
\label{subsec:owa_def}

At times, exchanges may exhibit price discrepancies across tradable pairs. For instance, directly exchanging BTC~$\rightarrow$~ETH might result in less ETH than passing through DOGE as an \emph{intermediary coin}, \ie BTC~$\rightarrow$~DOGE~$\rightarrow$~ETH.
This type of indirect conversion is called \emph{one-way arbitrage (OWA)}~\cite{deardorff1979one} and illustrated in Figure~\ref{fig:owa_schema}.
(In contrast to triangular arbitrage, OWA does not ``close the loop'' back to BTC.)

Profitable OWA opportunities arise from temporary price mismatches and depend on the depth of the order book.
Once orders are filled and prices adjust, the indirect exchange may become more expensive than the direct one, especially when accounting for exchange fees.
Due to expected competition and possible adverse market movements, speed is essential to profit from these opportunities.
Furthermore, the exchanges we consider do not support batch trading, \ie arbitrageurs must place two separate orders to form an \emph{OWA sequence}.
We hypothesize that they rely on automated bots that place both orders in quick succession (rather than waiting for the first one to finalize) to minimize the risk of price changes.

\emph{Triangular Arbitrage} involves three trades instead of two.
The third trade ``closes the loop'' and realizes the gain in the currency that started the loop (BTC in the example above).
In contrast, OWA realizes gains in a different currency (ETH).
While this may seem to expose OWA traders to additional long-term price risk, arbitrageurs likely hold multiple anchor coins regardless.
For example, it is likely that triangular arbitrageurs conduct arbitrage both in BTC and ETH to increase the number of profitable opportunities they can pursue.
If arbitrageurs are indifferent as to which anchor coins they hold, OWA opens up additional profitable opportunities between them compared to triangular arbitrage within them.

The additional trade in triangular arbitrage makes it less likely for profitable opportunities to arise because more prices need to be aligned, and the profits need to be large enough to cover the additional transaction fee.
Furthermore, an additional trade also increases execution risk by adding the potential for an additional adverse price change during execution (slippage).
In preliminary work, we measured both one-way and triangular arbitrage and detected considerably less triangular arbitrage, possibly because of these practical difficulties.
This paper focuses on OWA as the more common case for simplicity of exposure.

%% file: sections/03-data-collection.tex
\section{Trade Data Collection}
Today, there are hundreds of CEXes around the world.\footnote{\url{https://coinmarketcap.com/it/rankings/exchanges/}}
For our study, we selected two of them, Binance and Kraken, based on their size, reputation, age, and trade data availability.
As of August 2025, CoinMarketCap ranks them 1st and 13th, respectively~\cite{exchanges_rank}.
(Other CEXs either made no historical spot trade data available, or it was not usable for our purposes.)
We extract each exchange's historical spot trade data from their respective public API.

For each trade, the API returns a unique trade ID, the timestamp, price, quantity, and direction (\ie \textit{buy} or \textit{sell} from the taker's perspective).
A ``sell'' trade in $A/B$ with quantity $q$ at price $p$ corresponds to the following two logical \emph{operations}:
\begin{enumerate}
    \item A taker operation $A \rightarrow B$ with input quantity $q_A = q$ units of $A$ and price $p_{AB} = p$, which yields $q_B = q_A \cdot p_{AB} = q \cdot p$ units of $B$ before fees are deducted, \ie the true (and unknown) output is $\hat{q_B}$.
    \item A maker operation $B \rightarrow A$ with input quantity $q_B = q \cdot p$ units of $B$ and price $p_{BA} = 1/p$, which yields $q_A = q_B \cdot p_{BA} = q$ units of $A$ before fees are deducted, \ie the true (and unknown) output is $\hat{q_A}$.
\end{enumerate}
In a ``buy'' trade, the maker and taker roles above are swapped.
There is no attribute that directly links multiple operations to the same trader or OWA sequence.
We develop a methodology to infer OWA sequences from the order data in Section~\ref{sec:owa_detection}.

\mypara{Binance}, with its 24\,h trading volume of more than \$15\,B~\cite{binance_coinmarketcap}, is one of the largest CEXes with hundreds of tradable cryptocurrencies.
Due to differing regulations, Binance has several jurisdiction-specific entities.
We collect all spot trades for all pairs from Binance.com, the global exchange for traders not subject to a regional restriction, from its inception in July 2017 until May 2023
using the official Python wrapper.\footnote{\url{https://github.com/binance/binance-public-data/tree/master/python}}

\mypara{Kraken}, with a reported daily trading volume of around \$600\,M~\cite{kraken_coinmarketcap}, is among the oldest and most established centralized exchanges. Operational since 2013, it is widely used in North America and Europe, offering trading of more than 600 pairs.
Unlike Binance, Kraken operates through a single global platform.
We collect all Kraken spot trades from October 2013 to May 2023 using the official API.\footnote{\url{https://docs.kraken.com/api/\#operation/getRecentTrades}}

\subsection{Mark-to-Market}
\label{subsec:mark_to_market}
To compare cryptocurrency trade volumes and profits, we use the U.S. Dollar (USD) as a reference, given its widespread use in finance.
On Kraken, every token is directly traded against USD, allowing us to use the quoted price to determine the dollar value of each coin, a process known as mark-to-market.
We rely on the one-minute Volume-Weighted Average Price (VWAP) to mitigate short term volatility and outlier trades when determining prices.
The VWAP is a standard measure used to compute the average price of an asset over a given time interval, weighted by the traded volume.
It is defined as $\sum_{i=1}^n p_{i} \cdot q_{i} / \sum_{i=1}^n q_i$ and computed over all $n$ trades $i$ within the same wall clock minute with price $p_i$ and quantity $q_i$ of the coin to be marked to market.

In contrast, Binance does not provide direct USD trading pairs for all cryptocurrencies, requiring an additional conversion step.
Since most tokens on Binance are traded against BTC (92.5\%), we first convert all trade volumes to BTC based on the pair's quote coin (using the corresponding BTC VWAP), and then to USD using the historical BTC/USD VWAP from Kraken.
For tokens that are not directly traded against BTC, we use the utility token BNB, or the stable coins USDT or BUSD, depending on which has the longest trading history as a pair for that particular token.

\subsection{Spot Trade Data Set Overview}
\label{subsec:exchange_overview}

\begin{table}[t]
\small
    \centering
    \caption{Spot trade data set overview.}
    \label{tab:datasets_data}
    \begin{tabular}{l l l r r r}
        \toprule
         CEX & From & To  & \# Pairs & \# Trades & \$ Volume \\
        \midrule
        Binance & 2017-08 & 2023-05 & \binancepair & \binancetrades & 20.16T \\
        Kraken & 2013-09 & 2023-05  &\krakenpair & \krakentrades & 1.25T \\
        \bottomrule
    \end{tabular}
\end{table}

Our spot trade data set includes more than 36.5\,B trades on Binance across 2,236 pairs, corresponding to a total volume exceeding \$20\,T, and 881\,M trades on Kraken across 611 pairs, totaling more than \$1.25\,T, as summarized in Table~\ref{tab:datasets_data}.

\begin{figure*}[t]
  \centering
  \begin{subfigure}[t]{0.48\textwidth}
    \centering
    \includegraphics[width=\textwidth]{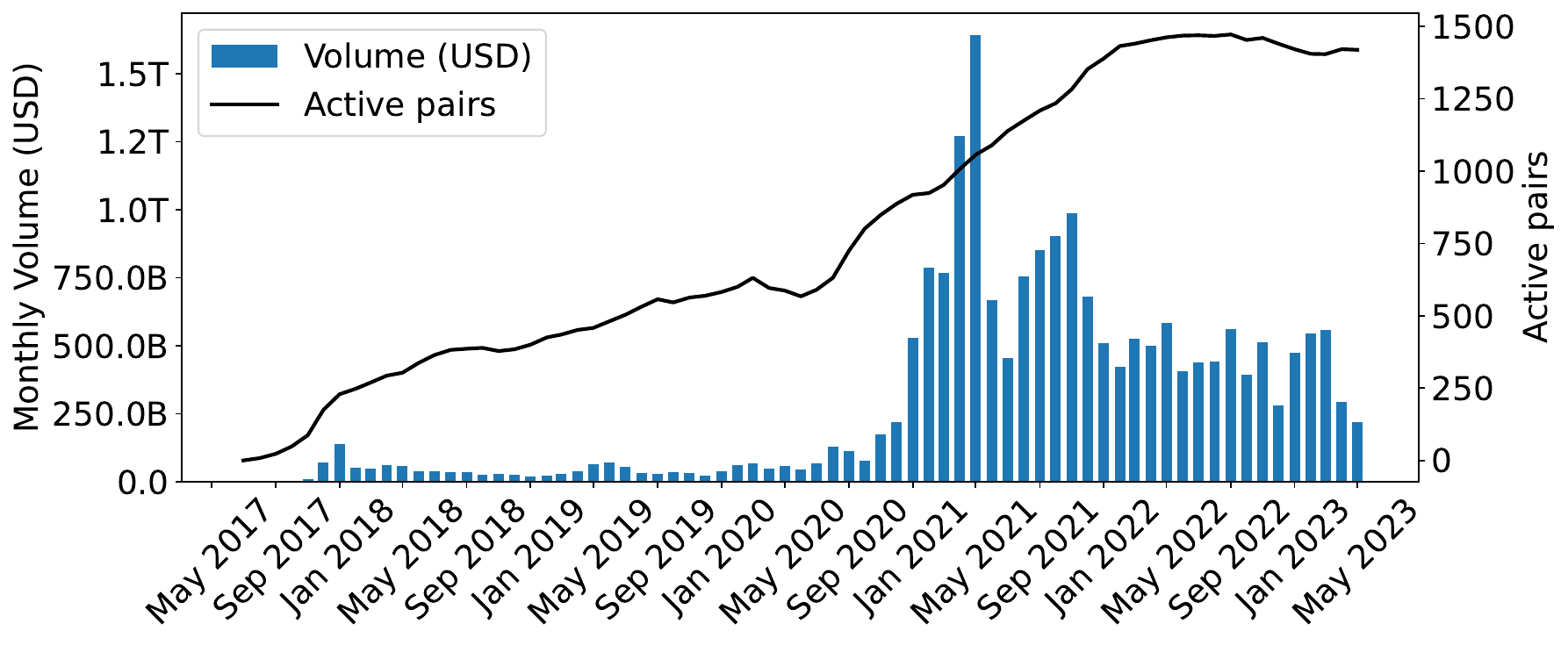}
    \caption{Binance}
    \label{fig:binance_volume_and_pairs}
  \end{subfigure}
  \hfill
  \begin{subfigure}[t]{0.48\textwidth}
    \centering
    \includegraphics[width=\textwidth]{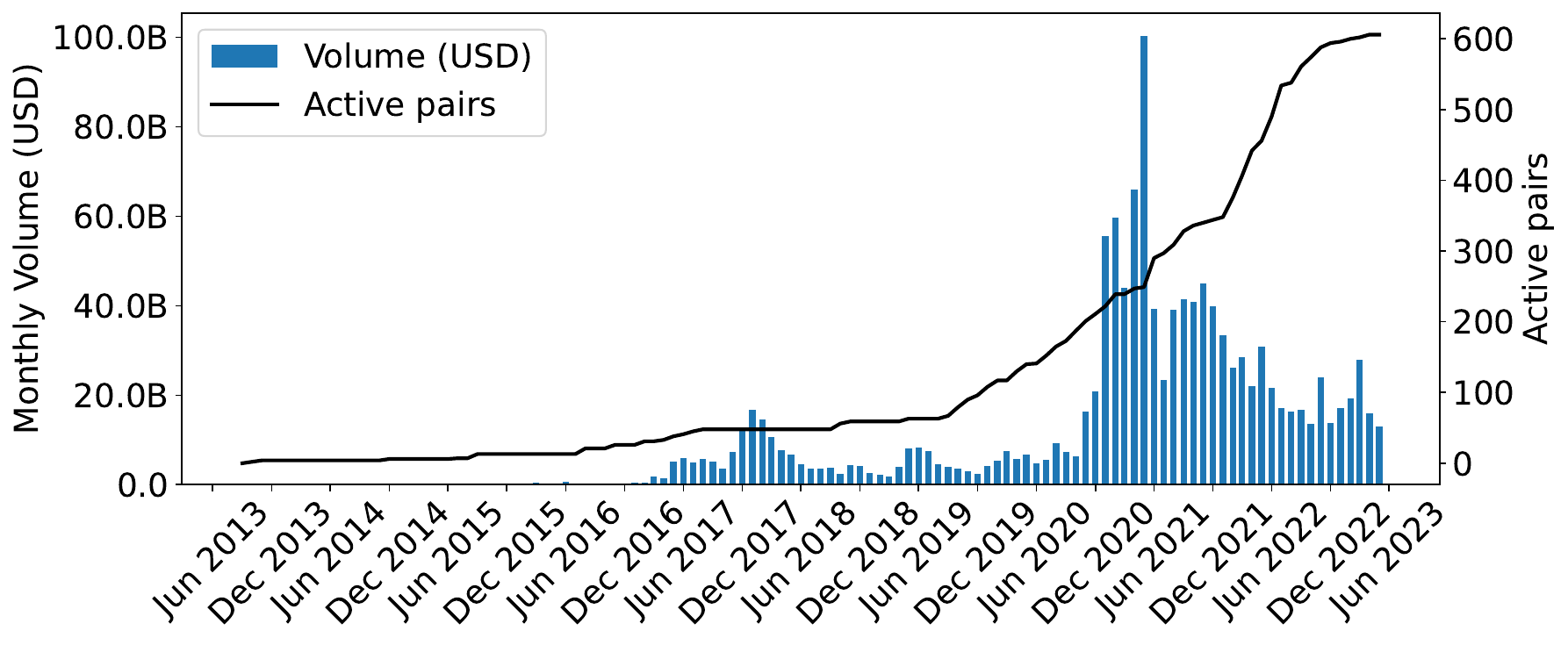}
    \caption{Kraken}
    \label{fig:kraken_volume_and_pairs}
  \end{subfigure}
  
  \caption{Monthly trade volume in USD (blue bars, left y axis) and active currency pairs (black line, right y axis).}
  \label{fig:exchanges_volume_and_pairs}
\end{figure*}

Figure~\ref{fig:exchanges_volume_and_pairs} shows the monthly trading volume together with the number of active pairs for the two exchanges.
Although Binance was launched several years after Kraken, it quickly surpassed it in both traded volume and number of exchangeable currency pairs, and has maintained this trend over time.
Over less than six years in operation, Binance has generated more than 1,500\,\% of the total volume traded on Kraken in almost 10 years.
Despite this large difference in scale, the two exchanges display similar temporal dynamics, with late 2020 to early 2021 marking the peak of trading activity.
In May 2021, both recorded their highest monthly volumes, reaching approximately \$100\,B on Kraken and \$1.64\,T on Binance.

%% file: sections/04-owa-detection.tex
\section{OWA Detection}
\label{sec:owa_detection}
The public spot trade data made available by Binance and Kraken is anonymized.
There is no information linking trades to specific users, and thus individual trades cannot be grouped to identify trading behaviors at the user level.
Since the OWA trading strategy consists of a sequence of two operations, we need to develop a methodology to infer which two operations likely belong together to form an OWA sequence.

\subsection{Trader Model}
\label{subsec:traders_model}
To motivate our OWA detection heuristics, we describe the type of trader we assume performs these operations.
Our main assumption is that OWA traders are motivated by profit.
Furthermore, we assume that such traders avoid holding volatile (non-anchor) coins  due to the associated market risk.
Third, we assume that OWA traders operate at high volumes, which qualifies them for the lowest published fee tiers on the exchanges.
Fourth, we assume that there is competition among OWA traders, which leads them to optimize their trading infrastructure for speed.
The traders we consider have the necessary financial resources, expertise, and technical setups to execute trades within milliseconds.
Lastly, we assume that OWA traders are opportunistic and attempt to capitalize on any profitable OWA sequence rather than seeking to convert one specific token into another.

\subsection{Detection Methodology Overview}
We detect OWA sequences through a two-step process.
First, we reduce the search space by broadly identifying all \emph{OWA candidates}, which are defined as pairs of operations that satisfy coarse heuristics based on our trader model (Section~\ref{subsec:owa_candidates_detection}).
This step filters out unlikely sequences while still identifying a broad set of possible OWA sequences.
In the second step, we refine this set by evaluating the hypothetical profitability of OWA candidates (Section~\ref{subsec:owa_refining}).
We derive tighter matching thresholds that maximize profits to reduce false positives and increase precision.
Since we do not have ground truth to evaluate our methodology, we prioritize precision over recall.
That is, our methodology detects OWA operations that have parameters making them likely true positives at the expense of not detecting more atypical OWA sequences or OWA attempts that failed due to adverse market movement.

\subsection{Detection of OWA Candidates}
\label{subsec:owa_candidates_detection}

An OWA sequence consists of two operations.
Operation $O_1$ at time $t_1$ converts between coins $A_1 \rightarrow I$ with output quantity $q_{I_1}$ before fees ($\hat{q}_{I_1}$ after fees, which is not visible in the data), and volume $v_1$ (converted to USD, as defined in Section~\ref{subsec:mark_to_market}).
Operation $O_2$ at time $t_2$ ($t_1 \leq t_2$) converts between coins $I \rightarrow A_2$ with input quantity $q_{I_2}$ and volume $v_2$.
The \emph{latency} of the sequence is $\Delta t = t_2 - t_1$, the time between the two operations' timestamps.
The \emph{relative quantity difference} of the sequence is $\Delta q = \frac{q_{I_2}}{q_{I_1}} - 1$, the relative difference between the output quantity of the first operation and the input quantity of the second operation.
The \emph{volume} of the sequence is $v = v_1 + v_2$, the combined volume of both operations.
With over 36.5\,B trades on Binance, the unrestricted search space for sequences is prohibitively large ($O(n^2)$).
To make the problem computationally tractable, we introduce four constraints based on our trader model and set coarse thresholds for initial matching of OWA candidate sequences.

\mypara{Coin types.}
We search for sequences of the form $A_1 \rightarrow I \rightarrow A_2$, where (1) $A_1$ and $A_2$ are anchor coins (BTC, ETH, BNB, stable coin, and on Kraken also USD), while (2) $I$ is a non-anchor coin.
The first requirement corresponds to our assumption that OWA traders avoid holding volatile and potentially illiquid non-anchor coins due to the associated financial risks.
We also note that non-anchor coins cannot appear next to each other in OWA sequences because CEXs usually do not offer trading pairs for direct conversions between non-anchor coins.
The second requirement is motivated mainly by the need to reduce the search space combined with an intuition that OWA opportunities might be less frequent between three anchor coins, as anchor pairs tend to be more liquid and therefore might be closer to market efficiency.
As a result of this search space reduction, we detect only a subset of possible OWA activity.

\mypara{Exchanged quantity difference.}
When searching for pairs of operations that could form an OWA sequence, we look for operations that closely match in their exchanged quantities ($q_{I_1} \approx q_{I_2}$).
This again follows the intuition that arbitrageurs do not wish to hold volatile or illiquid intermediary coins, and transfer as much out as they have transferred in.
In practice, $q_{I_1}$ and $q_{I_2}$ can differ
for a variety of reasons.
First, fees for $O_1$ can reduce the amount available for $O_2$ when the exchange deducts them in band.
Since the spot trade data only contains input quantities, the exact output quantity of each operation and the amount of fees paid are unknown; we need to approximate the quantity of $I$ that may be available to be converted in $O_2$.
Second, trading constraints may cause a small mismatch between the two quantities.
Each exchange-listed currency pair has different values for minimum tradable quantities; furthermore, quantities and prices change in discrete steps.
Therefore, a small quantity might not be convertible, leaving so-called \emph{dust}\footnote{https://www.binance.com/en/support/faq/detail/360003012371} in $I$.
This can cause a quantity mismatch even when fees are paid out of band.
Third, a larger mismatch in exchanged quantities could occur due to varying OWA trading strategies, \eg cases where the trader exchanges significantly less (or more) out of $I$ than the original quantity exchanged into $I$.
For tractability, we do not aim to detect OWA of this third kind.

To account for the first two effects, we consider two operations an OWA candidate when $\Delta q \in [-10, 1]$\,bps on Binance, and $\Delta q \in [-40, 1]$\,bps on Kraken.
We use a relative metric for the quantity difference because the biggest potential contributor (exchange fees) is relative to the exchanged quantity.
The lower end of the intervals, \ie a reduction in quantity, corresponds to exchange fees.
To fully explore the OWA candidate search space, we initially set it to the highest published fee charged by the respective CEX: 10\,bps on Binance, and 40\,bps on Kraken.
The upper 1\,bps end of the intervals (\ie a slight increase in quantity) is chosen to allow for dust left over from previous operations, or for potential negative fees offered to market makers.
We note that this relatively broad limit is used only for the initial reduction of the search space (for example, large-scale OWA traders likely pay discounted fees in practice); we analyze the resulting OWA candidates and apply a tighter constraint in Section~\ref{subsec:owa_refining}.

\mypara{Latency.}
Neither Binance nor Kraken support batch trading with atomic execution.
In order to win the race for profitable OWA opportunities and avoid getting stuck in a volatile intermediary coin, traders need to send both orders of the OWA sequence in quick succession.
For the initial filtering, we set a window $\Delta t \le 2$\,s between the timestamps of the two operations.
This threshold is a tradeoff between the need to be large enough to include as many actual OWA sequences as possible, while being small enough to still make our analysis computationally feasible.

\mypara{Order type.}
We only consider OWA sequences that end with a taker operation.
Maker orders do not execute immediately and would have the trader hold the volatile intermediary coin for an unpredictable amount of time (until the order book has reached the specified price level), effectively amounting to speculation rather than the more calculated risk of OWA.
In our filtering step, we do allow $O_1$ to be a maker order, however, because it allows the trader to take advantage of maker fee rebates to increase profitability.
If the market moves unfavorably, the trader can cancel the pending maker order and retain their holdings in the stable anchor coin, unlike the scenario where $O_2$ is a maker order.

\begin{figure*}[tp]
  \centering
  \begin{subfigure}[t]{0.48\textwidth}
    \centering
    \includegraphics[width=\textwidth]{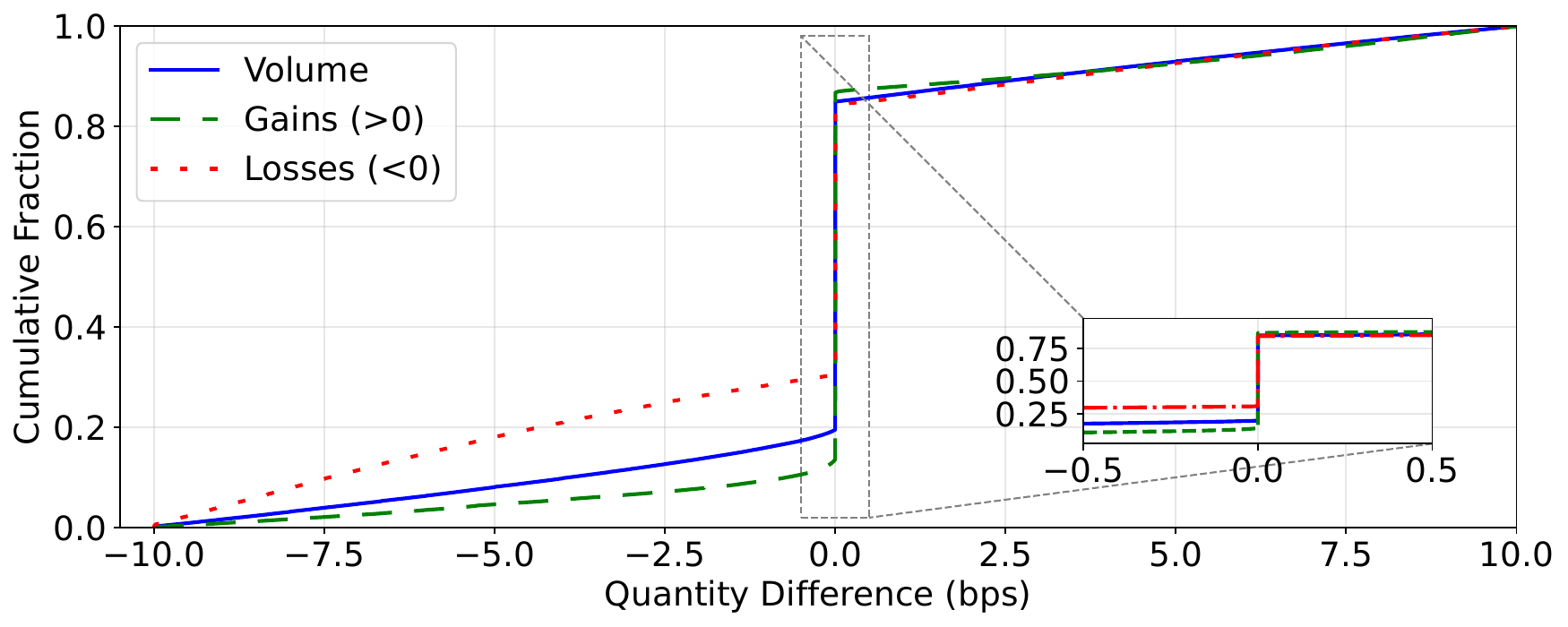}
    \caption{Binance \MT}
    \label{fig:binance_qtydiff_MT_cdf}
  \end{subfigure}
  \hfill
  \begin{subfigure}[t]{0.48\textwidth}
    \centering
    \includegraphics[width=\textwidth]{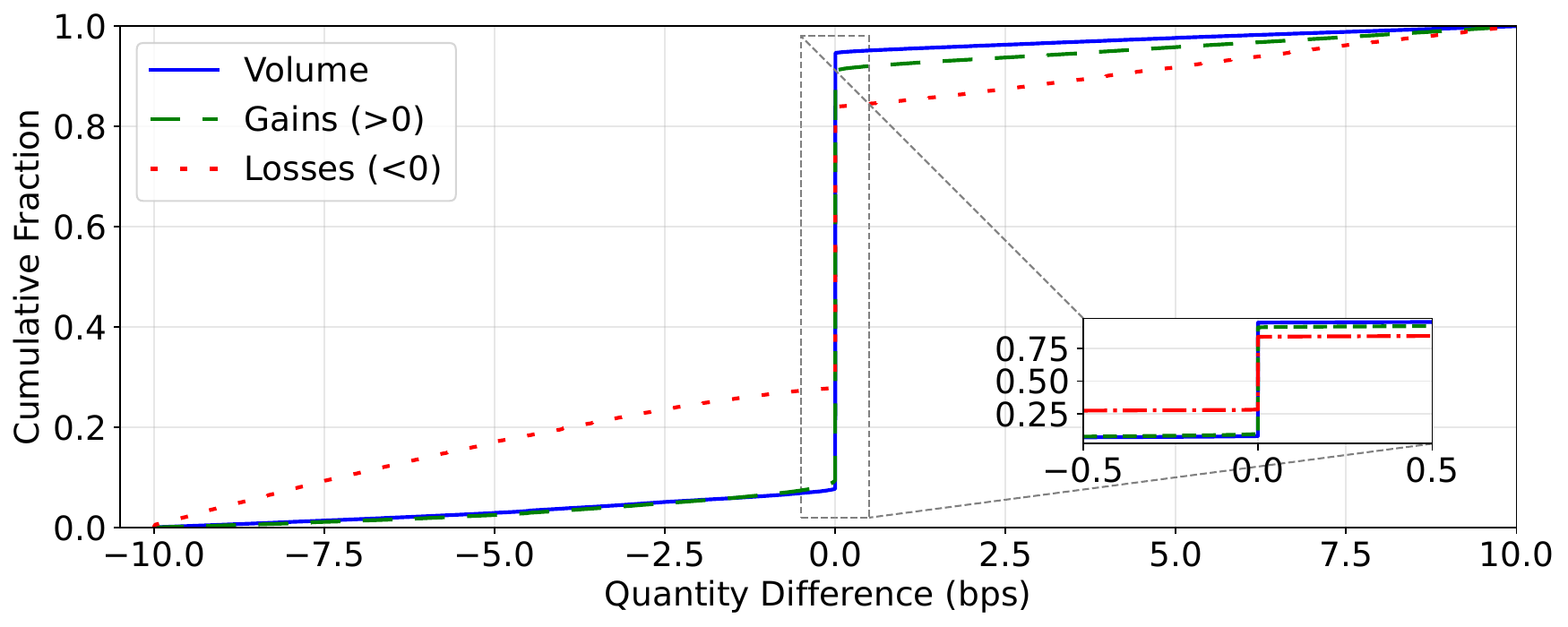}
    \caption{Binance \TT}
    \label{fig:binance_qtydiff_TT_cdf}
  \end{subfigure}

  \begin{subfigure}[t]{0.48\textwidth}
    \centering
    \includegraphics[width=\textwidth]{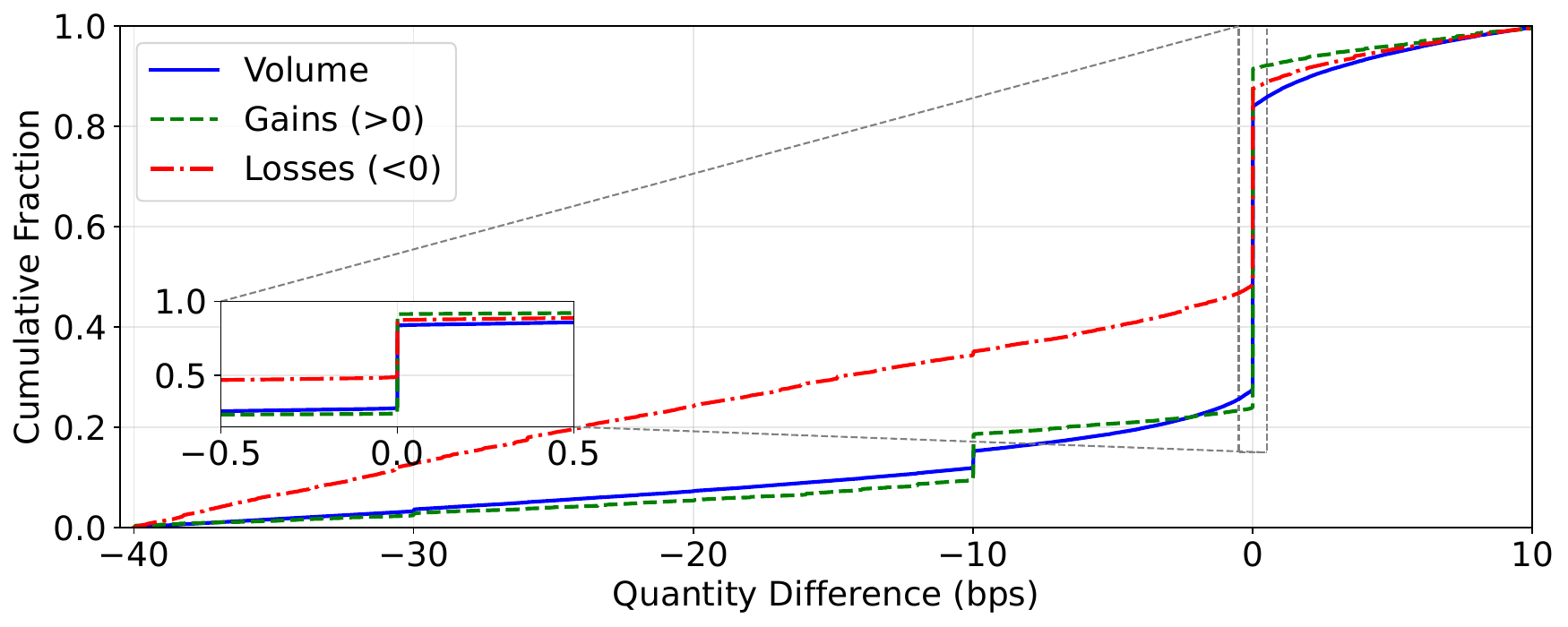}
    \caption{Kraken \MT}
    \label{fig:kraken_qtydiff_MT_cdf}
  \end{subfigure}
  \hfill
  \begin{subfigure}[t]{0.48\textwidth}
    \centering
    \includegraphics[width=\textwidth]{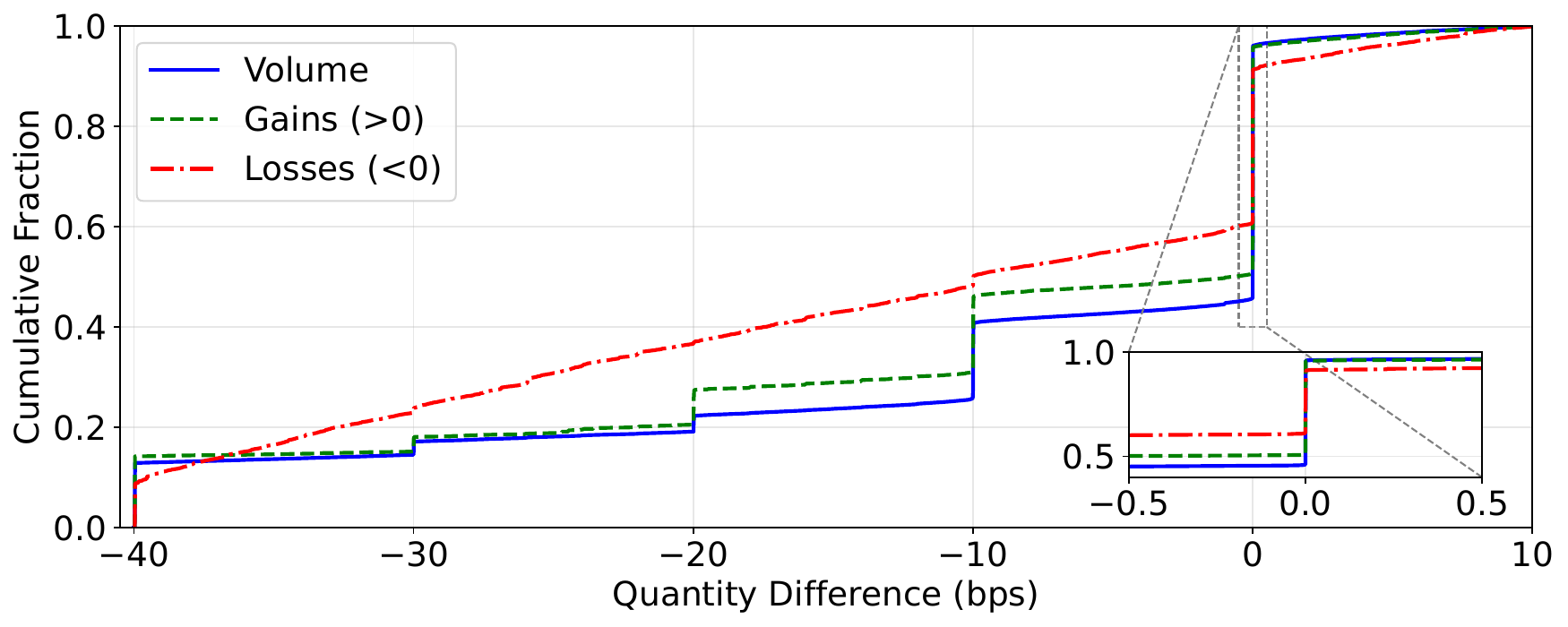}
    \caption{Kraken \TT}
    \label{fig:kraken_qtydiff_TT_cdf}
  \end{subfigure}
  
  \caption{CDFs of the accumulated volume and hypothetical gains and losses of OWA candidate sequences on Binance and Kraken as a function of $\Delta q$, the difference in exchanged quantity between the two operations in the sequence (aggregated over latencies up to 2\,s).
  Candidate sequences with $\Delta q = 0$ exhibit the largest density in all three metrics (likely actual OWA).
  Candidate sequences with $\Delta q \neq 0$ are nearly uniformly distributed (likely ``random'' background noise), with the exception of steps around published fee levels on Kraken (\eg -10\,bps).}
  \label{fig:qtydiff_cdf}
\end{figure*}

Conceptually, our matching algorithm consists of a double nested loop that considers all combinations of operations and tests whether they satisfy the four constraints described above, in which case they form a new OWA candidate sequence. In case of ambiguity, \ie multiple candidate sequences including the same operation, we retain only the one sequence with the best price conversion. Our implementation of this algorithm is optimized to avoid unnecessary comparisons, but its complexity still is $O(n^2)$, and executing it on a shared computing cluster took multiple months for Binance (three days for the smaller Kraken data).
We found 580\,M OWA candidates on Binance, and 2.7\,M on Kraken.

\subsection{Refining Matching Criteria}
\label{subsec:owa_refining}

After the purposefully loose initial filtering, we obtain a candidate set that is manageable in principle but still very large for Binance (580\,M candidates).
To refine our matching criteria, we construct a smaller ``development'' set.
For Binance, we randomly sample one month per year from 2018 to 2023, resulting in a subset containing 52\,M OWA candidates.
For Kraken, we use all 2.7\,M candidates without subsampling.

Our hypothesis is that we can separate likely OWA activity from unrelated background noise by looking for trading parameters co-occurring with a higher concentration of volume and profits, since we assume that rational traders conduct OWA under parameters that make it profitable in aggregate.
We explore (1) $\Delta q$, the relative difference in the quantity exchanged in the two operations of each candidate sequence, and (2) $\Delta t$, the latency between the two operations.
We plot these two parameters of OWA candidates in our development set against the corresponding volume and profitability to determine ranges with a higher incidence of volume and profits.
We look at maker/taker and taker/taker sequences separately on the two exchanges since all four combinations might require distinct parameter choices due to different fee levels, execution speed, and exchange-specific factors.

To quantify profitability, we compute the \emph{absolute gross return} $R = q_{A_2} - q^r_{A_2}$, the difference between the output of the OWA candidate sequence $A_1 \rightarrow I \rightarrow A_2$ and the output $q^r_{A_2}$ of a hypothetical direct conversion $A_1 \rightarrow A_2$ at the contemporaneous reference price $p^r_{A_1A_2}$, \ie the OWA profit.
We compute the reference price  $p^r_{A_1A_2}$ for the direct conversion as the VWAP of all operations $A_1 \rightarrow A_2$ within a 2\,s window \emph{before} the first operation of the OWA candidate sequence.
This reflects the assumption that an arbitrage opportunity should be visible to the arbitrageur prior to the execution of the initial operation.
For simplicity, we assume a ``best case'' scenario and assess profitability without consideration for fees.
(Section~\ref{sec:analysis:profits} will consider different fee scenarios.)

\subsubsection{Quantity Difference}

Figure~\ref{fig:qtydiff_cdf} shows CDFs for the accumulated volume and hypothetical gains and losses as a function of $\Delta q$, the relative quantity difference between the two trades in OWA candidate sequences.
We plot hypothetical gains ($R > 0$) and losses ($R < 0$) separately since they would cancel each other out in a combined metric.

On Binance, the vast majority of activity happens within a very small quantity difference range.
Around 69\,\% of the volume of \MT OWA candidates and 88\,\% of \TT candidates (across all latencies) occur with $\Delta q \in [-0.1, 0.1]$\,bps.
This suggests that these traders pay their exchange fees out of band, since the quantity exchanged into the intermediary coin is substantially the same as the quantity transferred out.
(On Binance, fees are lower when paid out of band using the exchange's utility token BNB.)
Gains are similarly concentrated around closely matching quantities.
While around 60\,\% of losses occur in candidate sequences with closely matching quantities (due to the high volume), losses are more widely spread overall.
This indicates that candidate sequences with larger quantity differences are more likely to result in a worse outcome than an equivalent direct conversion.
More generally, volume, gains and losses for candidate sequences outside the $[-0.1, 0.1]$\,bps range are nearly evenly distributed.
These are more likely ``background noise,'' \ie pairs of unrelated operations rather than coordinated OWA attempts.

These overall trends also hold on Kraken, but with a notable difference.
While the volume and gains of OWA candidates also increase in a large step around 0\,bps, there are additional steps, for example at $-40$, $-30$, $-20$, and $-10$\,bps for taker/taker sequences.
These steps correspond to a subset of published fee levels.
Kraken deducts fees in band,\footnote{https://www.kraken.com/en-gb/features/fee-schedule} yet the largest step is at 0\,bps.
We hypothesize that some traders have negotiated fee agreements and pay their taker fees out of band (or possibly do not pay any fees at all).

Since we have no certainty about which two operations constitute OWA by the same trader, we err on the side of caution and focus our analysis on the subset where it is the most likely to occur, \ie operations with a quantity difference in the range of $[-0.1, 0.1]$\,bps.
While the vast majority of volume occurs with an exact quantity match (\ie a difference of 0.0), we allow for some unconvertible dust being left behind between the two operations (up to $-0.1$\,bps) or for some previously leftover dust being added to the second operation (up to $+0.1$\,bps).
We apply the same range to Binance and Kraken, despite the evidence suggesting that a sizeable minority of OWA on Kraken was done with fees paid in band.
This decision keeps the two data sets more directly comparable and simplifies our analysis.
We leave a separate analysis of the subset of OWA sequences with in-band fees for future work, as their profitability and other characteristics may differ from OWA sequences with out-of-band fees.

\subsubsection{Latency}

\begin{figure*}[tp]
  \centering
  \begin{subfigure}[t]{0.48\textwidth}
    \centering
    \includegraphics[width=\textwidth]{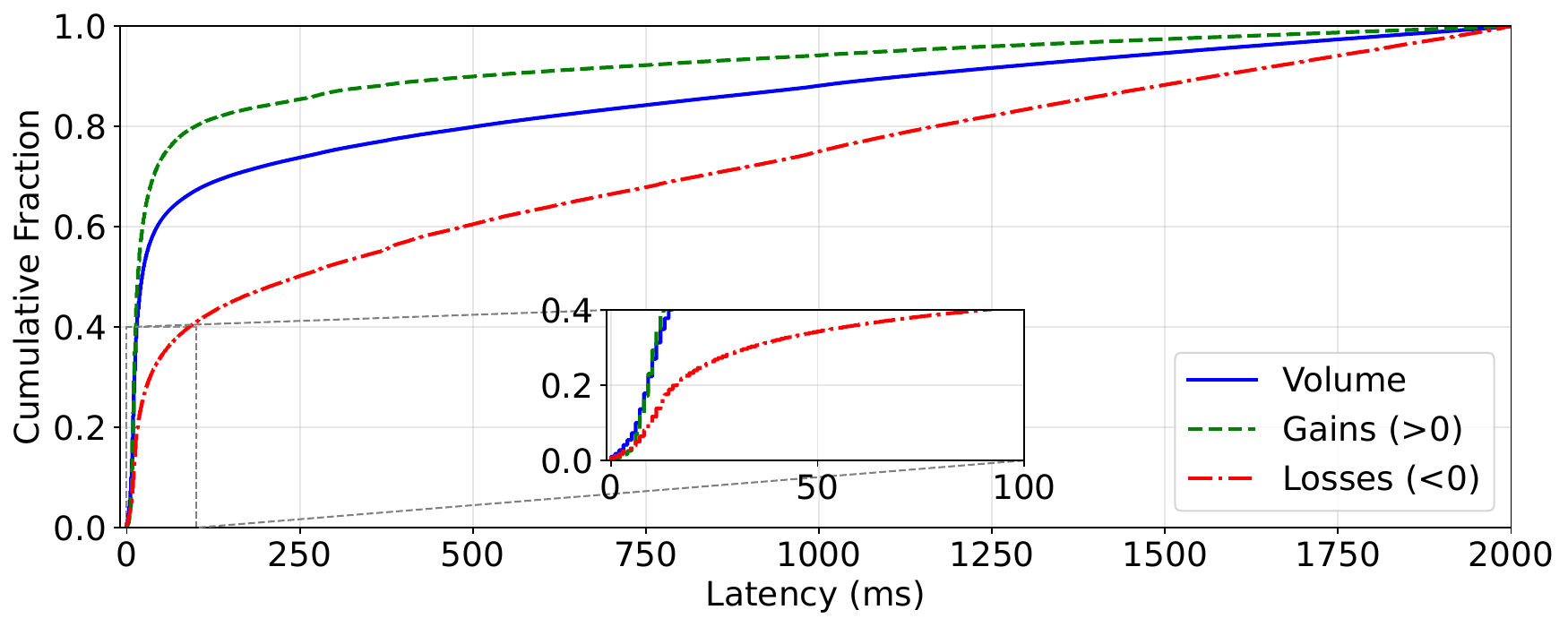}
    \caption{Binance \MT}
    \label{fig:binance_latency_MT_cdf}
  \end{subfigure}
  \hfill
  \begin{subfigure}[t]{0.48\textwidth}
    \centering
    \includegraphics[width=\textwidth]{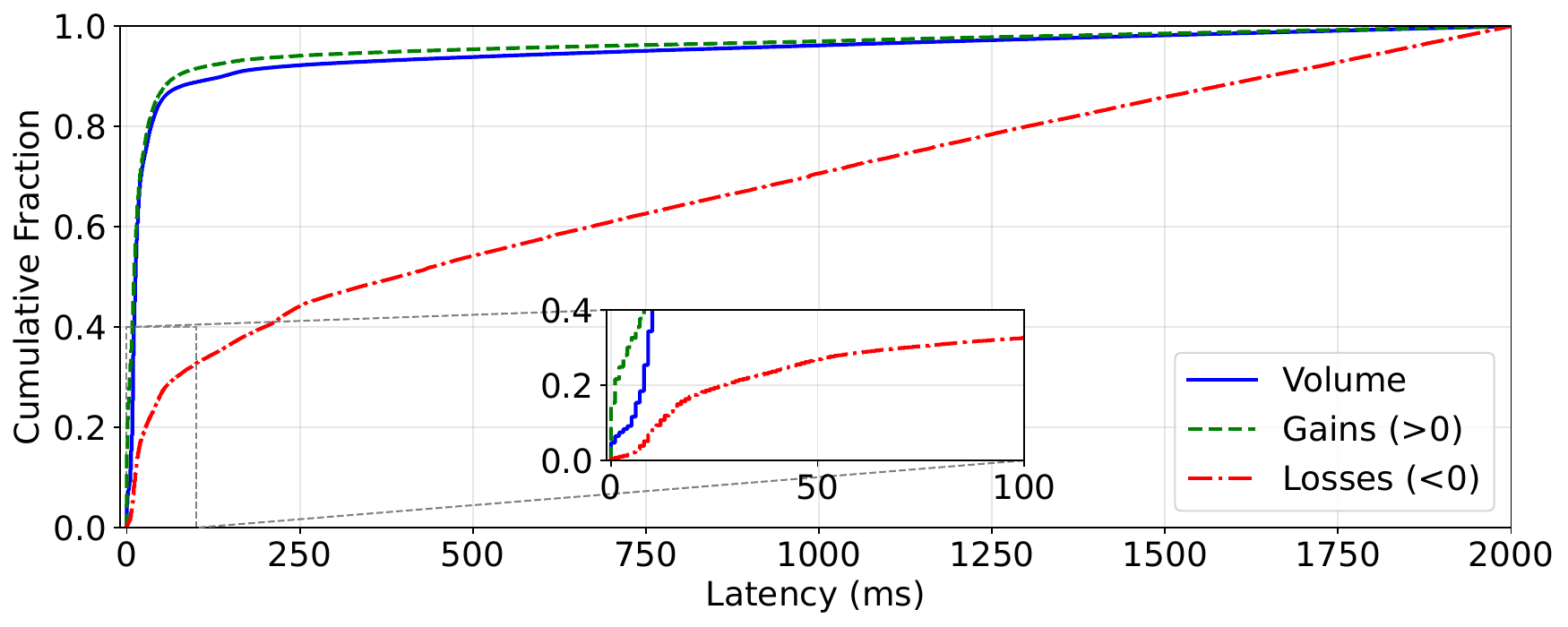}
    \caption{Binance \TT}
    \label{fig:binance_latency_TT_cdf}
  \end{subfigure}
  
  \begin{subfigure}[t]{0.48\textwidth}
    \centering
    \includegraphics[width=\textwidth]{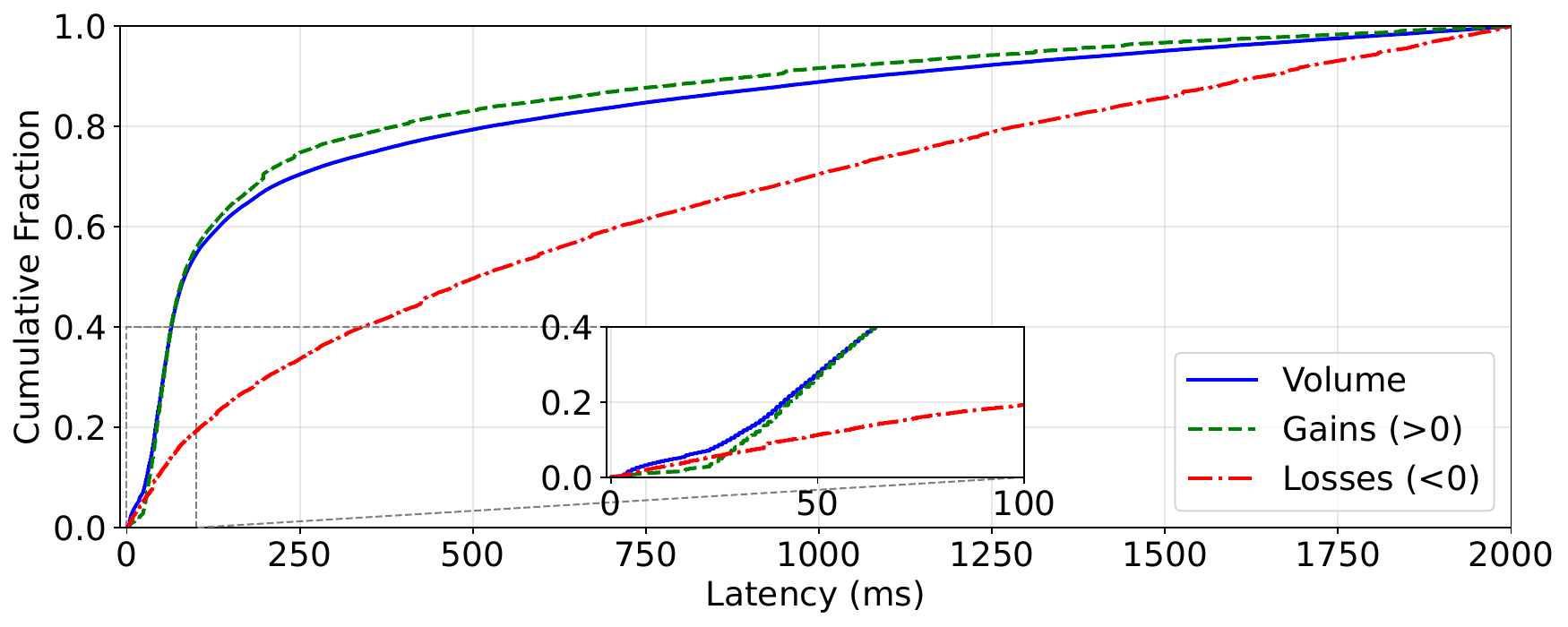}
    \caption{Kraken \MT}
    \label{fig:kraken_latency_MT_cdf}
  \end{subfigure}
  \hfill
  \begin{subfigure}[t]{0.48\textwidth}
    \centering
    \includegraphics[width=\textwidth]{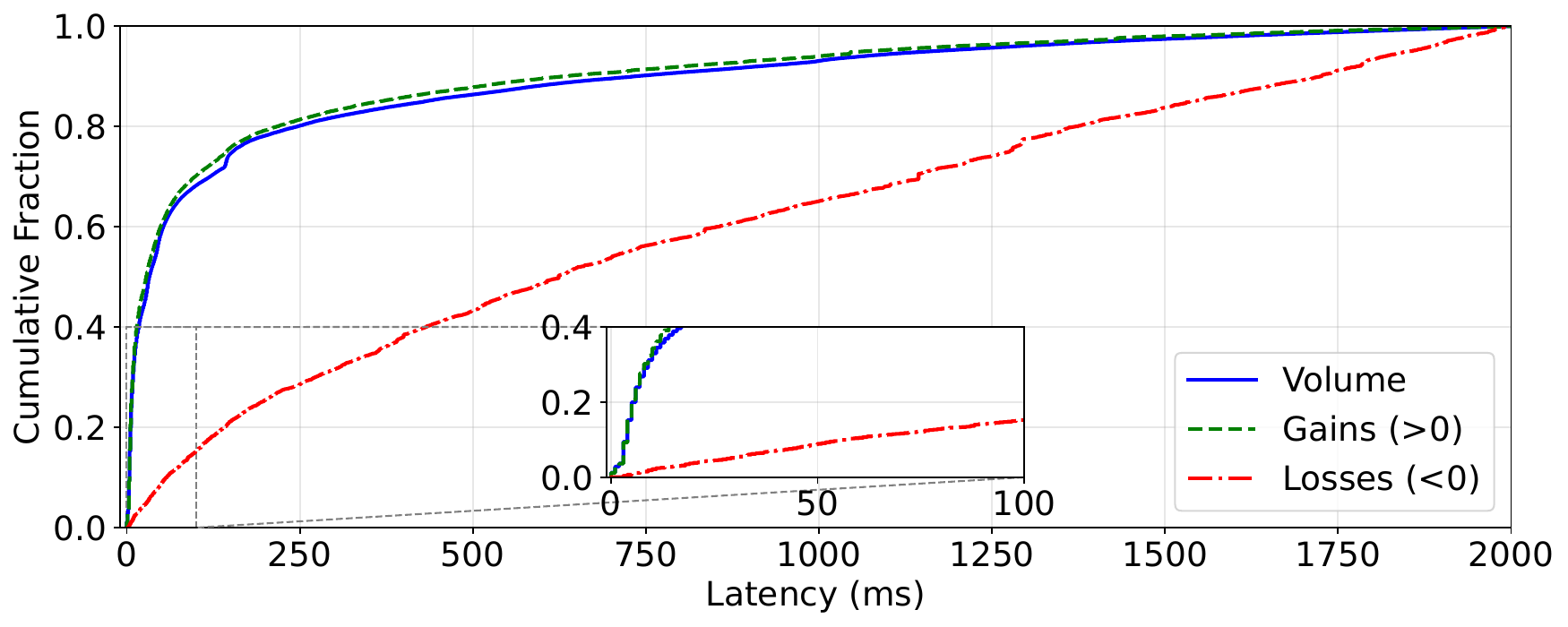}
    \caption{Kraken \TT}
    \label{fig:kraken_latency_TT_cdf}
  \end{subfigure}
  
  \caption{CDFs of the accumulated volume and hypothetical gains and losses of OWA candidate sequences on Binance and Kraken as a function of $\Delta t$, the latency between the two operations in the sequence (with $\Delta q \in [-0.1, 0.1]$).
  Gains tend to be concentrated in shorter latencies, while losses occur more for longer latencies.}
  \label{fig:latency_cdfs}
\end{figure*}

Just by restricting OWA candidates to quantity differences in $[-0.1, 0.1]$\,bps, we have already selected a subset that is ``profitable'' in aggregate, that is, the hypothetical total gains exceed total losses (not accounting for fees).
However, given that we assume OWA to be competitive, it is likely that OWA activity occurs faster than our computationally motivated initial latency threshold of 2\,s.
To determine a tighter threshold for latency that reduces false positives, Figure~\ref{fig:latency_cdfs} plots the three metrics (volume, gains and losses) as a function of the $\Delta t$ of any OWA candidate sequence with $\Delta q \in [-0.1, 0.1]$\,bps.
On Binance, the vast majority of volume (80.3\,\% for \TT and 57.8\,\% for \MT) and gains (82.6\,\% and 69\,\%, respectively) occur among OWA candidates with $\Delta t \leq 35$\,ms.
The majority of losses, however, accrues in OWA candidates with longer latencies (the halfway point being 300\,ms for \TT and 240\,ms for \MT).
That is, true OWA activity likely has low latency, which is consistent with the hypothesis that competition drives speed.
OWA candidates at higher latencies may constitute a mix of background noise and ``unsuccessful'' OWA, for example traders losing the race for a profitable second order out of the intermediary coin and taking a (measured) loss instead.

On Kraken, volume and gains are also more concentrated among shorter latencies (and losses among longer latencies), but less so than on Binance.
OWA on Kraken appears to be slower, and potentially less competitive than on Binance.

On both exchanges, \MT sequences tend to be slower than \TT sequences.
This is in line with the execution semantics of orders, given that maker orders are filled after an indeterminate amount of time, while taker orders are filled instantly.
The time taken to fill the first order in OWA sequences does not directly affect $\Delta t$, but the uncertainty of maker orders affects how quickly the second order can be sent (\eg after receipt of notification).

\begin{figure*}[tp]
  \centering
  \begin{subfigure}[t]{0.48\textwidth}
    \centering
    \includegraphics[width=\textwidth]{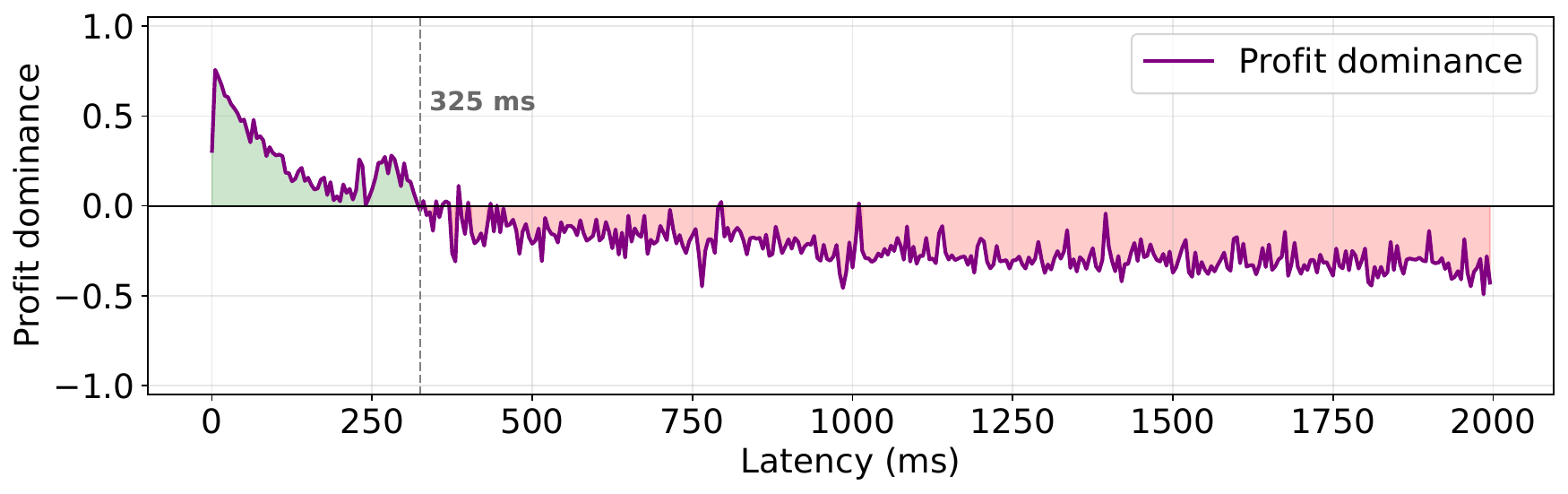}
    \caption{Binance \MT}
    \label{fig:binance_MT_profit_dominance}
  \end{subfigure}
  \hfill
  \begin{subfigure}[t]{0.48\textwidth}
    \centering
    \includegraphics[width=\textwidth]{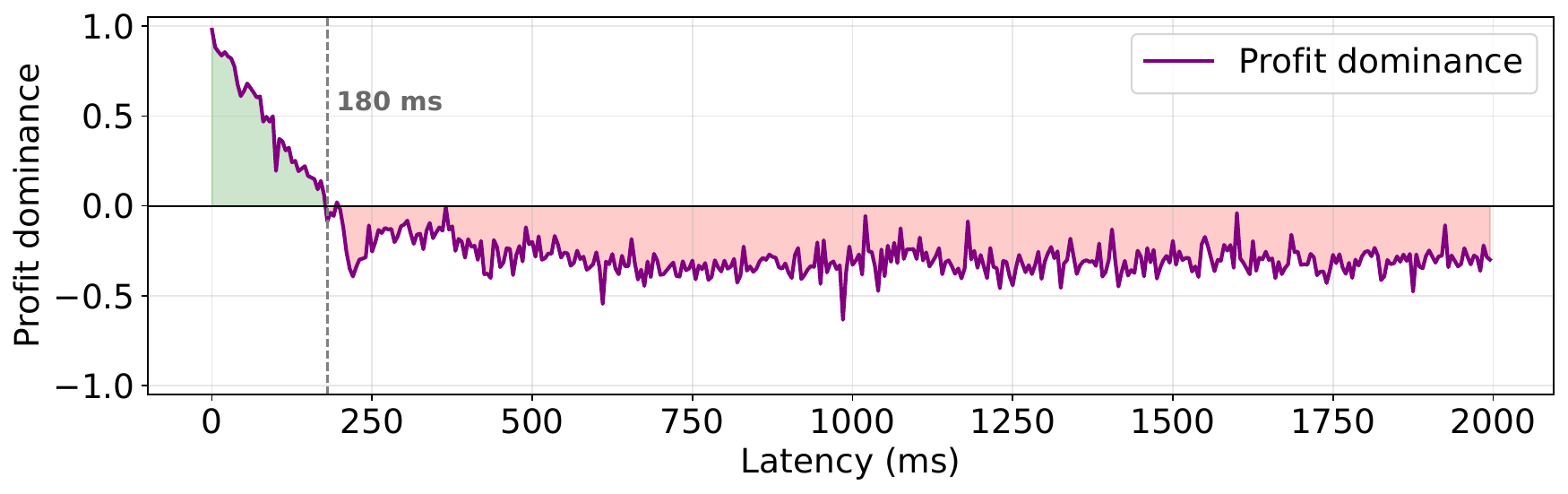}
    \caption{Binance \TT}
    \label{fig:binance_TT_profit_dominance}
  \end{subfigure}

  \begin{subfigure}[t]{0.48\textwidth}
    \centering
    \includegraphics[width=\textwidth]{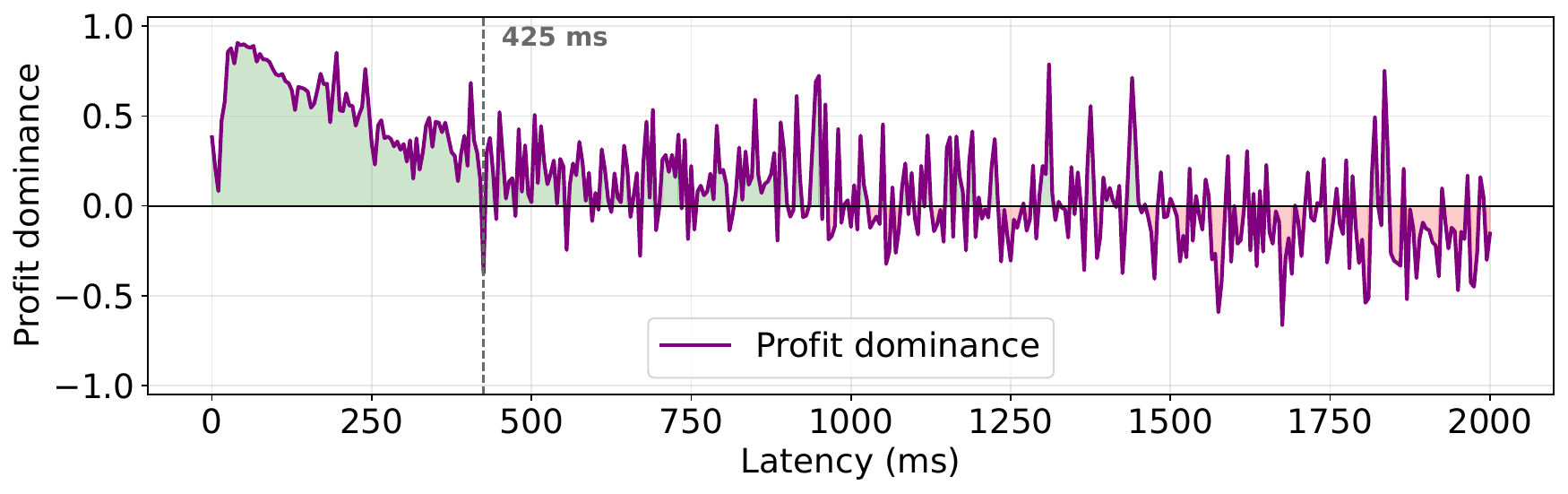}
    \caption{Kraken \MT}
    \label{fig:kraken_MT_profit_dominance}
  \end{subfigure}
  \hfill
  \begin{subfigure}[t]{0.48\textwidth}
    \centering
    \includegraphics[width=\textwidth]{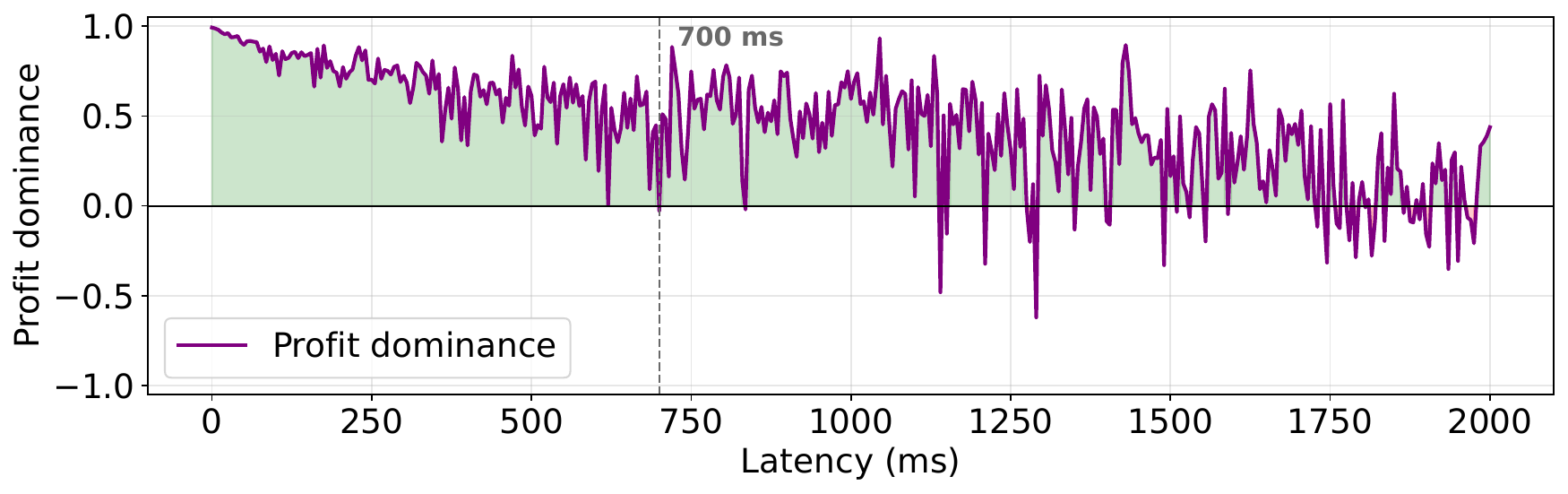}
    \caption{Kraken \TT}
    \label{fig:kraken_TT_profit_dominance}
  \end{subfigure}

  \caption{Profit dominance of OWA candidate sequences with $\Delta q \in [-0.1, 0.1]$ (in 5\,ms latency bins).
  Positive areas shaded green, negative red.
  On Binance, sequences with $\Delta t \leq 180$\,ms (\TT) and 325\,ms (\MT) are positive, meaning that in aggregate there are more gains than losses, whereas for longer latencies, losses begin to dominate.
  On Kraken, the picture is murkier, as there is more oscillation between profitable and lossy latency bins.}
  \label{fig:profit_dominance}
\end{figure*}

The CDFs above allow us to measure when gains (and losses) accrue over latencies, but they do not reveal the proportions between these two metrics.
To select a latency threshold, ultimately we would like to know whether hypothetical gains exceed losses in specific latency bins (since we already know that gains exceed losses overall).
To that end, we introduce a new \emph{profit dominance} metric that combines gains and losses.
It is defined as $\frac{\sum_i \Delta R_i}{\sum_i |\Delta R_i|}$ for all candidate sequences $i$, \ie the (signed) gross return normalized for a more readable scale.
It ranges from -1 (100\,\% losses) to +1 (100\,\% gains), with 0 being an equal amount of gains and losses (not accounting for fees).
We calculate it in 5\,ms latency intervals.
Figure~\ref{fig:profit_dominance} shows a very clear picture for Binance: All 5\,ms latency bins up to 180\,ms for taker/taker (and 325\,ms for maker/taker) are positive, whereas nearly all remaining bins are negative. That is, there are clear thresholds after which hypothetical losses outweigh gains.
Since we assume that OWA is done for financial gain, and that arbitrageurs would not continue to operate under parameters that are unprofitable in the long run, we discard OWA candidates with higher latencies on Binance.
To avoid overfitting, we round these thresholds
up to 200\,ms and 400\,ms, respectively.

On Kraken, the picture is not as clear.
While there are thresholds until which all latency bins are positive (425\,ms for \MT and 700\,ms for \TT), negative bins are initially the exception, and so we select larger thresholds.
For \MT on Kraken, we select 1\,s as the threshold because most of the following bins are negative.
For \TT, hypothetical gains exceed losses for most latency instants up to at least 2\,s.
In line with our previous findings of relatively less and slower OWA activity on Kraken, we do not introduce a stricter latency threshold for \TT on Kraken, and retain the initial computationally motivated 2\,s window.

\subsection{Selected Thresholds}

\begin{table}
\small
    \centering
    \caption{Selected thresholds for OWA sequences.}
    \label{tab:selected_thresholds}
    \begin{tabular}{l l r r}
        \toprule
         CEX & Mode & Qty diff (bps)  & Latency (ms) \\
        \midrule
        Binance & \MT & $[-0.1, 0.1]$ & $[0, 400]$ \\
        
        Binance & \TT & $[-0.1, 0.1]$ & $[0, 200]$ \\ 

        \midrule
        
        Kraken & \MT & $[-0.1,0.1]$ & $[0, 1000]$ \\ 

        Kraken & \TT & $[-0.1,0.1]$ & $[0, 2000]$ \\ 
      
        \bottomrule
    \end{tabular}
\end{table}

Table~\ref{tab:selected_thresholds} summarizes the selected thresholds for quantity difference and latency in OWA sequences.
We apply these thresholds to the full OWA candidate sets on Binance and Kraken to derive the OWA data sets used in the following analysis.
This results in more than 402\,M likely OWA sequences on Binance, and almost 2\,M on Kraken.

%% file: sections/05-analysis.tex
\section{Analysis}
\label{sec:analysis}

\mypara{Our data sets show that OWA is an actively used trading strategy.}
The volume of the detected OWA sequences reaches nearly \$190\,B over the 5~years and 10~months on Binance, and \$1.7\,B over the 9~years and 9~months on Kraken.
This corresponds to approximately 0.94\,\% and 0.13\,\% of the total traded volume on Binance and Kraken, respectively.

On Binance, \TT makes up the majority of OWA volume (72.4\,\%) and OWA sequences (68.3\,\%).
Kraken exhibits a more balanced pattern with \TT accounting for approximately 52\,\% of OWA sequences and volume.

\mypara{Longitudinally, the counts and volume of OWA sequences follow a similar pattern as the overall market} (comparing Figures~\ref{fig:longitudinal_operations_all} and~\ref{fig:longitudinal_volume_all} in the appendix to Figure~\ref{fig:exchanges_volume_and_pairs}), with some outliers in the beginning of the data set when there was less overall trading activity.
On both exchanges, during the major bull run\footnote{https://www.weforum.org/stories/2022/01/top-cryptocurrencies-performance-2021/} between the end of 2020 and the first half of 2021, there was an increase in OWA activity, with the number of sequences increasing between approximately 100\,\% and 140\,\%, and volume increasing around 200\,\%.

\mypara{The dollar amounts traded in individual OWA sequences are relatively low.}
On Binance, median volumes are \$97 per \MT sequence and \$160 for \TT, while the means are \$411 and \$500, respectively.
(By definition, the capital used in an OWA sequence is approximately half that.)
On Kraken, median volumes are higher at \$291 for \MT and \$216 for \TT, while the means are \$907 and \$844.
This is likely a reflection of the size of available OWA opportunities, \ie the amount of liquidity with a pricing discrepancy, which appears to be larger on Kraken (though less frequently available or exploited).

\mypara{The top intermediary coins} in terms of OWA volume are ADA, DOGE and SOL on Binance, and DOGE, XRP and ADA on Kraken.
To determine the intermediary coins with the largest price discrepancies (independent of volume), we first define the \emph{relative gross return} of an OWA sequence as $\Delta p = \frac{p_{A_1IA_2}}{p^r_{A_1A_2}} - 1$, with $p_{A_1IA_2} = p_{A_1I} \cdot p_{IA_2}$ the (gross) price of the OWA sequence relative to the reference price $p^r_{A_1A_2}$ of a direct conversion (using the 2\,s VWAP as defined in Section~\ref{subsec:owa_refining}).
It is exclusive of fees because the focus here is on market conditions.
The \emph{cumulative price imbalance} of an intermediary coin $I$ is computed as the summed absolute values of $\Delta p$ across all observed OWA sequences passing through $I$.
The top intermediary coins by exploited cumulative price imbalance are DOGE, SOL and ADA on Binance, and LUNA, DOGE and XRP on Kraken.
(Table~\ref{appendinx:tab:intermediary_coins} in the appendix shows the top~10.)
While Binance and Kraken share five intermediary coins in the top~10 volume ranking, they share only three in the cumulative price imbalance ranking.
This could indicate that price fluctuations are, at least in part, specific to an exchange, likely reflecting differences in liquidity conditions and micro trading activity, whereas volume is more similar due to both exchanges mirroring the macro trading activity of the broader cryptocurrency market.

\subsection{Profits}
\label{sec:analysis:profits}

\begin{figure*}[tp]
  \centering
  \begin{subfigure}[t]{0.48\textwidth}
    \centering
    \includegraphics[width=\textwidth]{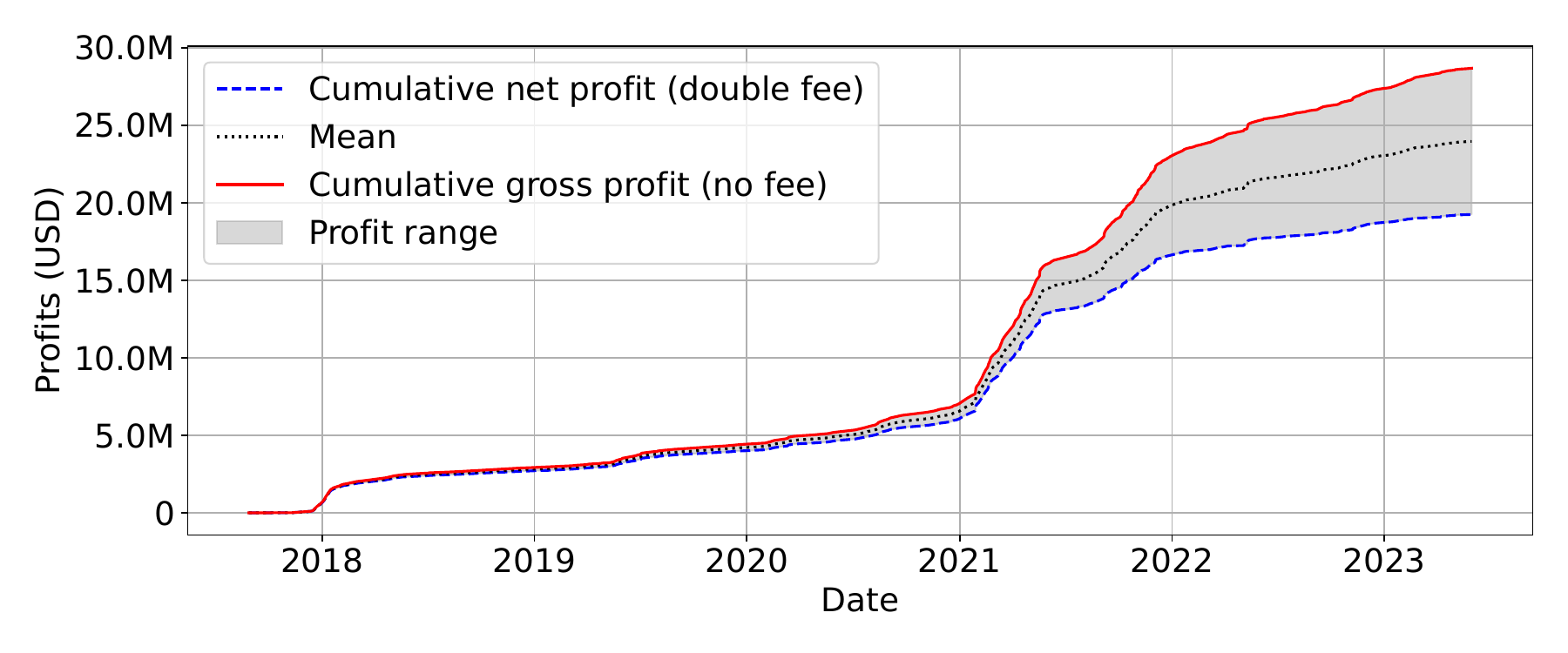}
    \caption{Binance \MT}
    \label{fig:binance_owa_profits_comparison_mt}
  \end{subfigure}
  \hfill
  \begin{subfigure}[t]{0.48\textwidth}
    \centering
    \includegraphics[width=\textwidth]{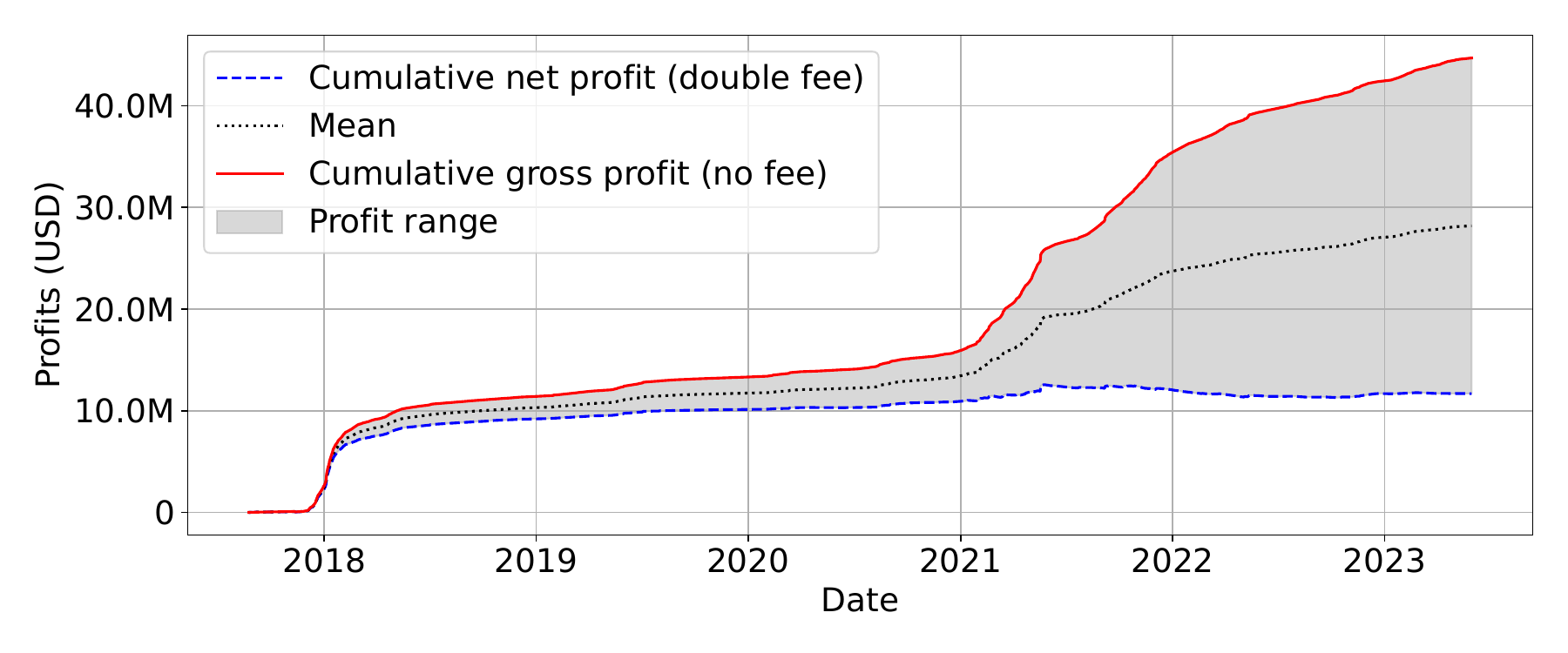}
    \caption{Binance \TT}
    \label{fig:binance_owa_profits_comparison_tt}
  \end{subfigure}

  \vspace{0.5em}

  \begin{subfigure}[t]{0.48\textwidth}
    \centering
    \includegraphics[width=\textwidth]{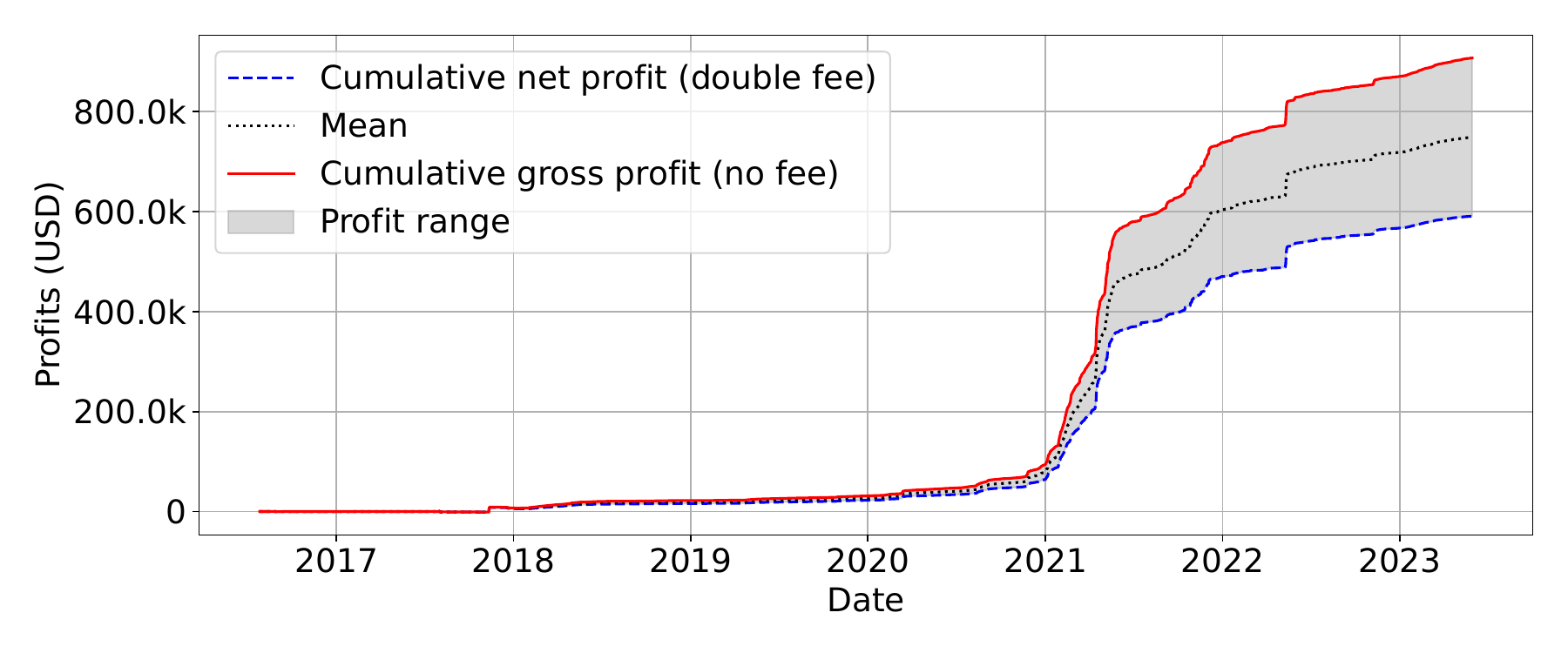}
    \caption{Kraken \MT}
    \label{fig:kraken_owa_profits_comparison_mt}
  \end{subfigure}
  \hfill
  \begin{subfigure}[t]{0.48\textwidth}
    \centering
    \includegraphics[width=\textwidth]{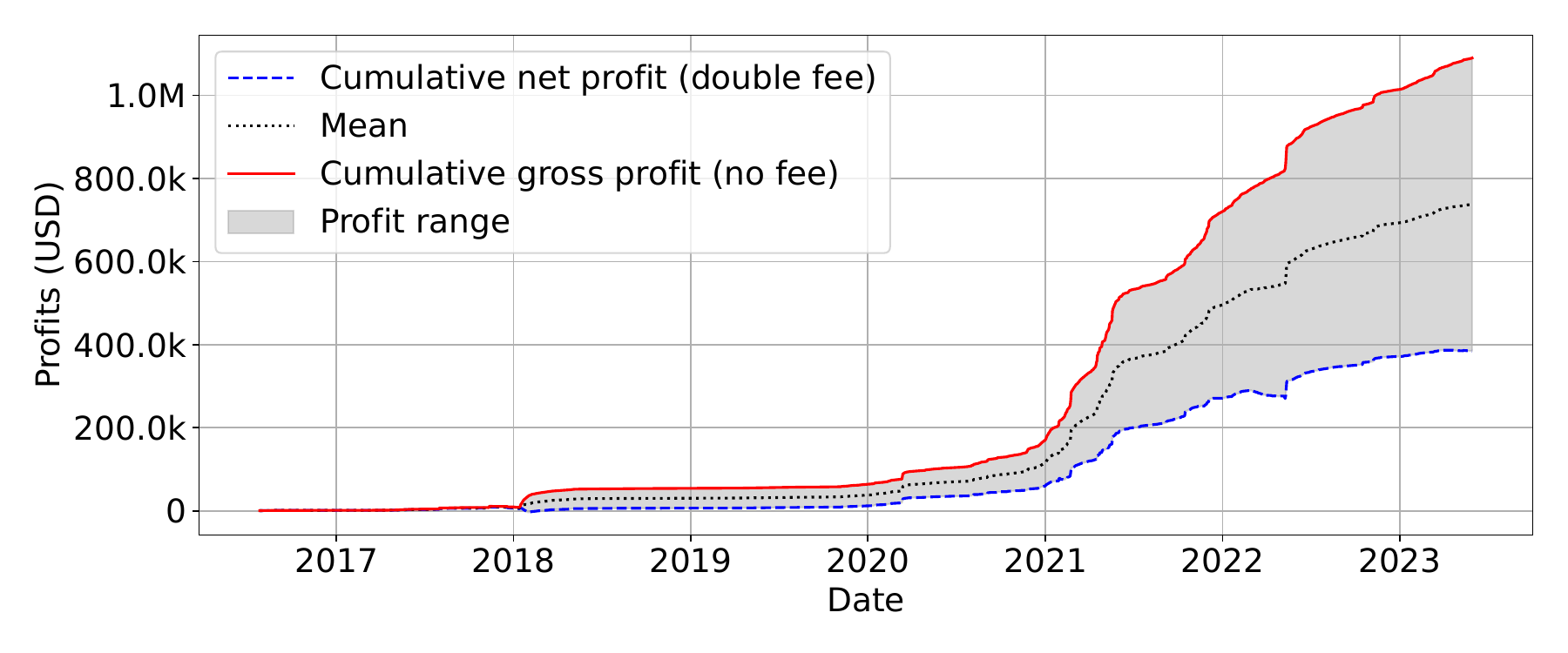}
    \caption{Kraken \TT}
    \label{fig:kraken_owa_profits_comparison_tt}
  \end{subfigure}

  \caption{Longitudinal estimated OWA profits (cumulative), with the ``no fee'' scenario as the upper bound, and the ``double fee'' scenario as the lower bound. The gray dotted line represents the mean of the two scenarios.
  On both exchanges, \TT ``double fee'' cumulative profits experienced prolonged periods of stagnation and even brief decline (\ie losses), which suggests a more difficult environment than \MT, but could also indicate the ``double fee'' assumption is too conservative.
  (Non-cumulative gross profits are shown in Figure~\ref{fig:longitudinal_profits_all} in the appendix.)
  }
  \label{fig:longitudinal_profits_comparison_all}
\end{figure*}

To estimate OWA profits, \ie the difference between the output of the indirect conversion and a hypothetical direct conversion, we initially consider a ``best case'' scenario in which traders do not pay any fees.
We estimate gross profits of \$73.3\,M on Binance and below \$2.0\,M on Kraken over the entirety of our data sets.
On both exchanges, \TT sequences contribute the largest share, accounting for 60.9\,\% (\$44.6\,M) of gross profits on Binance, and 56.9\,\% (\$1.1\,M) on Kraken.
Longitudinally, gross profits broadly mirror the trends in overall trading volume (Figure~\ref{fig:exchanges_volume_and_pairs}), with a marked increase beginning in 2021, as shown in Figure~\ref{fig:longitudinal_profits_comparison_all} (red lines).

\mypara{Incorporating fees.}
While these gross estimates suggest substantial potential returns, they are unrealistic in practice as trades on a CEX usually incur transaction costs.
Unfortunately, the public API data does not include explicit information on the transaction fees paid.
We approximate these costs based on each exchange’s published fee schedule, and on the assumption that OWA traders are large-volume participants (such as liquidity providers) who qualify for the most favorable publicly available fee tiers on the exchanges (\ie taker fees of 2.4\,bps on Binance and 8\,bps on Kraken, and maker fees of 1.2\,bps and 0.0\,bps, respectively).
While each OWA sequence incurs two fees, we do not include any fee adjustment when calculating the reference price for direct conversions.
This choice reflects a conservative modeling assumption that OWA traders are not primarily interested in converting assets, but rather in opportunistically capitalizing on transient mispricings between asset pairs.
Thus, OWA sequences need to yield sufficient profit margins to
offset two transaction fees in order to be considered net profitable.

\mypara{Accounting for fees leads to a sharp reduction in profits.}
Total profits drop from over \$73\,M without fees to \$31.2\,M when two fees are paid on Binance, and from \$2\,M to \$975\,k on Kraken.
Because taker fees are higher, the distribution of profits across order-type combinations also shifts: \MT sequences become dominant, contributing \$19.4\,M (62.1\,\% of net profits) on Binance, and \$590\,k (60.5\,\%) on Kraken.

\mypara{On Binance, cumulative net profits from \TT sequences plateaued after Q1~2021} and even declined between the end of November~2021 and October~2022, indicating losses under the conservative ``double fee'' scenario (Figure~\ref{fig:binance_owa_profits_comparison_tt}, blue line).
Yet, \TT volume remained high during these two years, which would be unlikely if profit-oriented actors consistently encountered actual losses.
It is more likely that some arbitrageurs pay less than the lowest published fee, or have found other ways to offset these fees.
For example, large liquidity providers operating on CEXes might benefit from private arrangements granting lower fees or even rebates for market-making activities.\footnote{https://www.binance.com/en/vip-portal/liquidity-hub}
Furthermore, if arbitrageurs actually intend to convert (\eg on behalf of a customer who will pay a conversion fee), the profitability calculation becomes less conservative than our ``double fee'' assumption.
This highlights the sensitivity of OWA profitability to fee structures, where even small differences can change aggregate profits by the order of millions of dollars.

While Kraken exhibits a 10-week period of overall losses in H1~2022, in contrast to Binance, \TT OWA still appears to be profitable under the ``double fee'' scenario, despite exchange fees that are significantly higher than on Binance.
As we will see below, this is driven by larger price discrepancies of individual \TT sequences, which make them more profitable than on Binance after deducting fees.

\mypara{In absolute terms, the profits associated with individual OWA sequences are generally small}, and they are higher on Kraken than on Binance, consistent with the larger per-sequence volume observed in the beginning of the analysis.
Excluding fees, the mean profit per sequence is \$0.22 for \MT and \$0.16 for \TT on Binance, while on Kraken both configurations yield around \$1 on average.
When trading fees are deducted, per-sequence profits decrease across all configurations: on Binance, means drop to \$0.15 for \MT and remain slightly positive at \$0.04 for \TT, while on Kraken they decrease to \$0.68 and \$0.36, respectively.
Arbitrageurs need to identify and execute large numbers of OWA sequences in order to accumulate profits that can offset their fixed costs (\eg labor, infrastructure).

\begin{figure*}[t]
  \centering
  \begin{subfigure}[t]{0.48\textwidth}
    \centering
    \includegraphics[width=\textwidth]{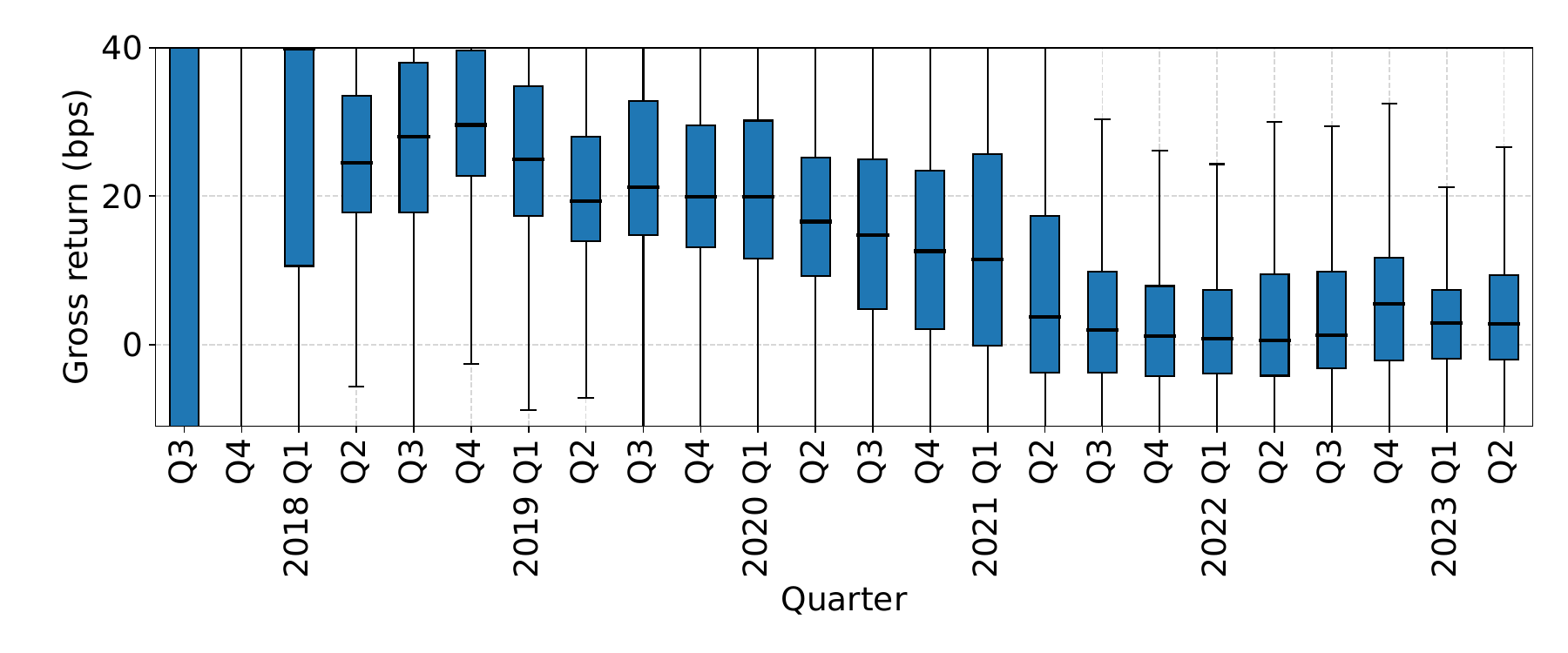}
    \caption{Binance \MT}
    \label{fig:boxplot_bps_returns_MT}
  \end{subfigure}
  \hfill
  \begin{subfigure}[t]{0.48\textwidth}
    \centering
    \includegraphics[width=\textwidth]{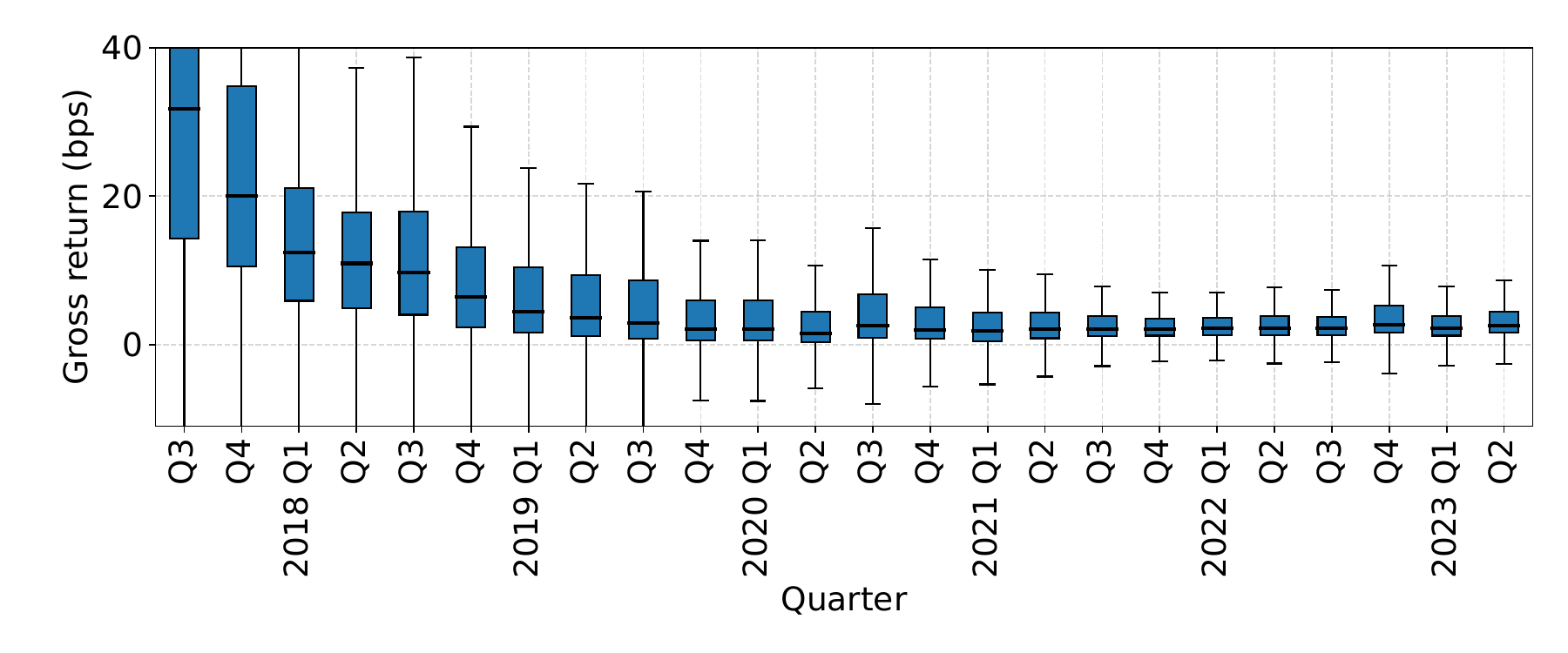}
    \caption{Binance \TT}
    \label{fig:boxplot_bps_returns_TT}
  \end{subfigure}
  
  \caption{Quarterly distribution of relative gross returns per OWA sequence on Binance.
  The discrepancy between the indirect price and the reference price has decreased and become less variable over time, suggesting the market has become more balanced.
  The range of returns for \MT remains higher than for \TT, which may be related to the higher uncertainty inherent in maker orders.
  (Kraken data in Figure~\ref{fig:boxplot_bps_returns_kraken} in the appendix.)
}
  \label{fig:boxplot_bps_returns_binance}
\end{figure*}

\mypara{On Binance, the median of relative price discrepancies (gross returns) has decreased over time for both \MT and \TT sequences, indicating that they have become less profitable}, especially when considering that the still-to-be-deducted fees have likely remained more stable.
This is visible in Figure~\ref{fig:boxplot_bps_returns_binance}, which plots the quarterly distribution of per-sequence gross returns $\Delta p$ as defined in the beginning of Section~\ref{sec:analysis}.
We also observe a general, though not strictly continuous, decrease in extreme price discrepancies, \ie instances where indirect and direct prices diverge sharply.
Large mispricings have become increasingly rare;
the minimum and maximum gross returns observed in \TT sequences have remained similar since Q3~2021.
We note that \MT sequences tend to have larger variation in terms of gross returns compared to \TT sequences.
This might be related to the higher uncertainty inherent in maker orders, as arbitrageurs cannot precisely predict which price the market will accept.
Overall, the beginning flattening of the distribution and lowering variability coincide with the peak of OWA operations and increased market activity around the first quarter of 2021.
On Kraken, the longitudinal data (Figure~\ref{fig:boxplot_bps_returns_kraken} in the appendix) exhibits no clear trend, possibly because of the smaller size of the data set.
Overall, per-sequence gross returns on Kraken are higher than on Binance, with a mean of 28\,bps for Kraken and 6\,bps on Binance.
This gap persists even in the most recent months of data (Q2 2023), where Kraken still exhibits larger gross returns of 20.4\,bps on average, while Binance averages only 3.8\,bps, potentially due to Kraken's lower overall volume.

\subsection{Timing}

\begin{figure*}[tp]
  \centering
  \begin{subfigure}[t]{0.48\textwidth}
    \centering
    \includegraphics[width=\textwidth]{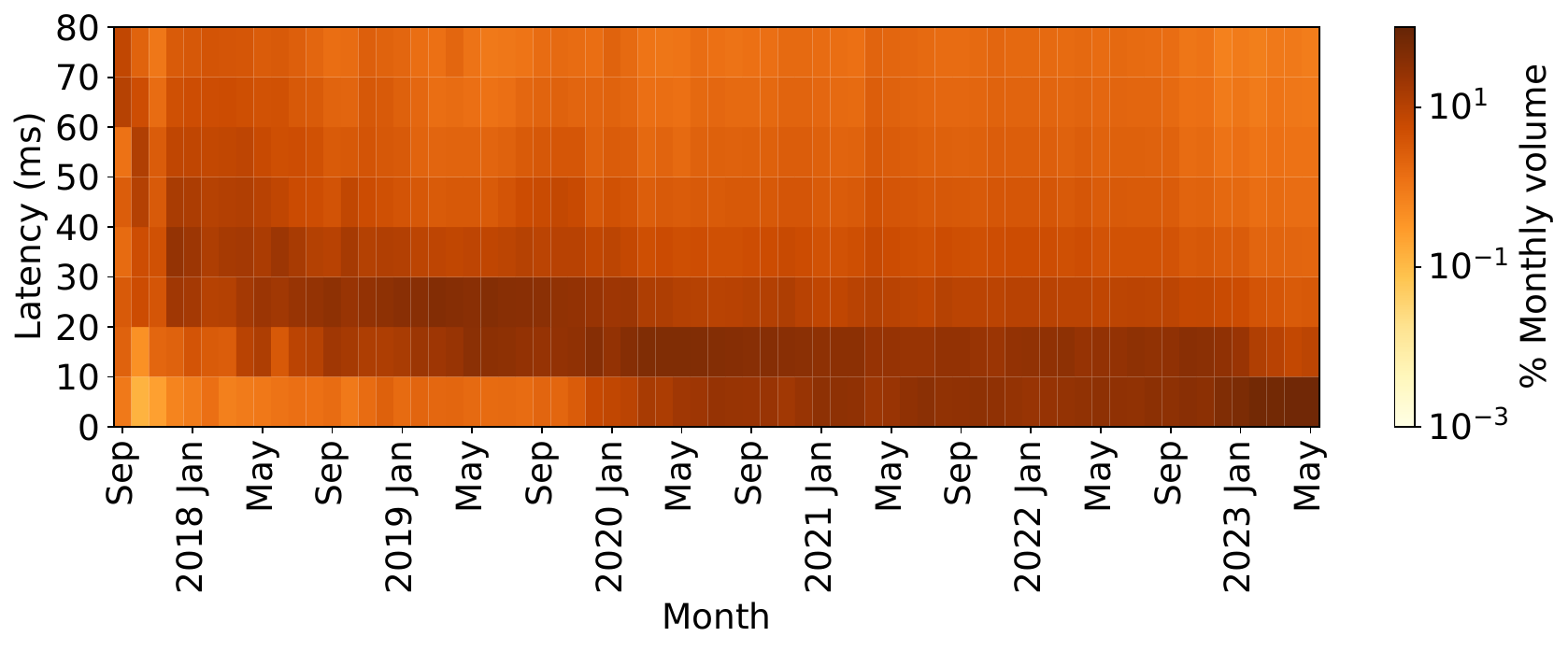}    \caption{Binance \MT}
    \label{fig:binance_owa_volume_per_bin_mt_zoom}
  \end{subfigure}
  \hfill
  \begin{subfigure}[t]{0.48\textwidth}
    \centering
    \includegraphics[width=\textwidth]{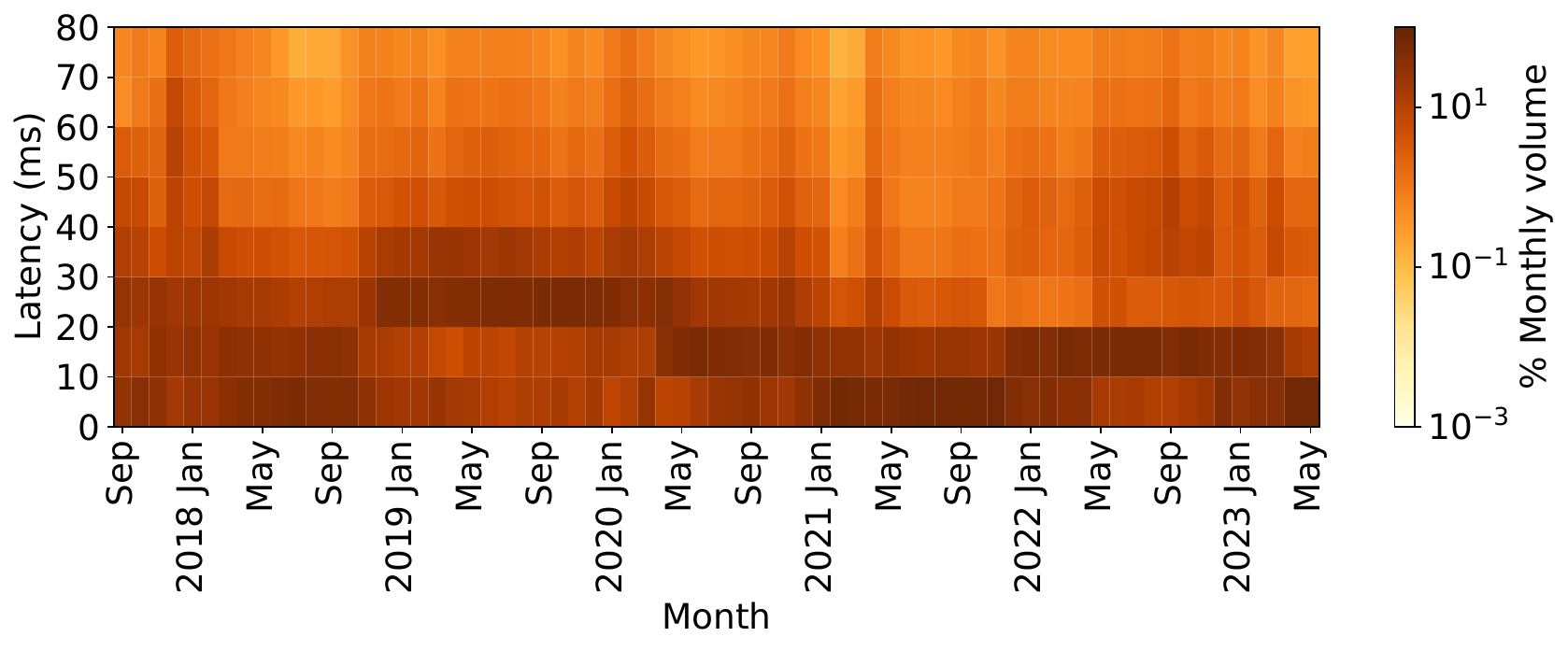}
    \caption{Binance \TT}
    \label{fig:binance_owa_volume_per_bin_tt_zoom}
  \end{subfigure}

  \vspace{0.5em}

  \begin{subfigure}[t]{0.48\textwidth}
    \centering
    \includegraphics[width=\textwidth]{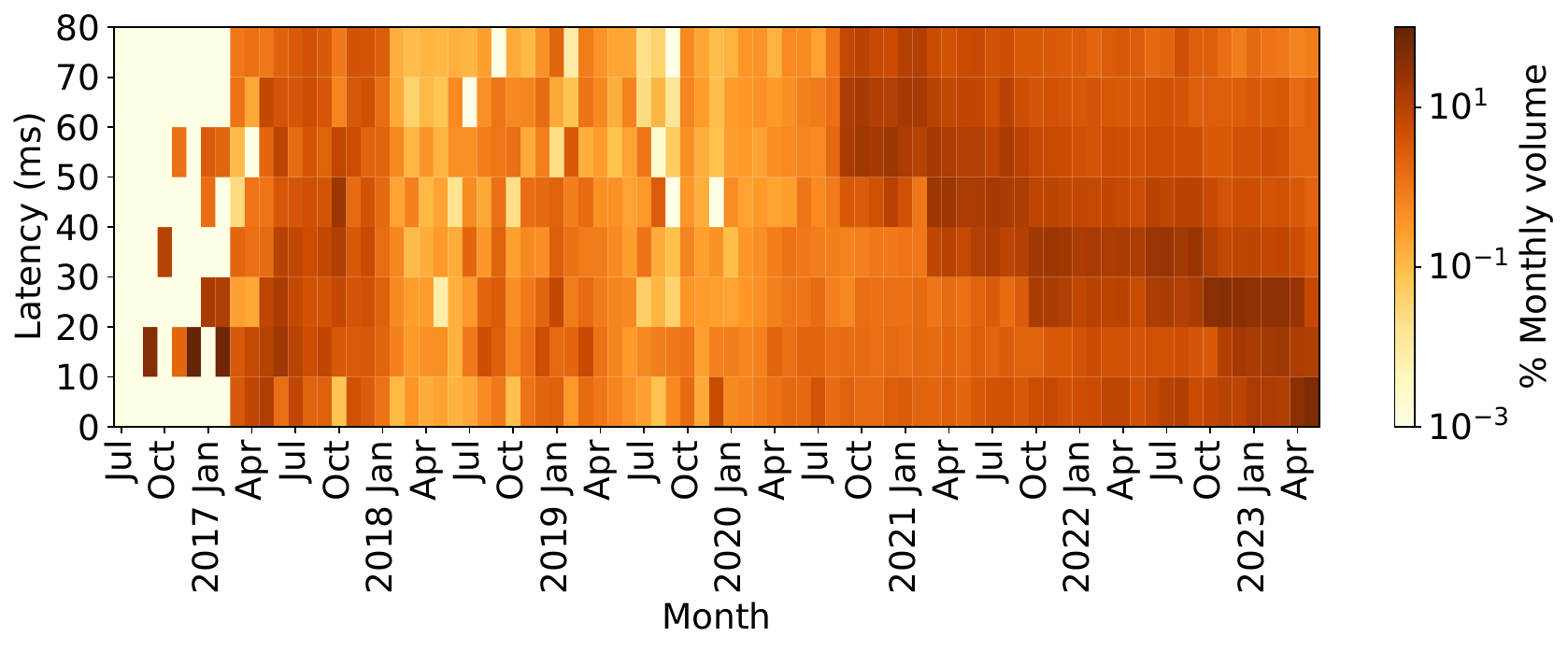}
    \caption{Kraken \MT}
    \label{fig:kraken_owa_volume_per_bin_mt_zoom}
  \end{subfigure}
  \hfill
  \begin{subfigure}[t]{0.48\textwidth}
    \centering
    \includegraphics[width=\textwidth]{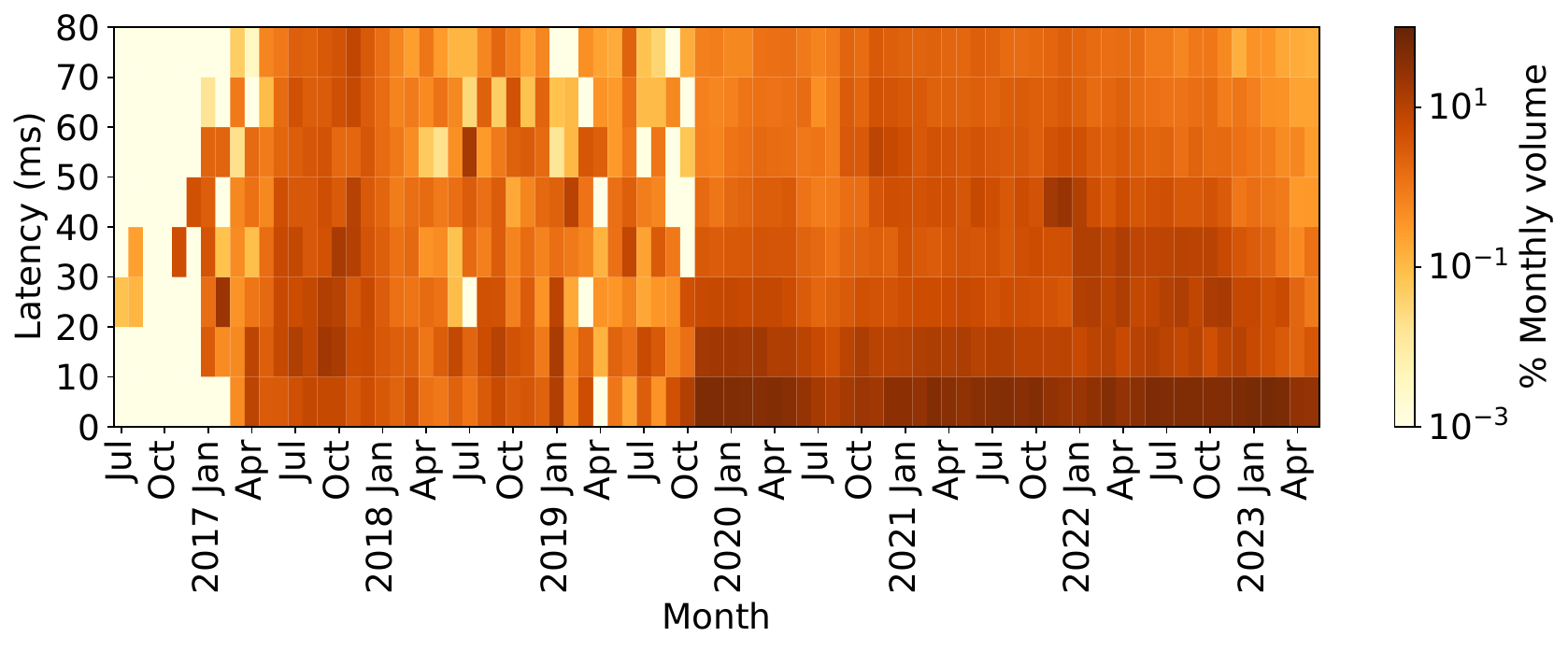}
    \caption{Kraken \TT}
    \label{fig:kraken_owa_volume_per_bin_tt_zoom}
  \end{subfigure}

  \caption{Heatmaps of monthly OWA volume by latency between the first and second operation of the sequence, for the first 80\,ms (full range in Figure~\ref{fig:heatmaps_owa_volume_latency_all} in the appendix; color in log scale).
  In the later months of the data sets, OWA volume has moved to lower latency bins, suggesting increased competition for profitable OWA opportunities.}
  \label{fig:heatmaps_owa_volume_latency_zoom}
\end{figure*}

In light of the decreasing per-sequence profits, we hypothesize that competition between arbitrageurs may have increased.
Competition drives arbitrageurs to become faster because profitable OWA opportunities are limited and first come, first served.
Figure~\ref{fig:heatmaps_owa_volume_latency_zoom} plots the latency between the first and second operation of OWA sequences as a heatmap.
While the two exchanges and \MT and \TT combinations differ in the earlier, low-volume months of the respective data set, they all show a long-term trend of OWA volume moving towards lower latency bins.
(OWA sequence counts, plotted in Figure~\ref{fig:heatmaps_owa_operations_latency_all} in the appendix, show the same trend.)
The largest concentration of \MT volume on Binance shifted from initially 70-80\,ms to 0-10\,ms, where we record almost the entire OWA volume during the last few months.
Prior work~\cite{muck2025wish} observed latencies of 4\,ms from Binance to a server located in Tokyo, \ie arbitrageurs have barely enough time for one network round-trip time between the two operations.
It is also conceivable that arbitrageurs have developed strategies that do not require them to wait until they are notified by the exchange that their maker order has been filled before they send the second order for the OWA sequence (by sending the second order proactively).

%% file: sections/06-discussion.tex
\section{Discussion}

Our analysis of trading patterns on Binance and Kraken has shown that OWA is carried out at a significant volume, for example almost \$190\,B during the first 5~years and 10~months of Binance, making up 0.94\,\% of the total traded volume on the spot exchange.
In other words, OWA is not merely a theoretical possibility.
Over time, OWA has become faster, likely due to increased competition among arbitrageurs, all the while profitability has declined.
A single instance of OWA yields a net return in the order of pennies on average---in order to turn meaningful profits, arbitrageurs need to execute OWA at scale.
At this point, OWA on Binance (and to a lesser degree on Kraken) appears to have become professional enough that it is out of reach of hobbyist traders, since it takes effort and upfront investment to get started: arbitrageurs need to execute profitable opportunities at high speed and high volume, and possibly even negotiate lower trading fees with the exchange because of the slim margins between gross returns and even the lowest public fee levels.
Our data contains hints that this may already be the case for some arbitrageurs, which in turn indicates that the exchange sees value in the arbitrageurs' activity.
A full discussion of what that value may be is beyond the scope of this measurement paper, but a look at scenarios that may give rise to profitable OWA opportunities can provide some context.

OWA exploits pricing discrepancies of an asset in one trading pair compared to other pairs.
In other words, out of all order books in which an asset is traded, one has the asset at a price that is inconsistent with the remaining order books.
It is unclear yet when precisely such circumstances arise.
Some examples include (1) irrational, abnormal, or non-optimal trading behavior, for example a trader quickly purchasing (or selling) a quantity so large that it depletes a specific order book, causing its prices to become more expensive in one direction (and cheaper in the other), \eg pump and dump schemes, and (2) technical reasons such as the granularity of price steps supported by the exchange, when the value of an asset has become so large relative to the supported step size that granular price changes are no longer possible in one trading pair (but still possible in other pairs).
We leave an in-depth exploration of these hypotheses for future work.
For now, we note that exploiting price discrepancies for OWA effectively removes them from the affected order books, and the prices of affected assets (in our case, intermediary coins) become more balanced, \ie consistent across order books, which is generally seen as a sign for a healthy market~\cite{gromb2010limits}.
As such, arbitrageurs may be providing a service that the exchange cannot ethically carry out itself, \ie balancing order books and adding liquidity where it is needed.

\subsection{Limitations}
A key constraint of our analysis lies in the anonymized nature of the trade data, which prevents us from directly grouping trades executed by the same user.
As a result, our detection of OWA sequences relies on heuristics; while the data exhibit patterns that suggest a deliberate trading strategy, we do not have ground truth for confirmation.
Furthermore, our heuristics only capture the most straightforward form of OWA; we have excluded sequences with fees paid in band, and cannot detect sequences combining multiple inputs or splitting into multiple outputs, or conducting OWA with only anchor coins, for instance.
Due to our threshold selection being based on profitability, we cannot quantify unprofitable sequences or failed OWA attempts (\eg due to slippage), which prevents us from estimating losses.
Fees are another unknown, since they are not directly represented in the trade data.
As a result, we can only estimate a range of possible fee scenarios without knowing precisely which costs arbitrageurs face in practice.
For example, smaller-scale arbitrageurs might be trading at a fee level more costly than the lowest public fee.
Our analysis is based on actually executed trades; due to the lack of order book data, we cannot fully reconstruct arbitrageurs' view of market conditions, such as available depth or slippage.
Given these limitations, it is clear that our data sets cover only a subset of the multiple forms of OWA that may exist in the wild.
We envision our methodology as a foundation to be extended in future work to detect different forms of algorithmic trading.

%% file: sections/07-related-work.tex
\section{Related Work}
\label{sec:related_work}
Prior work has studied market inefficiencies on~\cite{pieters2017financial} and across~\cite{makarov2020trading} CEXes, \ie the frequency of price differences that lead to ``theoretical'' arbitrage opportunities.
In line with our findings, prior work has reported that these arbitrage opportunities have become less frequent and less profitable over time as markets have become more efficient~\cite{crepelliere2023arbitrage}.
Yet, exchange outages can create substantial price discrepancies and result in arbitrage opportunities across markets~\cite{morin2023cryptocurrency}.
In contrast to our data sets spanning 5--9~years and including thousands of tradable pairs (Table~\ref{tab:datasets_data}), prior studies were limited in scale or duration~\cite{muck2025wish,nan2019bitcoin,czaplinski2019using}.
For example, Muckl et al.~\cite{muck2025wish} calculated triangular arbitrage opportunities on Binance between Bitcoin, Litecoin, and the U.S.\ Dollar during a one-week window and found that most theoretical opportunities were unprofitable after accounting for exchange fees (\eg totaling a net return of \$170.76 for a trader paying the lowest level of fees).
All of these studies on CEXes have left open whether arbitrage opportunities are actually exploited by traders, and none considered OWA as an arbitrage strategy.
Based on our data of settled trades, we estimate OWA to occur at a larger scale and with higher profits, possibly because OWA is easier to exploit than triangular arbitrage, requiring only two price-aligned trades instead of three.

In contrast to the line of work on arbitrage opportunities on CEXes, prior work on crypto arbitrage on DEXes has measured actual on-chain and cross-chain trades and profits observed in public ledgers (\eg \cite{daian2020flash,qin2022quantifying,wang2022cyclic,heimbach2024non,mclaughlin2023large}).
While these works studied various arbitrage strategies, OWA was not considered.
Furthermore, decentralized trading occurs at a much smaller scale than on centralized exchanges~\cite{cexes_vs_dexes}.
As a result, reported decentralized arbitrage volume tended to be lower than our observations of OWA on CEXes.

Similar to our methodology inferring OWA activity from anonymized CEX spot trade data, prior studies in the cryptocurrency community have encountered the need to develop heuristics that can link transactions in the absence of hard identifiers, \eg by combining temporal proximity and value matching to infer when two transfers belong to the same cross-ledger flow~\cite{yousaf2019tracing}, or more generally to study privacy leakages and trace payment flows in privacy-focused cryptocurrencies such as Monero~\cite{moser2017empirical,kumar2017traceability} and Zcash~\cite{quesnelle2017linkability,kappos2018empirical}.
To the best of our knowledge, our work is the first to apply such techniques at scale to CEX spot trade data.

While these studies have provided important insights into potential mispricings, DEX-visible arbitrage, and inference on anonymized transaction data, none analyzed executed arbitrage at scale within a CEX. Our work fills this gap by reconstructing indirect conversion paths from spot trades.
This is an updated version of~\cite{grimberg2020empirical}.

%% file: sections/08-conclusion.tex
\section{Conclusion}
In this paper, we have presented a novel methodology to measure OWA in anonymized spot trade data from large CEXes.
It matches trades based on their timing and exchanged quantity, and then refines these thresholds based on the profitability of the resulting OWA sequences.
We have produced two data sets of OWA sequences on Binance and Kraken.
Our analysis of these data sets shows that OWA is an actively used trading strategy accounting for 0.94\,\% of the total traded volume on Binance and 0.13\,\% on Kraken.
Net profits from OWA are relatively low on an individual basis (in the order of cents), but scale up to \$31.2\,M on Binance and \$975\,k on Kraken over the course of 70 and 117~months, respectively.
OWA has become faster, likely the result of increased competition, while pricing discrepancies (\ie profitability) have declined, which can indicate these markets are maturing.

%% file: sections/AA-appendix.tex
\begin{table*}[htbp]
  \noindent
  \begin{minipage}[t]{0.475\textwidth}

\section{Appendix}
\subsection{Ethics}
This paper is based on publicly available and anonymized trade data collected from Binance and Kraken’s public APIs.
The data does not contain any personally identifiable information, and it is not possible to reconstruct or infer any linkage to individual users.

  \end{minipage}%
  \hfill
  \begin{minipage}[t]{0.475\textwidth}

\centering
\caption{Lowest public maker and taker fees on Binance and Kraken. (Fees for low-volume traders are higher.)}
\label{tab:appendix:fees_table}
\small
\begin{tabular}{lrr}
\toprule
\textbf{Exchange} & \textbf{Maker Fee} & \textbf{Taker Fee} \\
\midrule
Binance & 0.012\,\% & 0.024\,\% \\
Kraken  & 0.000\,\%  & 0.080\,\%  \\
\bottomrule
\end{tabular}

  \end{minipage}
\end{table*}

\begin{table*}[htbp]
\vspace{2em}
\centering
\caption{Top 10 intermediary coins by total volume (in USD) and gross cumulative return or ``price imbalance'' (in bps) on Binance and Kraken. Shared coins between exchanges for the same metric are shown in \textbf{bold}. Combines \MT and \TT sequences.}
\label{appendinx:tab:intermediary_coins}
\renewcommand{\arraystretch}{1.1}

\begin{tabular}{rlr@{\hspace{1cm}}rlr}
\toprule
\multicolumn{3}{c}{\textbf{Binance}} & \multicolumn{3}{c}{\textbf{Kraken}} \\
\cmidrule(lr{1cm}){1-3} \cmidrule(r){4-6}
\textbf{\#} & \textbf{Coin} & \textbf{Volume (USD)} & \textbf{\#} & \textbf{Coin} & \textbf{Volume (USD)} \\
\midrule
1 & \textbf{ADA} & 9.66$\times$10$^9$ & 1 & \textbf{DOGE} & 1.80$\times$10$^8$ \\
2 & \textbf{DOGE} & 9.27$\times$10$^9$ & 2 & XRP & 1.47$\times$10$^8$ \\
3 & \textbf{SOL} & 7.53$\times$10$^9$ & 3 & \textbf{ADA} & 1.36$\times$10$^8$ \\
4 & LUNA & 5.66$\times$10$^9$ & 4 & \textbf{LTC} & 7.92$\times$10$^7$ \\
5 & \textbf{LTC} & 4.93$\times$10$^9$ & 5 & BCH & 6.44$\times$10$^7$ \\
6 & \textbf{LINK} & 4.90$\times$10$^9$ & 6 & \textbf{SOL} & 5.90$\times$10$^7$ \\
7 & DOT & 4.79$\times$10$^9$ & 7 & ETC & 5.57$\times$10$^7$ \\
8 & AVAX & 3.73$\times$10$^9$ & 8 & UST & 5.53$\times$10$^7$ \\
9 & MATIC & 3.70$\times$10$^9$ & 9 & FLOW & 5.31$\times$10$^7$ \\
10 & FTM & 3.20$\times$10$^9$ & 10 & \textbf{LINK} & 5.04$\times$10$^7$ \\
\bottomrule
\end{tabular}

\vspace{3em}

\begin{tabular}{rlr@{\hspace{1cm}}rlr}
\toprule
\multicolumn{3}{c}{\textbf{Binance}} & \multicolumn{3}{c}{\textbf{Kraken}} \\
\cmidrule(lr{1cm}){1-3} \cmidrule(r){4-6}
\textbf{\#} & \textbf{Coin} & \textbf{Return (bps)} & \textbf{\#} & \textbf{Coin} & \textbf{Return (bps)} \\
\midrule
1 & \textbf{DOGE} & 9.47$\times$10$^7$ & 1 & \textbf{LUNA} & 4.35$\times$10$^6$ \\
2 & SOL & 6.53$\times$10$^7$ & 2 & \textbf{DOGE} & 3.68$\times$10$^6$ \\
3 & \textbf{ADA} & 6.32$\times$10$^7$ & 3 & XRP & 3.64$\times$10$^6$ \\
4 & MATIC & 4.80$\times$10$^7$ & 4 & \textbf{ADA} & 2.75$\times$10$^6$ \\
5 & MANA & 4.61$\times$10$^7$ & 5 & BCH & 2.20$\times$10$^6$ \\
6 & \textbf{LUNA} & 4.44$\times$10$^7$ & 6 & NANO & 1.96$\times$10$^6$ \\
7 & FTM & 4.35$\times$10$^7$ & 7 & EWT & 1.89$\times$10$^6$ \\
8 & SAND & 4.01$\times$10$^7$ & 8 & UST & 1.79$\times$10$^6$ \\
9 & LINK & 3.86$\times$10$^7$ & 9 & FLOW & 1.76$\times$10$^6$ \\
10 & AVAX & 3.78$\times$10$^7$ & 10 & LTC & 1.75$\times$10$^6$ \\
\bottomrule
\end{tabular}
\end{table*}

\begin{figure*}[t]
  \centering
    \begin{subfigure}[t]{0.48\textwidth}
    \centering
    \includegraphics[width=\textwidth]{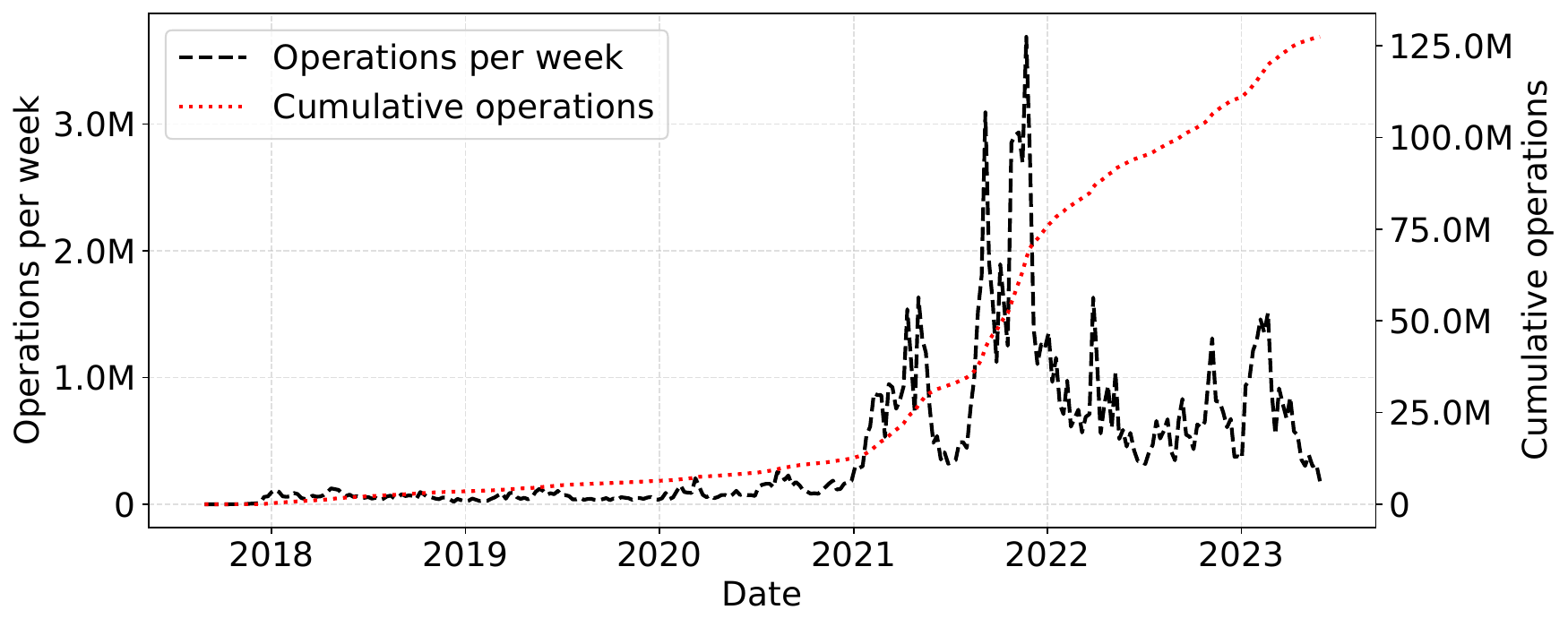}
    \caption{Binance \MT operations over time}
    \label{fig:binance_owa_operations_mt}
  \end{subfigure}
  \hfill
  \begin{subfigure}[t]{0.48\textwidth}
    \centering
    \includegraphics[width=\textwidth]{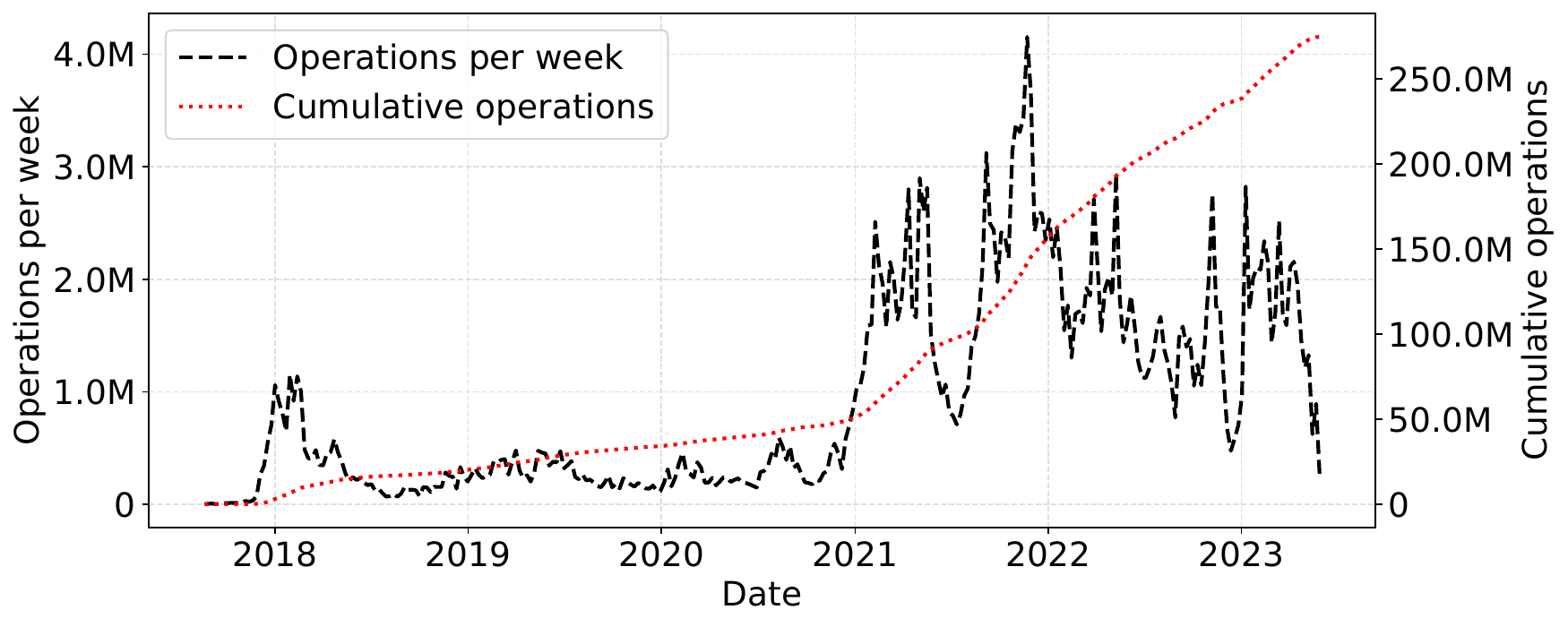}
    \caption{Binance \TT operations over time}
    \label{fig:binance_owa_operations_tt}
  \end{subfigure}

  \vspace{1.5em}

  \begin{subfigure}[t]{0.48\textwidth}
    \centering
    \includegraphics[width=\textwidth]{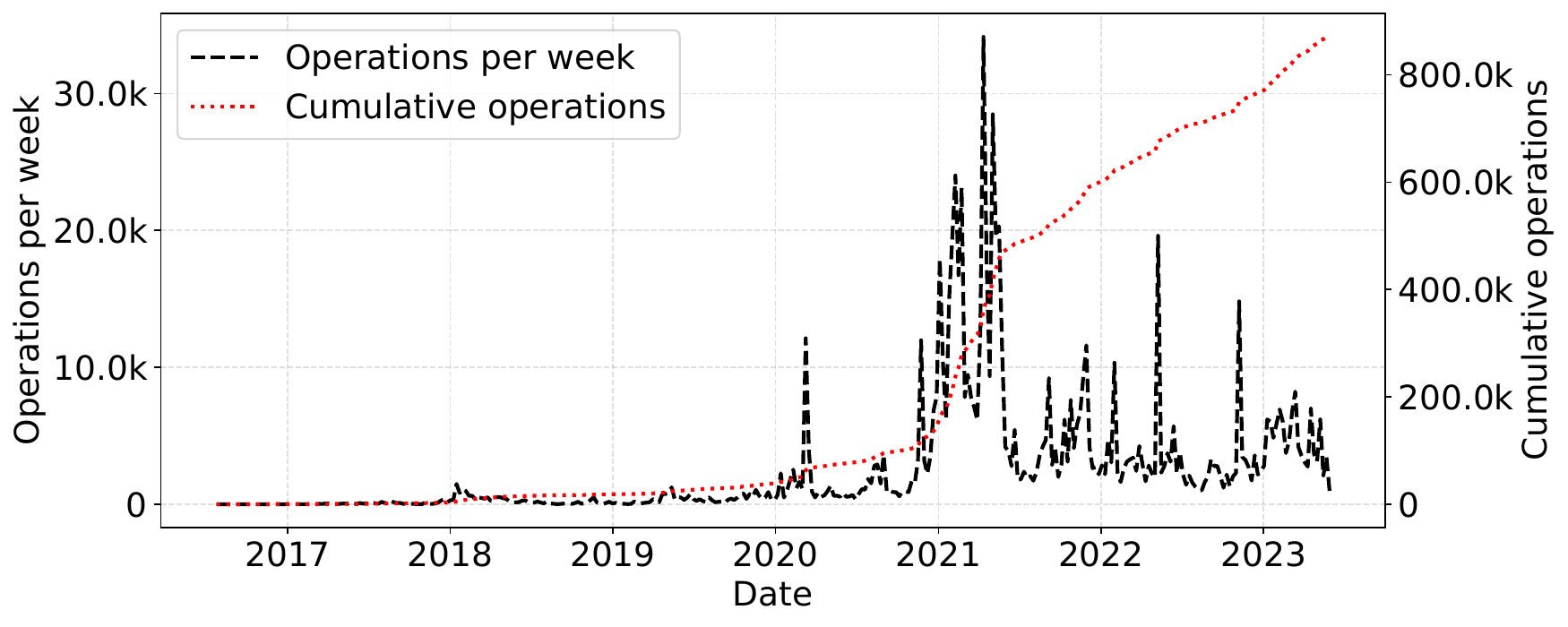}
    \caption{Kraken \MT operations over time}
    \label{fig:kraken_owa_operations_mt}
  \end{subfigure}
  \hfill
    \begin{subfigure}[t]{0.48\textwidth}
    \centering
    \includegraphics[width=\textwidth]{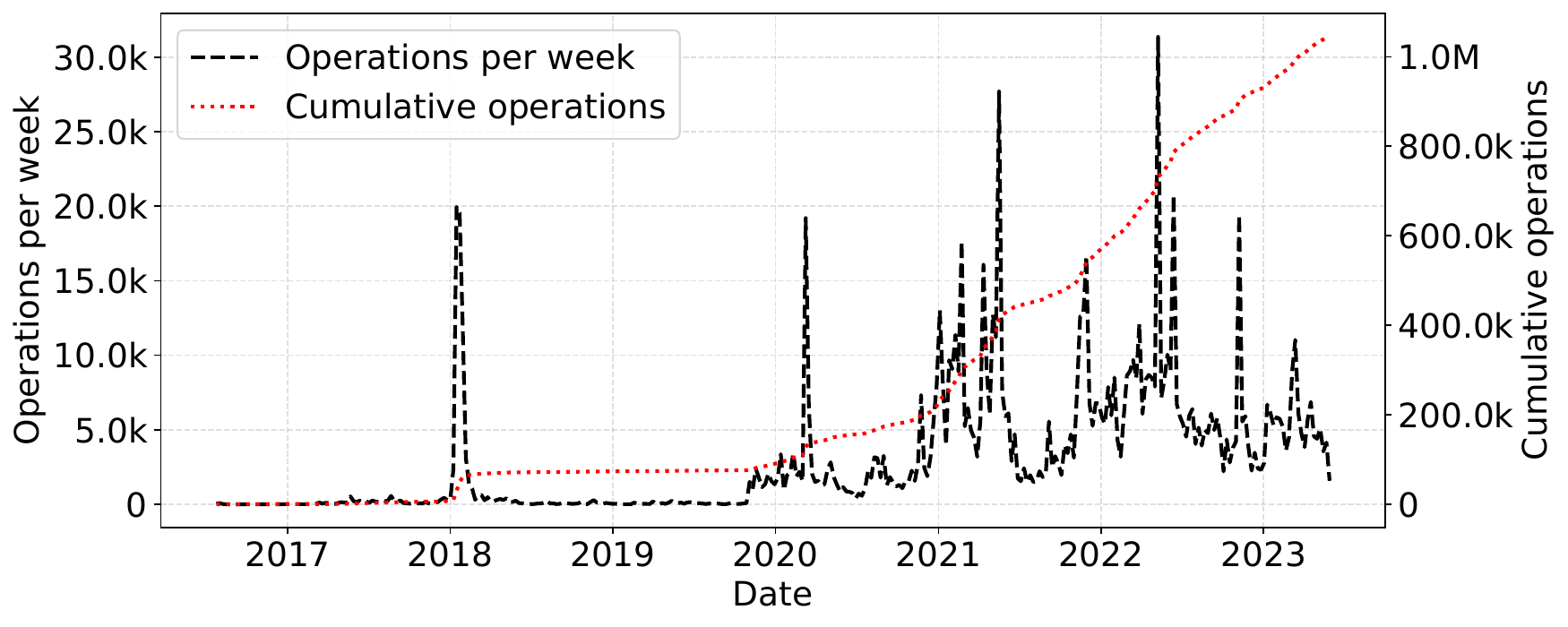}
    \caption{Kraken \TT operations over time}
    \label{fig:kraken_owa_operations_tt}
  \end{subfigure}

  \caption{Longitudinal OWA operations on Binance and Kraken grouped by \MT configurations.}
  \label{fig:longitudinal_operations_all}
  \vspace{3em}
\end{figure*}

\begin{figure*}[t]
  \centering

  \begin{subfigure}[t]{0.48\textwidth}
    \centering
    \includegraphics[width=\textwidth]{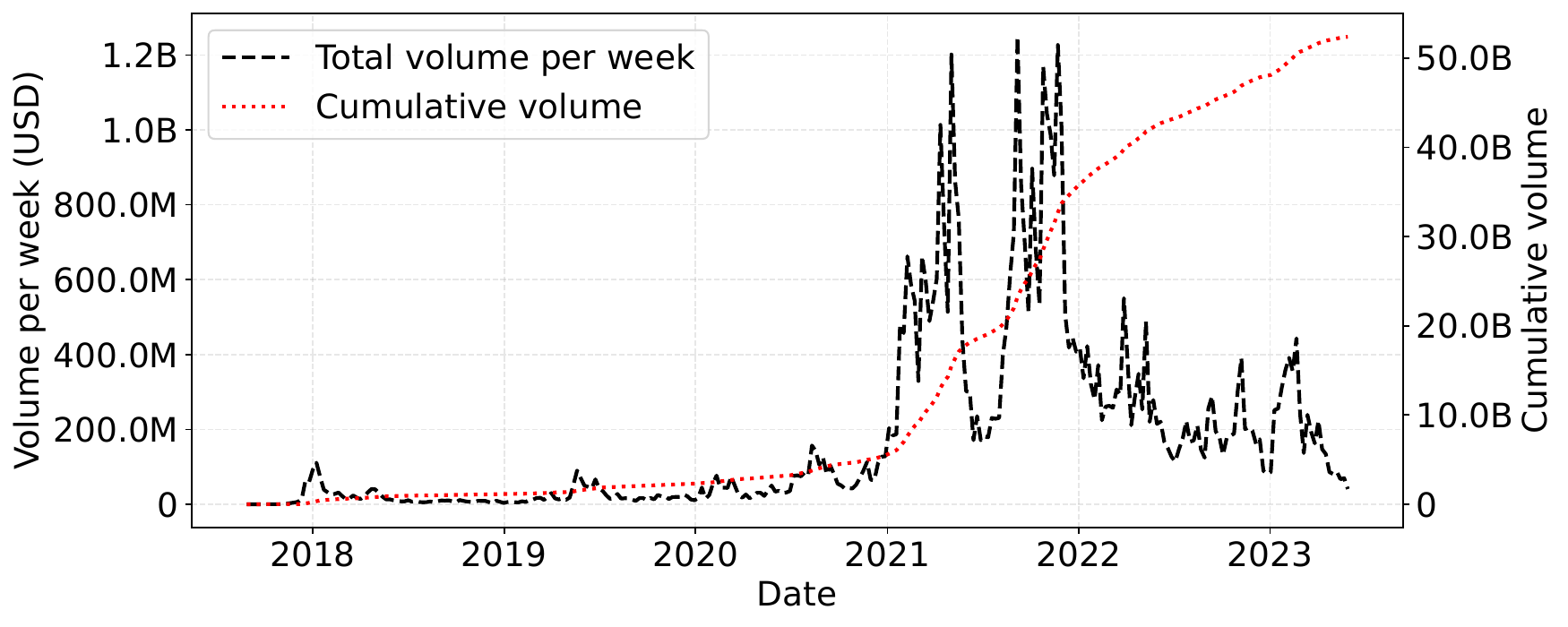}
    \caption{Binance \MT volume over time}
    \label{fig:binance_owa_volume_mt}
  \end{subfigure}
  \hfill
    \begin{subfigure}[t]{0.48\textwidth}
    \centering
    \includegraphics[width=\textwidth]{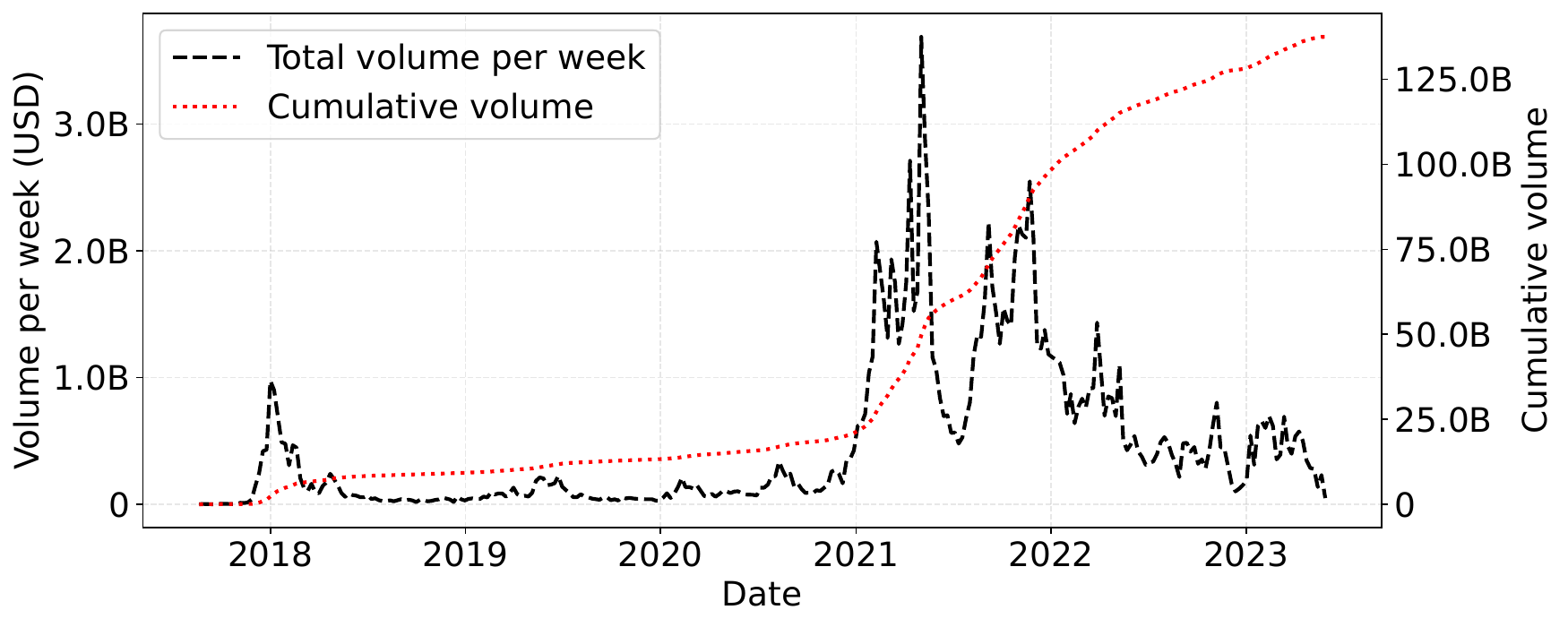}
    \caption{Binance \TT volume over time}
    \label{fig:binance_owa_volume_tt}
  \end{subfigure}

  \vspace{1.5em}

    \begin{subfigure}[t]{0.48\textwidth}
    \centering
    \includegraphics[width=\textwidth]{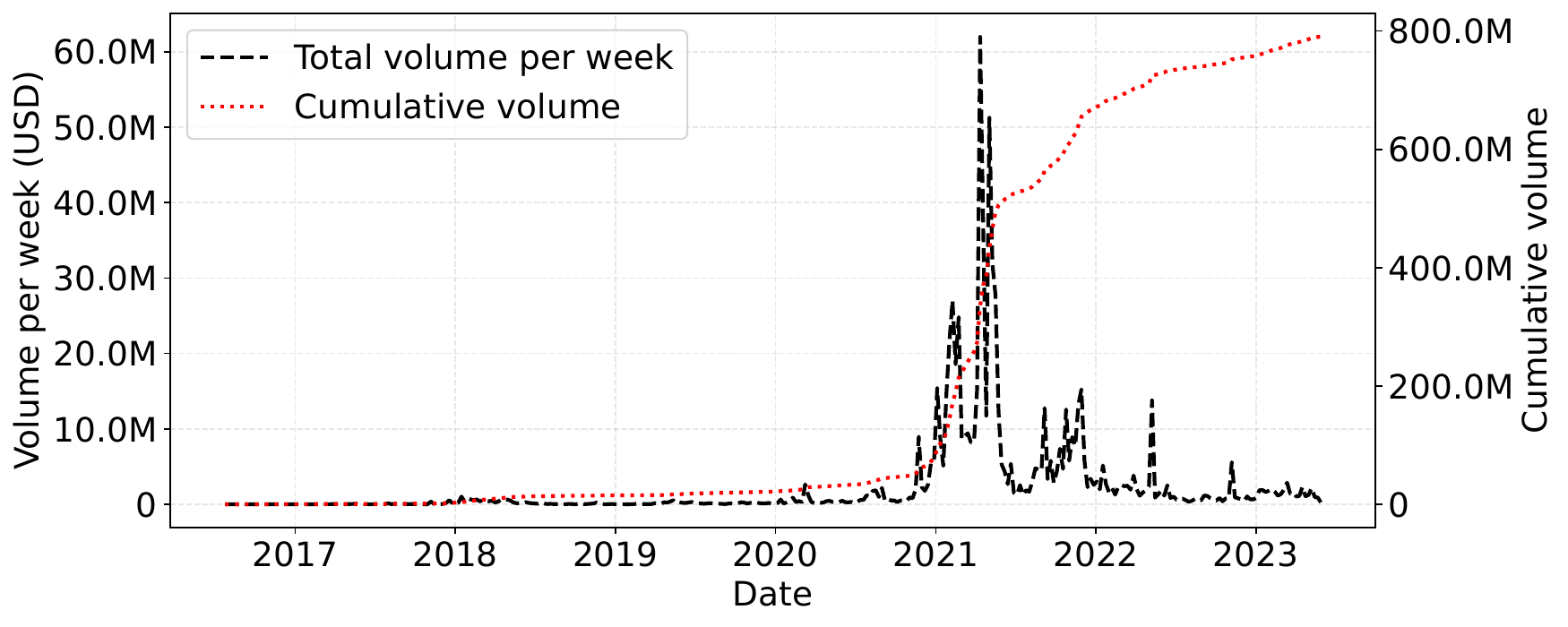}
    \caption{Kraken \MT volume over time}
    \label{fig:kraken_owa_volume_mt}
  \end{subfigure}
  \hfill
  \begin{subfigure}[t]{0.48\textwidth}
    \centering
    \includegraphics[width=\textwidth]{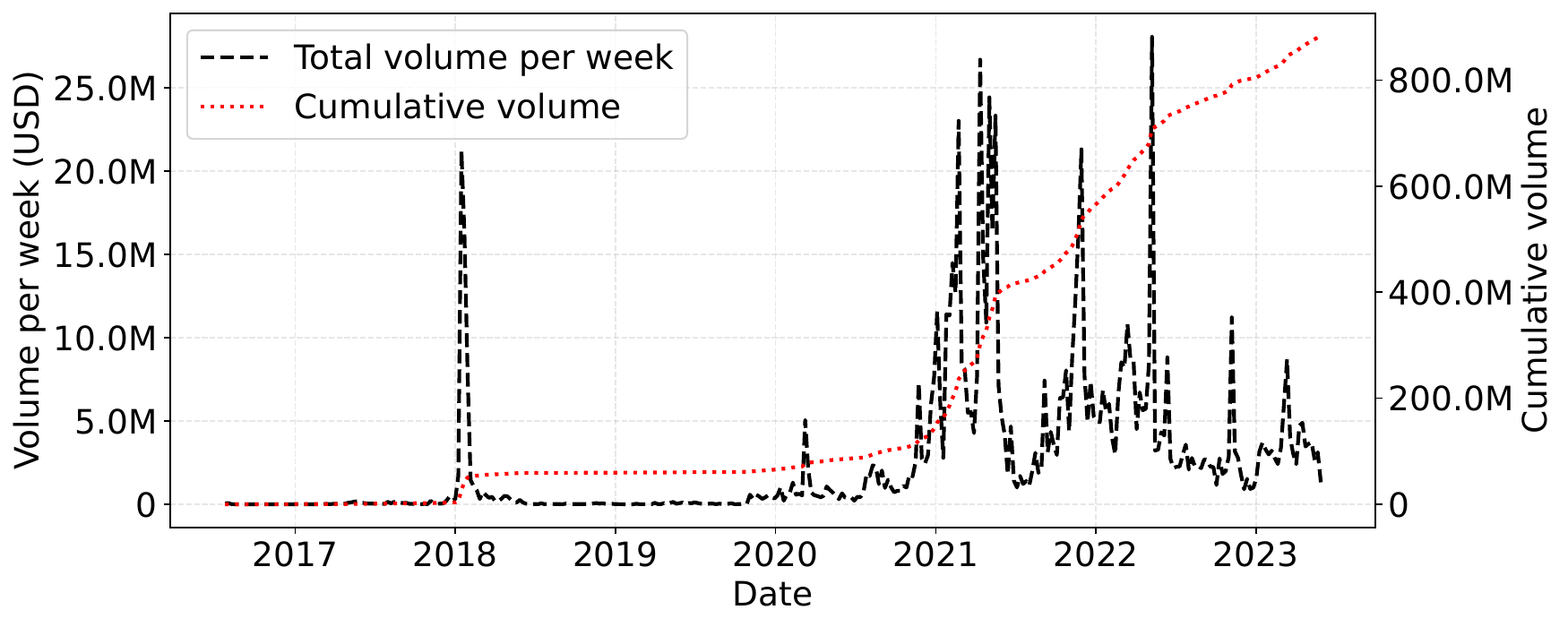}
    \caption{Kraken \TT volume over time}
    \label{fig:kraken_owa_volume_tt}
  \end{subfigure}

  \caption{Longitudinal OWA volume on Binance and Kraken grouped by \MT configurations.}
  \label{fig:longitudinal_volume_all}
\end{figure*}

\begin{figure*}[t]
  \centering

    \begin{subfigure}[t]{0.48\textwidth}
    \centering
    \includegraphics[width=\textwidth]{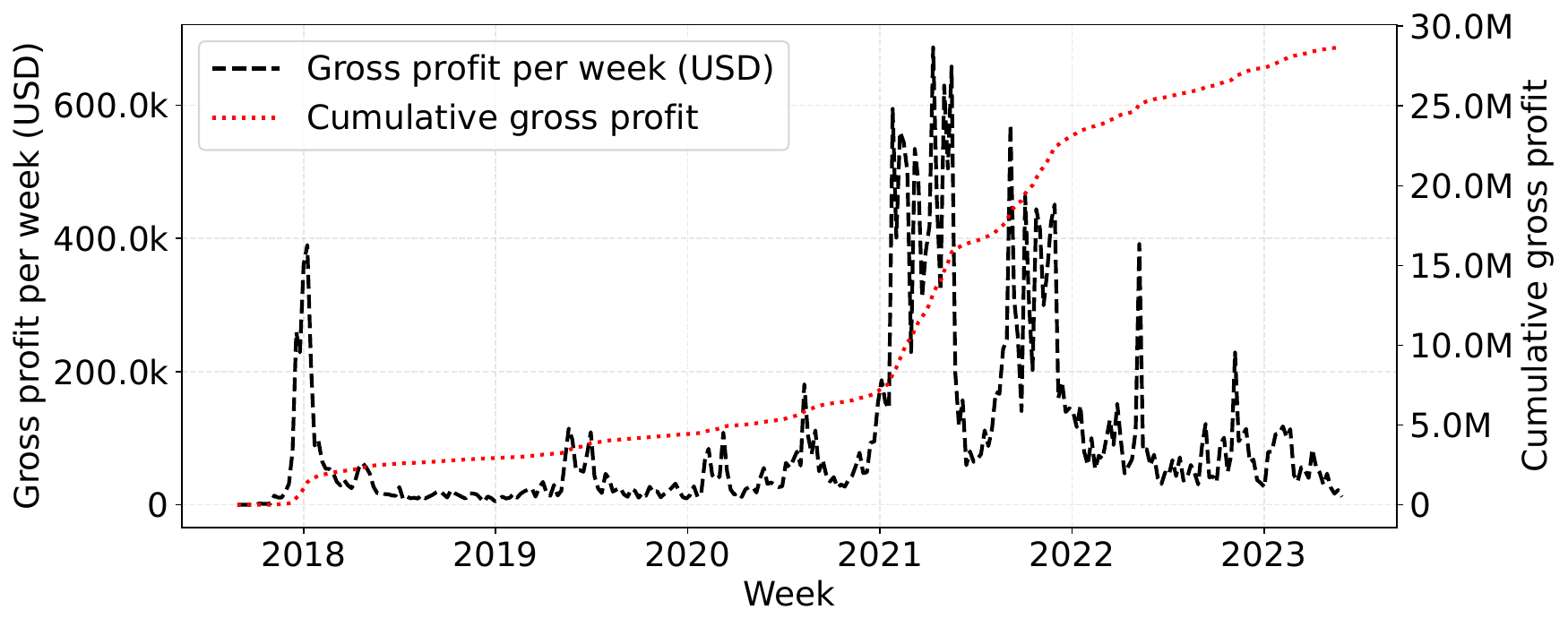}
    \caption{Binance \MT profits over time}
    \label{fig:binance_owa_profits_mt}
  \end{subfigure}
  \hfill
  \begin{subfigure}[t]{0.48\textwidth}
    \centering
    \includegraphics[width=\textwidth]{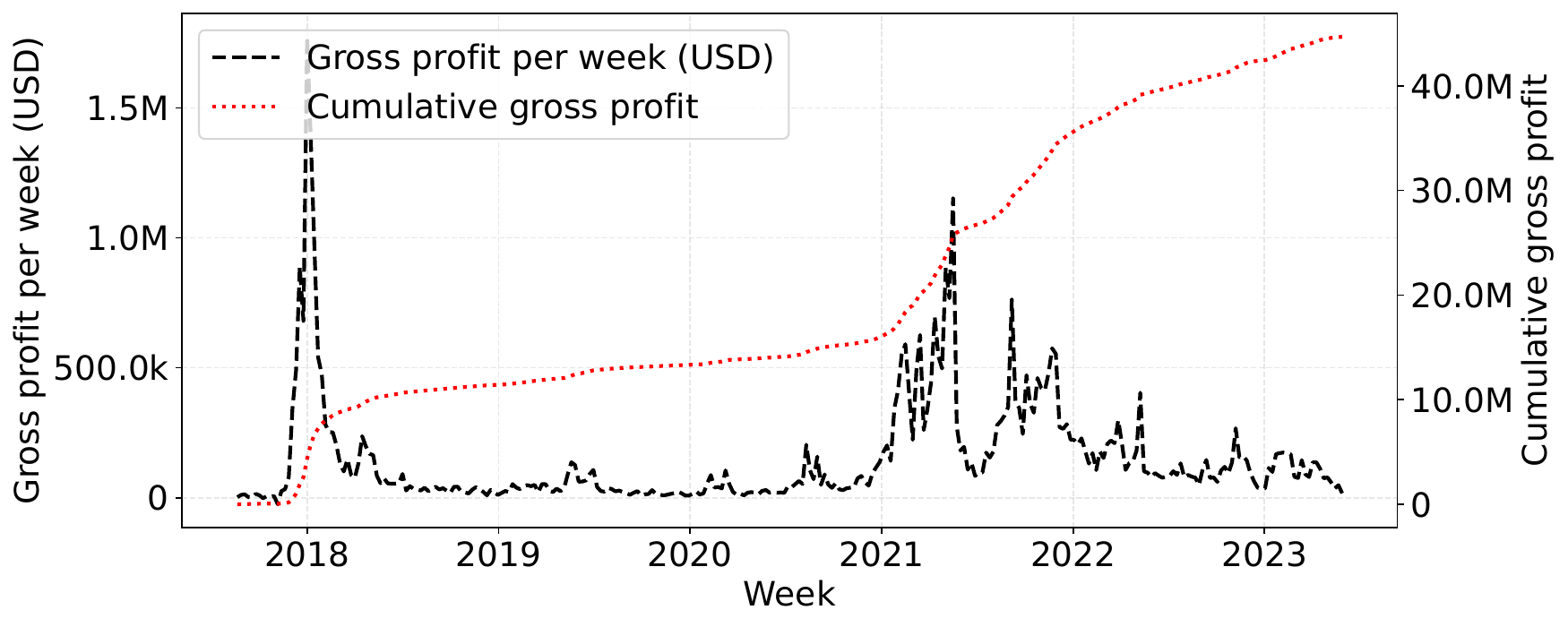}
    \caption{Binance \TT profits over time}
    \label{fig:binance_owa_profits_tt}
  \end{subfigure}

  \vspace{1.5em}

     \begin{subfigure}[t]{0.48\textwidth}
    \centering
    \includegraphics[width=\textwidth]{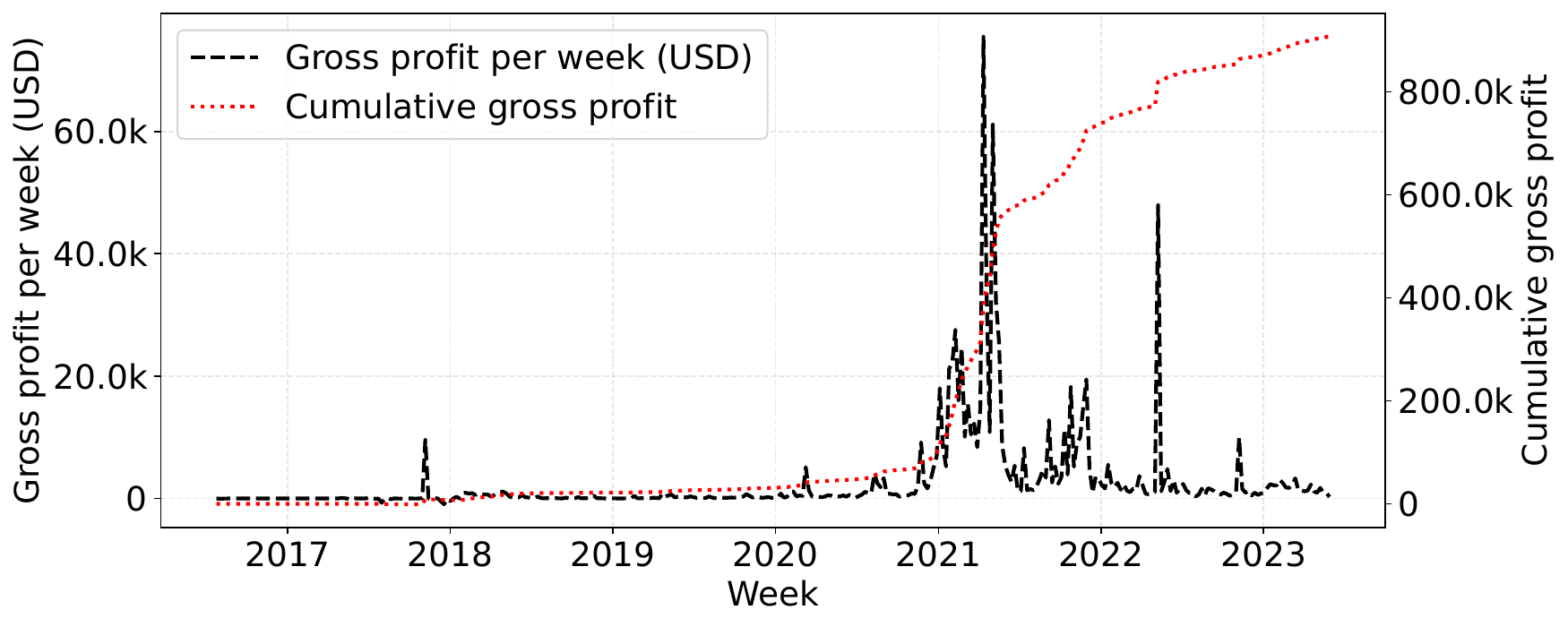}
    \caption{Kraken \MT profits over time}
    \label{fig:kraken_owa_profits_mt}
  \end{subfigure}
  \hfill
  \begin{subfigure}[t]{0.48\textwidth}
    \centering
    \includegraphics[width=\textwidth]{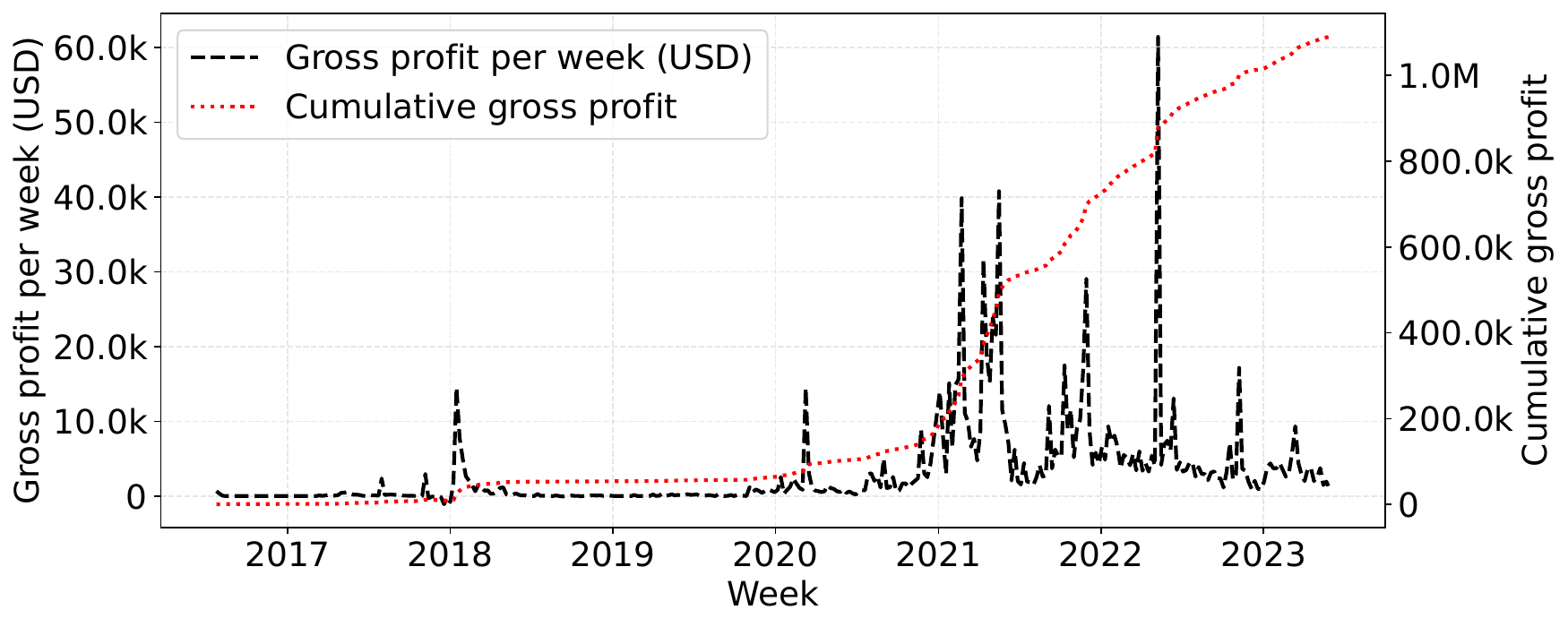}
    \caption{Kraken \TT profits over time}
    \label{fig:kraken_owa_profits_tt}
  \end{subfigure}

  \caption{Longitudinal daily (black line) and cumulative (red dotted line) gross profits over time on Binance and Kraken grouped by \MT configurations.}
  \label{fig:longitudinal_profits_all}
  \vspace{3em}
\end{figure*}

\begin{figure*}[t]
  \centering
  \begin{subfigure}[t]{0.48\textwidth}
    \centering
    \includegraphics[width=\textwidth]{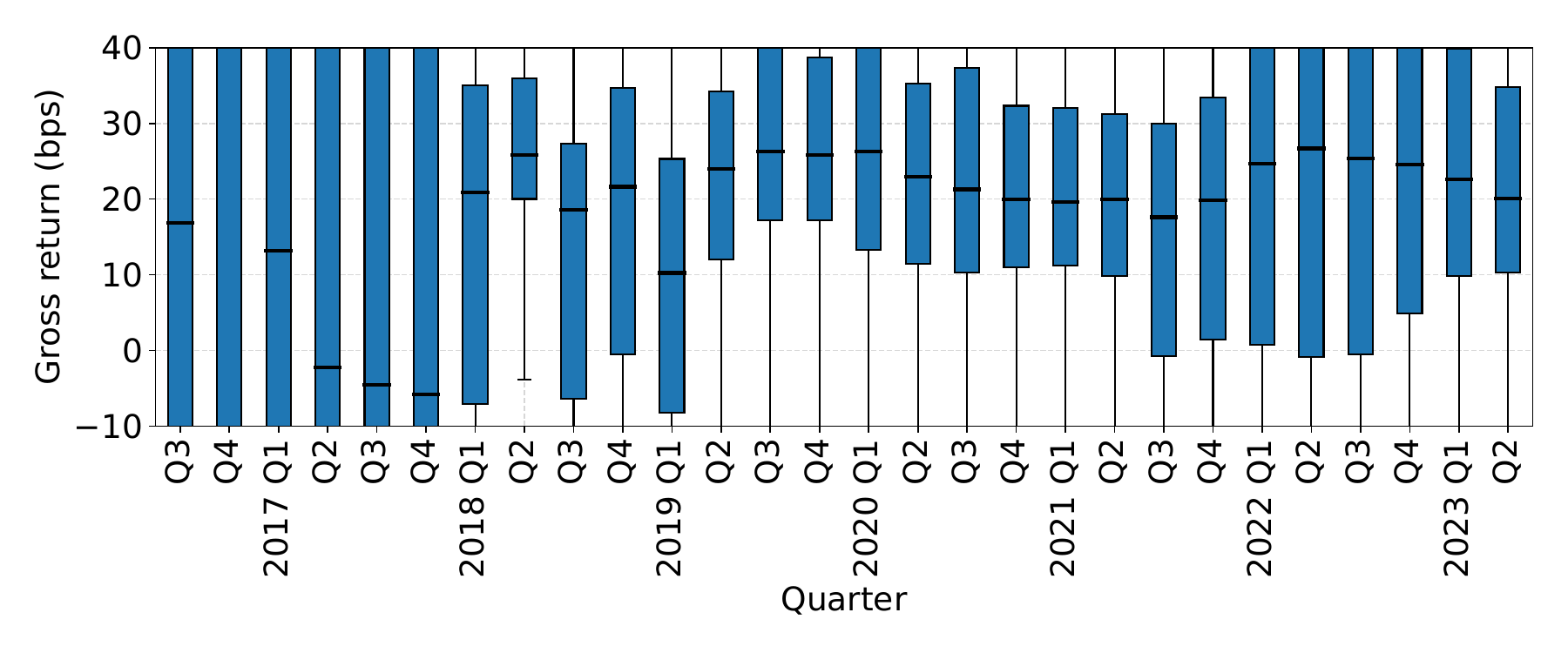}
    \caption{Kraken \MT}
    \label{fig:boxplot_bps_returns_kraken_MT}
  \end{subfigure}
  \hfill
  \begin{subfigure}[t]{0.48\textwidth}
    \centering
    \includegraphics[width=\textwidth]{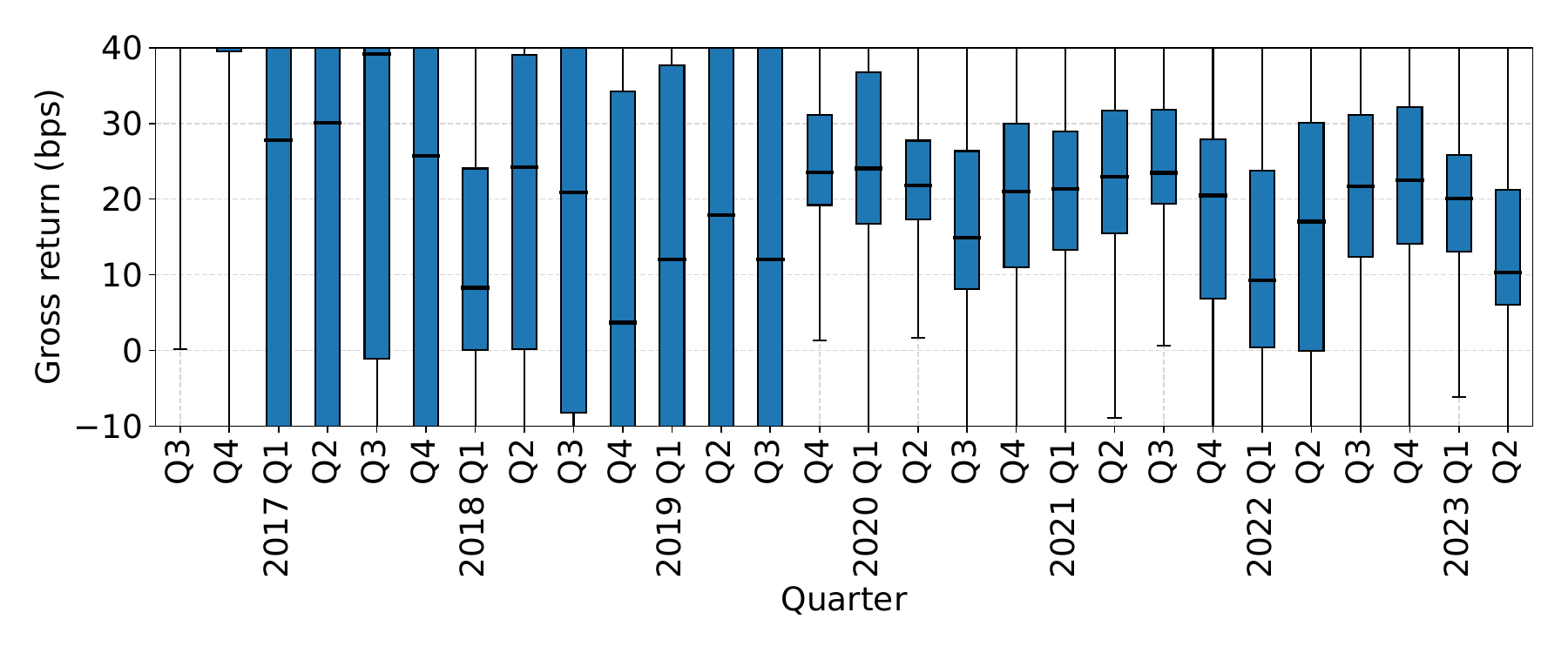}
    \caption{Kraken \TT}
    \label{fig:boxplot_bps_returns_kraken_TT}
  \end{subfigure}
  
  \caption{Quarterly distribution of price discrepancy per OWA sequence on Kraken (excluding fees). Contrary to Binance (Figure~\ref{fig:boxplot_bps_returns_binance}), Kraken does not show a clear pattern over time.
}
  \label{fig:boxplot_bps_returns_kraken}
\end{figure*}

\begin{figure*}[t]
  \centering

    \begin{subfigure}[t]{0.48\textwidth}
    \centering
    \includegraphics[width=\textwidth]{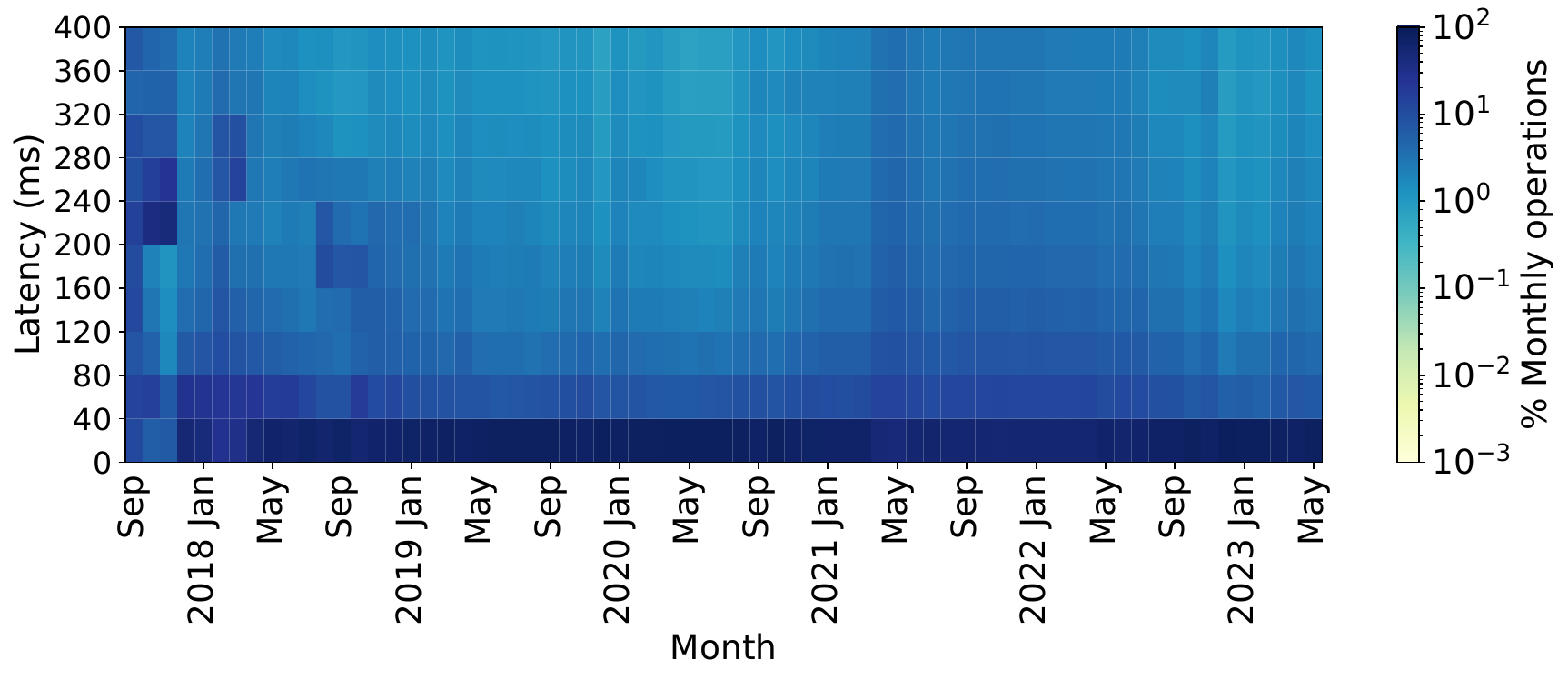}
    \caption{Binance \MT operations over time}
    \label{fig:binance_owa_operations_per_bin_mt}
  \end{subfigure}
  \hfill
  \begin{subfigure}[t]{0.48\textwidth}
    \centering
    \includegraphics[width=\textwidth]{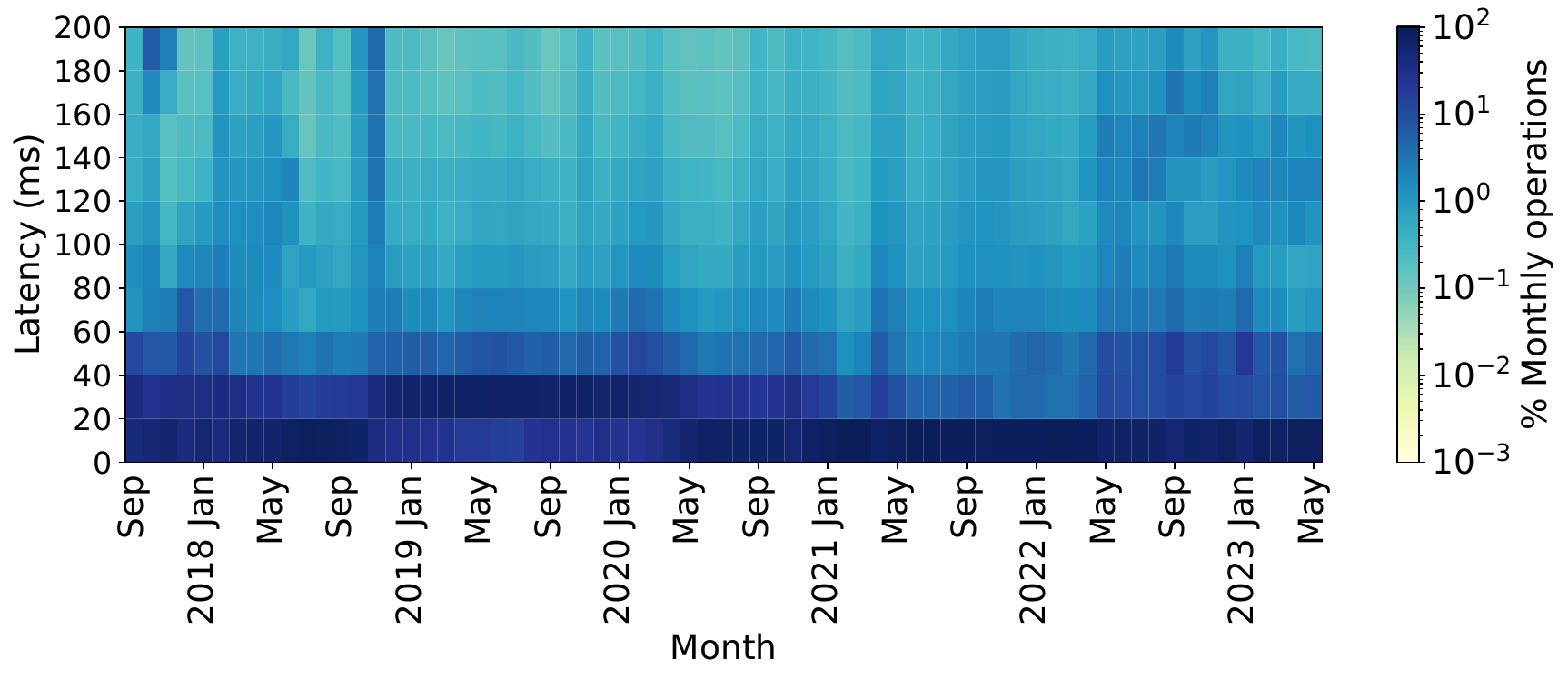}
    \caption{Binance \TT operations over time}
    \label{fig:binance_owa_operations_per_bin_tt}
  \end{subfigure}

  \vspace{1.5em}

  \begin{subfigure}[t]{0.48\textwidth}
    \centering
    \includegraphics[width=\textwidth]{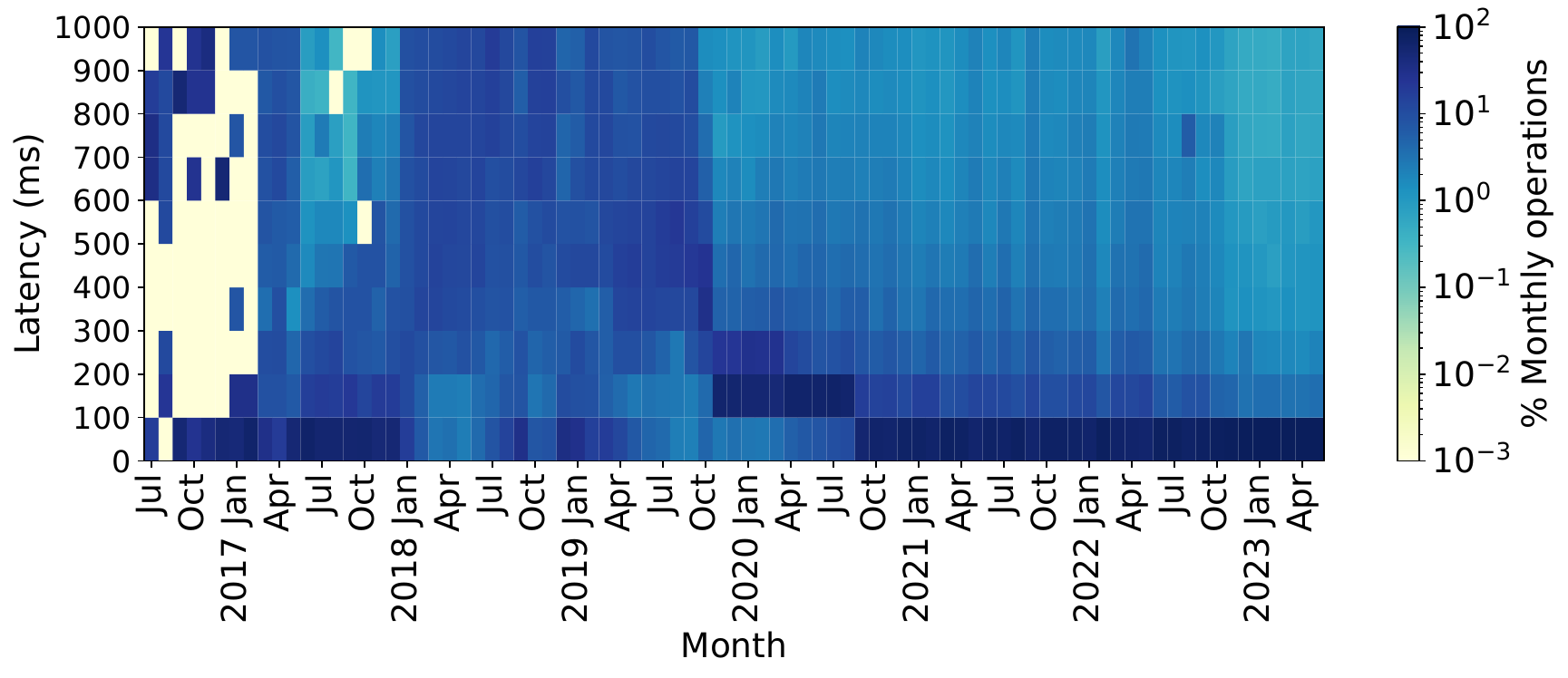}
    \caption{Kraken \MT operations over time}
    \label{fig:kraken_owa_operations_per_bin_mt}
  \end{subfigure}
  \hfill
    \begin{subfigure}[t]{0.48\textwidth}
    \centering
    \includegraphics[width=\textwidth]{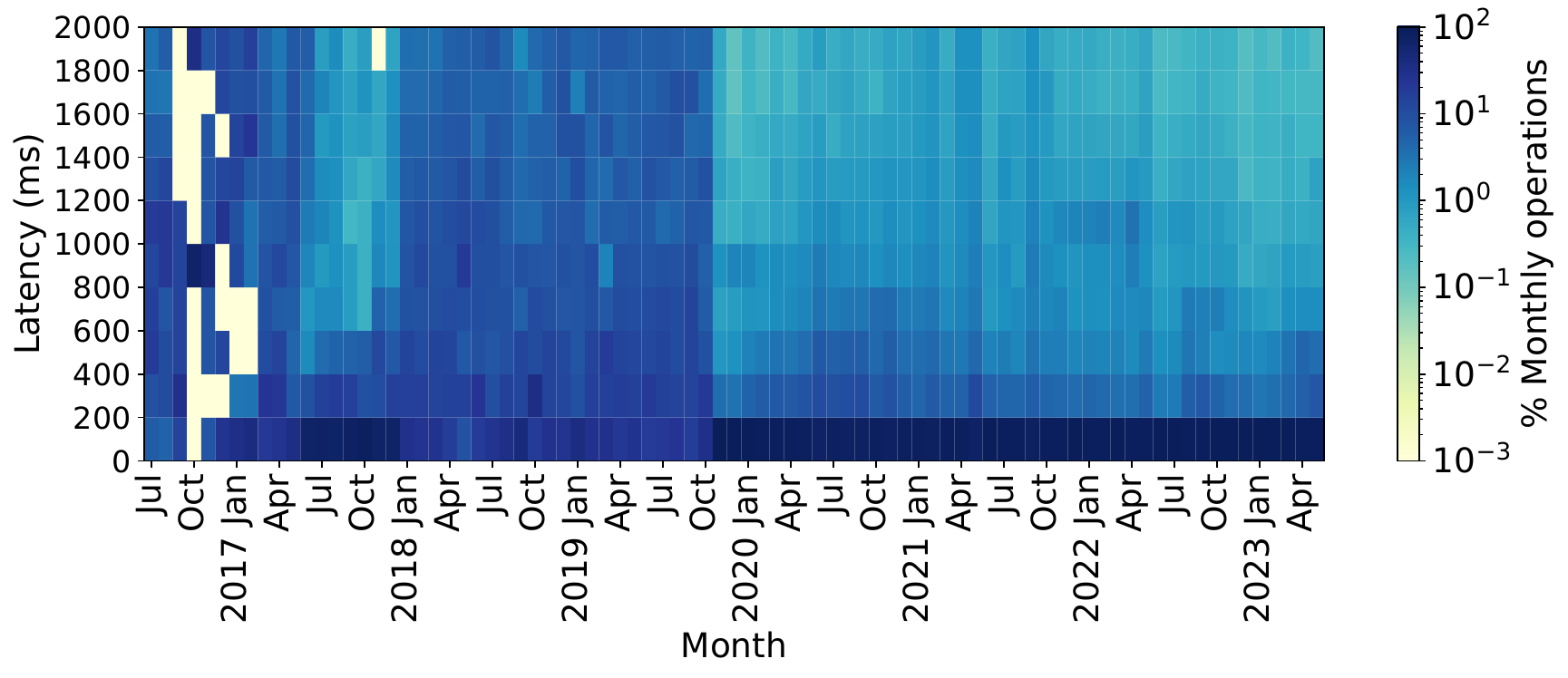}
    \caption{Kraken \TT operations over time}
    \label{fig:kraken_owa_operations_per_bin_tt}
  \end{subfigure}

  \caption{Heatmaps of monthly OWA operations by latency between the first and second operation of the sequence. }
  \label{fig:heatmaps_owa_operations_latency_all}
  \vspace{3em}
\end{figure*}

\begin{figure*}[t]
  \centering
  \begin{subfigure}[t]{0.48\textwidth}
    \centering
    \includegraphics[width=\textwidth]{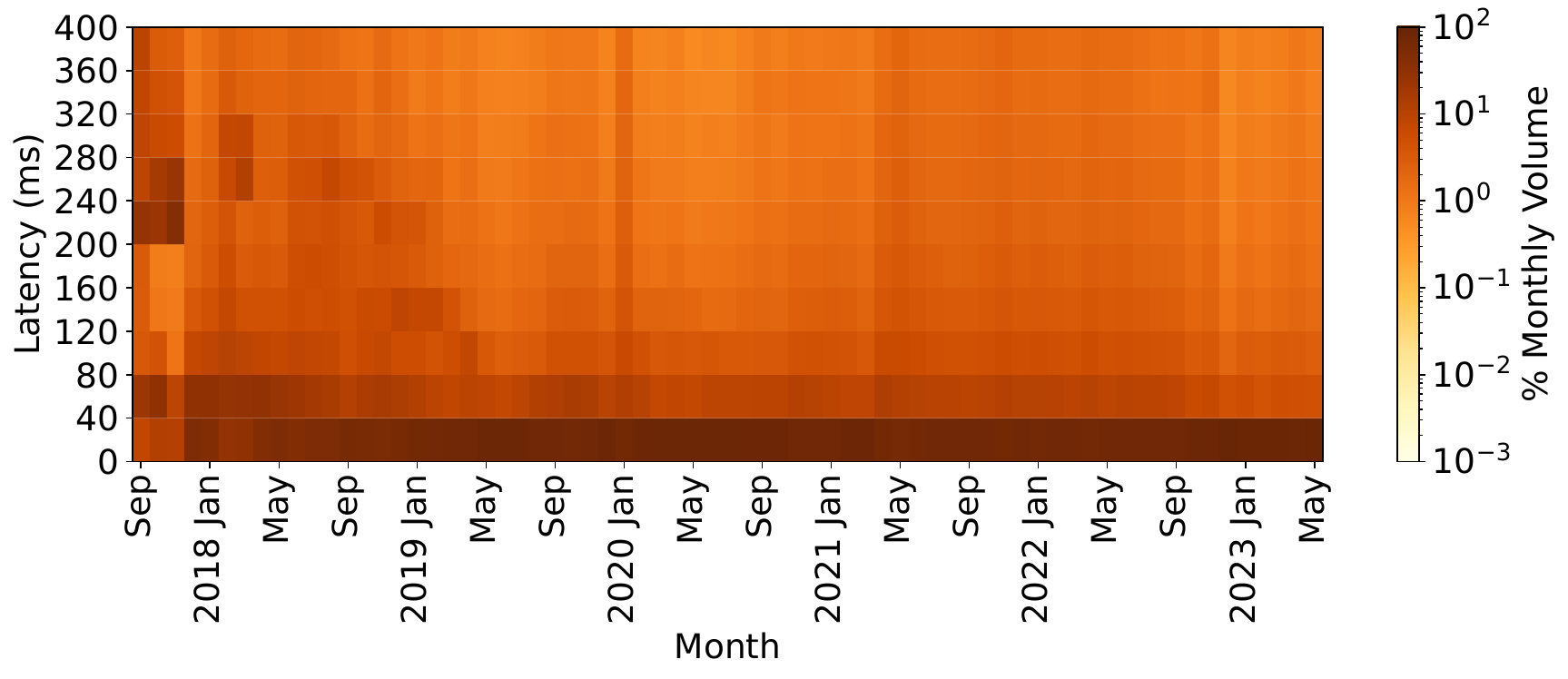}
    \caption{Binance \MT}
    \label{fig:binance_owa_volume_per_bin_mt}
  \end{subfigure}
  \hfill
  \begin{subfigure}[t]{0.48\textwidth}
    \centering
    \includegraphics[width=\textwidth]{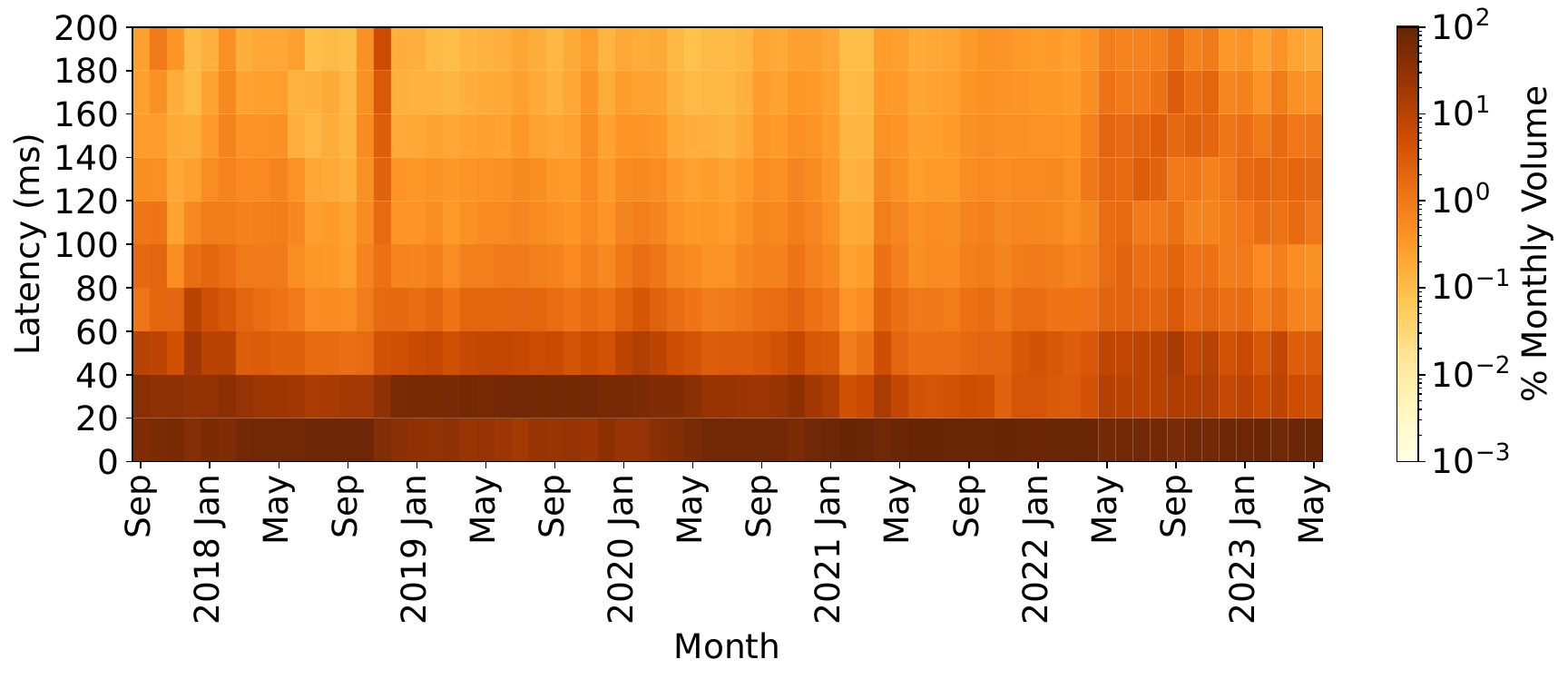}
    \caption{Binance \TT}
    \label{fig:binance_owa_volume_per_bin_tt}
  \end{subfigure}

  \vspace{1.5em}

  \begin{subfigure}[t]{0.48\textwidth}
    \centering
    \includegraphics[width=\textwidth]{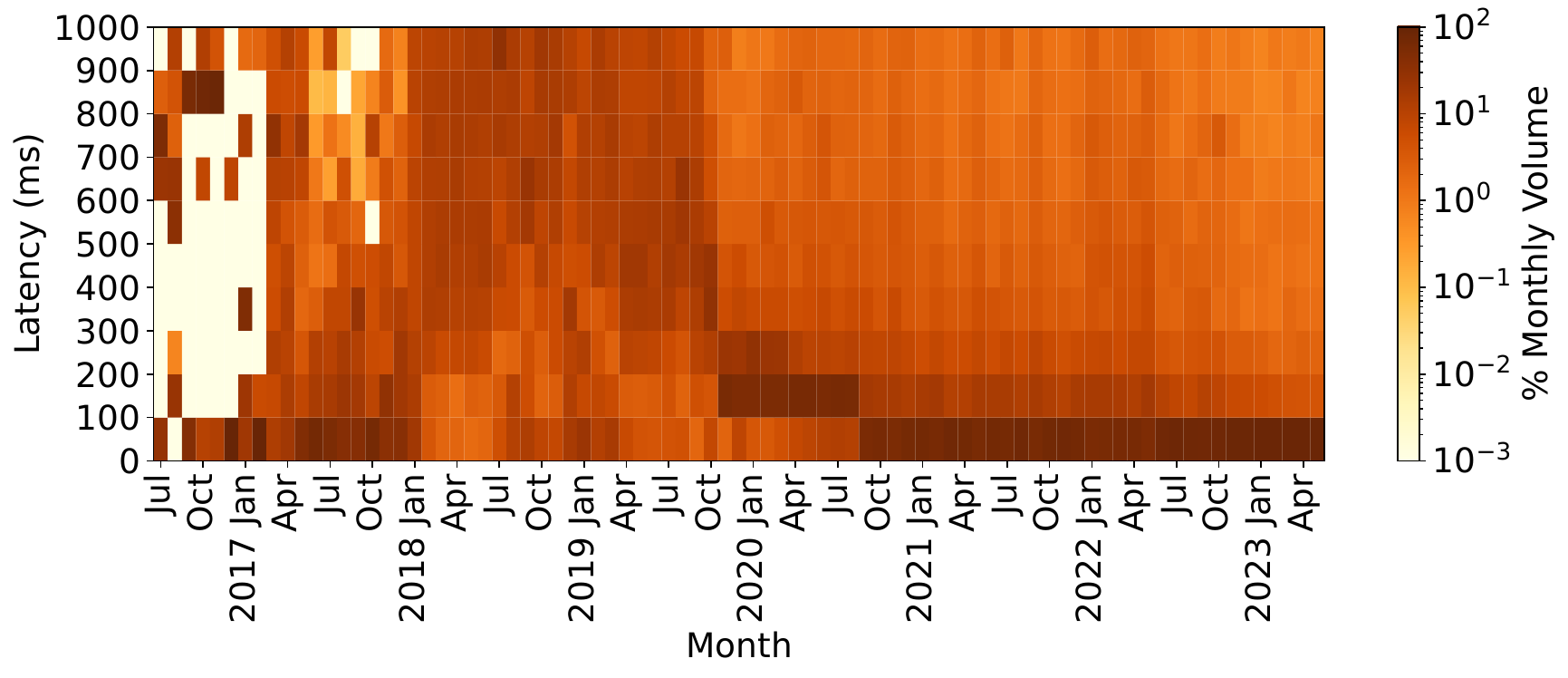}
    \caption{Kraken \MT}
    \label{fig:kraken_owa_volume_per_bin_mt}
  \end{subfigure}
  \hfill
  \begin{subfigure}[t]{0.48\textwidth}
    \centering
    \includegraphics[width=\textwidth]{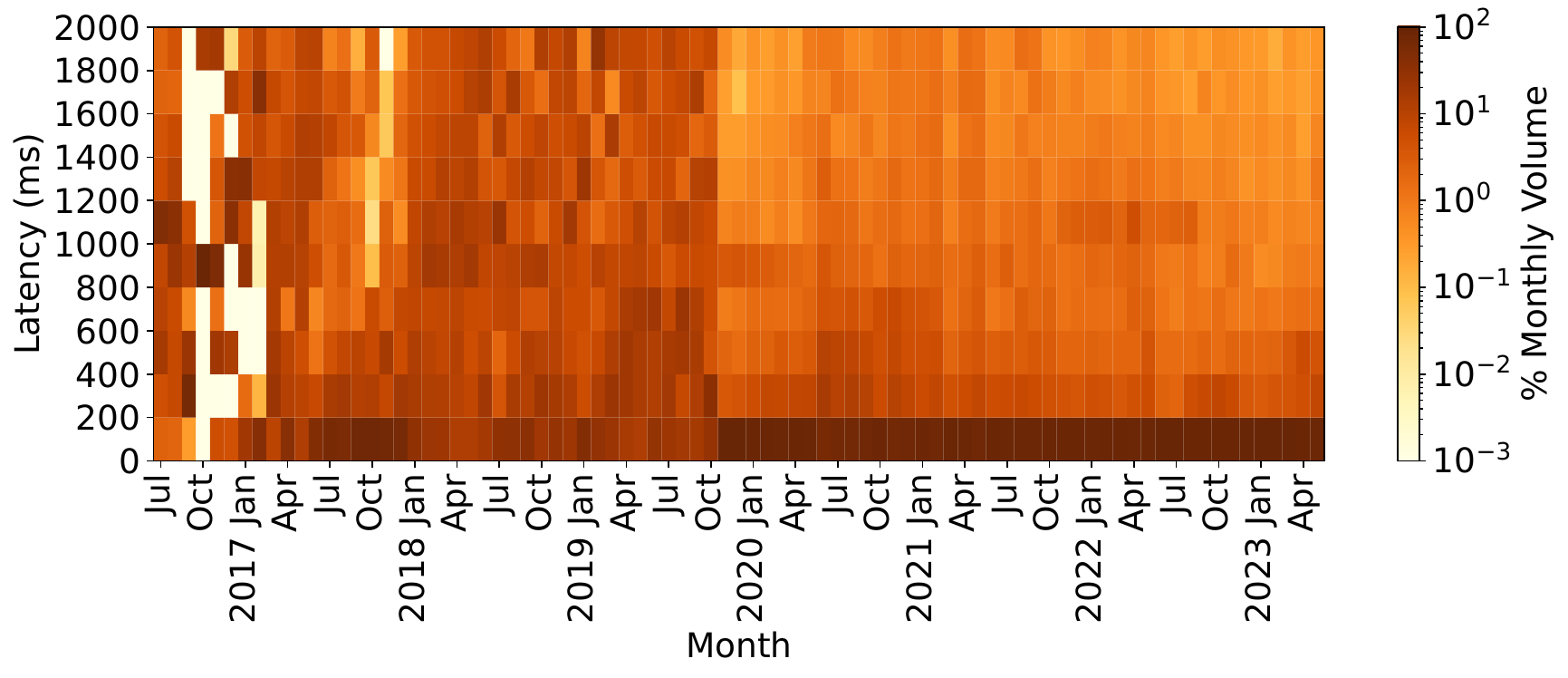}
    \caption{Kraken \TT}
    \label{fig:kraken_owa_volume_per_bin_tt}
  \end{subfigure}

  \caption{Heatmaps of monthly OWA volume by latency between the first and second operation of the sequence.}
  \label{fig:heatmaps_owa_volume_latency_all}
\end{figure*}

%% file: biblio.bib
@article{deardorff1979one,
  title={One-way arbitrage and its implications for the foreign exchange markets},
  author={Deardorff, Alan V},
  journal={Journal of Political Economy},
  volume={87},
  number={2},
  pages={351--364},
  year={1979},
  publisher={The University of Chicago Press}
}

@online{binance_coinmarketcap,
  author = {CoinMarketCap},
  title = {Binance exchange},
  year = {2025},
  url = {https://coinmarketcap.com/exchanges/binance/},
  lastaccessed = {2025-04-02}
}

@online{kraken_coinmarketcap,
  author = {CoinMarketCap},
  title = {Kraken exchange},
  year = {2025},
  url = {https://coinmarketcap.com/exchanges/kraken/},
  lastaccessed = {2025-04-02}
}

@online{cexes_vs_dexes,
  author = {OAK Research},
  title = {CEX vs DEX: Report from 2024 and forecasts for 2025},
  year = {2025},
  url = {https://oakresearch.io/en/reports/sectors/cex-vs-dex-report-2024-forecasts-2025},
  lastaccessed = {2025-08-25}
}

@online{exchanges_rank,
  author = {CoinMarketCap},
  title = {Top cryptocurrency spot exchanges},
  year = {2025},
  url = {https://coinmarketcap.com/rankings/exchanges/},
  lastaccessed = {2025-08-25}
}

@article{makarov2020trading,
  title={Trading and arbitrage in cryptocurrency markets},
  author={Makarov, Igor and Schoar, Antoinette},
  journal={Journal of Financial Economics},
  volume={135},
  number={2},
  pages={293--319},
  year={2020},
  publisher={Elsevier}
}

@article{crepelliere2023arbitrage,
  title={Arbitrage in the market for cryptocurrencies},
  author={Cr{\'e}pelli{\`e}re, Tommy and Pelster, Matthias and Zeisberger, Stefan},
  journal={Journal of Financial Markets},
  volume={64},
  pages={100817},
  year={2023},
  publisher={Elsevier}
}

@article{muck2025wish,
  title={Wish or reality? On the exploitability of triangular arbitrage in cryptocurrency markets},
  author={Muck, Matthias and Schmidl, Thomas and Wolf, Julian},
  journal={Finance Research Letters},
  volume={73},
  pages={106508},
  year={2025},
  publisher={Elsevier}
}

@article{nan2019bitcoin,
  title={Bitcoin-based triangular arbitrage with the Euro/US dollar as a foreign futures hedge: Modeling with a bivariate GARCH model},
  author={Nan, Zheng and Kaizoji, Taisei and Kaizoji, T},
  journal={Quantitative Finance and Economics},
  volume={3},
  number={2},
  pages={347--365},
  year={2019},
  publisher={American Institute of Mathematical Sciences (AIMS)}
}

@inproceedings{qin2022quantifying,
  title={Quantifying blockchain extractable value: How dark is the forest?},
  author={Qin, Kaihua and Zhou, Liyi and Gervais, Arthur},
  booktitle={Symposium on Security and Privacy (S\&P)},
  year={2022}
}

@inproceedings{wang2022cyclic,
  title={Cyclic arbitrage in decentralized exchanges},
  author={Wang, Ye and Chen, Yan and Wu, Haotian and Zhou, Liyi and Deng, Shuiguang and Wattenhofer, Roger},
  booktitle={Web Conference (WWW)},
  year={2022}
}

@inproceedings{heimbach2024non,
  title={Non-atomic arbitrage in decentralized finance},
  author={Heimbach, Lioba and Pahari, Vabuk and Schertenleib, Eric},
  booktitle={Symposium on Security and Privacy (S\&P)},
  year={2024}
}

@inproceedings{mclaughlin2023large,
  title={A large scale study of the {Ethereum} arbitrage ecosystem},
  author={McLaughlin, Robert and Kruegel, Christopher and Vigna, Giovanni},
  booktitle={USENIX Security Symposium},
  year={2023}
}

@inproceedings{yousaf2019tracing,
  title={Tracing transactions across cryptocurrency ledgers},
  author={Yousaf, Haaroon and Kappos, George and Meiklejohn, Sarah},
  booktitle={USENIX Security Symposium},
  year={2019}
}

@inproceedings{kappos2018empirical,
  title={An empirical analysis of anonymity in {Zcash}},
  author={Kappos, George and Yousaf, Haaroon and Maller, Mary and Meiklejohn, Sarah},
  booktitle={USENIX Security Symposium},
  year={2018}
}

@article{quesnelle2017linkability,
  title={On the linkability of Zcash transactions},
  author={Quesnelle, Jeffrey},
  journal={arXiv preprint arXiv:1712.01210},
  year={2017}
}

@article{moser2017empirical,
  title={An empirical analysis of traceability in the {Monero} blockchain},
  author={M{\"o}ser, Malte and Soska, Kyle and Heilman, Ethan and Lee, Kevin and Heffan, Henry and Srivastava, Shashvat and Hogan, Kyle and Hennessey, Jason and Miller, Andrew and Narayanan, Arvind and others},
  journal={arXiv preprint arXiv:1704.04299},
  year={2017}
}

@inproceedings{kumar2017traceability,
  title={A traceability analysis of {Monero's} blockchain},
  author={Kumar, Amrit and Fischer, Cl{\'e}ment and Tople, Shruti and Saxena, Prateek},
  booktitle={European Symposium on Research in Computer Security (ESORICS)},
  year={2017}
}

@inproceedings{morin2023cryptocurrency,
  title={How cryptocurrency exchange interruptions create arbitrage opportunities},
  author={Morin, Andrew and Moore, Tyler},
  booktitle={European Symposium on Security and Privacy Workshops (EuroS\&PW)},
  year={2023}
}

@inproceedings{daian2020flash,
  title={Flash boys 2.0: Frontrunning in decentralized exchanges, miner extractable value, and consensus instability},
  author={Daian, Philip and Goldfeder, Steven and Kell, Tyler and Li, Yunqi and Zhao, Xueyuan and Bentov, Iddo and Breidenbach, Lorenz and Juels, Ari},
  booktitle={Symposium on Security and Privacy (S\&P)},
  year={2020}
}

@article{gromb2010limits,
  title={Limits of arbitrage},
  author={Gromb, Denis and Vayanos, Dimitri},
  journal={Annu. Rev. Financ. Econ.},
  volume={2},
  number={1},
  pages={251--275},
  year={2010},
  publisher={Annual Reviews}
}

@article{czaplinski2019using,
  title={Using FIAT currencies to arbitrage on cryptocurrency exchanges.},
  author={Czapli{\'n}ski, Tomasz and Nazmutdinova, Elena},
  journal={Journal of International Studies (2071-8330)},
  volume={12},
  number={1},
  year={2019}
}

@article{pieters2017financial,
  title={Financial regulations and price inconsistencies across Bitcoin markets},
  author={Pieters, Gina and Vivanco, Sofia},
  journal={Information Economics and Policy},
  volume={39},
  pages={1--14},
  year={2017},
  publisher={Elsevier}
}

@article{chang2002one,
  title={One-way arbitrage-based interest parity: An application of the Fletcher-Taylor approach in short-date markets},
  author={Chang, Rosita P and Lee, Sang-Hyop and Reid, Sean F and Rhee, S Ghon},
  journal={Tinbergen Institute Discussion Papers},
  number={2002-115},
  pages={2},
  year={2002}
}

@inproceedings{meiklejohn2013fistful,
  title={A fistful of Bitcoins: Characterizing payments among men with no names},
  author={Meiklejohn, Sarah and Pomarole, Marjori and Jordan, Grant and Levchenko, Kirill and McCoy, Damon and Voelker, Geoffrey M and Savage, Stefan},
  booktitle={Internet Measurement Conference (IMC)},
  year={2013}
}

@article{grimberg2020empirical,
  title={Empirical analysis of indirect internal conversions in cryptocurrency exchanges},
  author={Grimberg, Paz and Lauinger, Tobias and McCoy, Damon},
  journal={arXiv preprint arXiv:2002.12274},
  year={2020}
}
